\documentclass[superscriptaddress,preprintnumbers,amsmath,amssymb,prb,longbibliography,twocolumn]{revtex4-1}
\usepackage[utf8]{inputenc}
\usepackage{enumitem}
\usepackage{graphicx,amsmath,color,multirow}
\def\k{\kappa}

\def\a{\alpha}
\def\w{\omega}

\def\bk{{\bf k}}

\def\bq{{\bf q}}
\def\bQ{{\bf Q}}
\def\bg{{\bf g}}

\def\ve{\varepsilon}

\def\<{\langle}
\def\>{\rangle}
\def\D{\partial}

\let\hide\iffalse
\usepackage{braket} 

\usepackage[caption=false]{subfig}

\begin{document}

\title{Roadmap for electronic structure, anharmonicity, and electron-phonon calculations in locally disordered 
inorganic and hybrid halide perovskites} 

\author{Marios Zacharias}
\email{zachariasmarios@gmail.com}
\affiliation{Univ Rennes, INSA Rennes, CNRS, Institut FOTON - UMR 6082, F-35000 Rennes, France}
\affiliation{Computation-based Science and Technology Research Center, The Cyprus Institute, Aglantzia 2121, Nicosia, Cyprus}
\author{George Volonakis}
\affiliation{Univ Rennes, ENSCR, INSA Rennes, CNRS, ISCR - UMR 6226, F-35000 Rennes, France}
\author{Laurent Pedesseau}
\affiliation{Univ Rennes, INSA Rennes, CNRS, Institut FOTON - UMR 6082, F-35000 Rennes, France}
\author{Claudine Katan}
\affiliation{Univ Rennes, ENSCR, INSA Rennes, CNRS, ISCR - UMR 6226, F-35000 Rennes, France}
\author{Feliciano Giustino}
\affiliation{ Oden Institute for Computational Engineering and Sciences, The University of Texas at Austin,
Austin, Texas 78712, USA
}%
\affiliation{Department of Physics, The University of Texas at Austin, Austin, Texas 78712, USA}
\author{Jacky Even}
\email{jacky.even@insa-rennes.fr}
\affiliation{Univ Rennes, INSA Rennes, CNRS, Institut FOTON - UMR 6082, F-35000 Rennes, France}

\date{\today}

\begin{abstract}
The role of data in modern materials science becomes more valuable and accurate when
effects such as electron-phonon coupling and anharmonicity are included, providing a more
realistic representation of finite-temperature material behavior. Furthermore, positional
polymorphism, characterized by correlated local atomic disorder usually not reported by standard
diffraction techniques, is a critical yet underexplored factor in understanding the
electronic structure and transport properties of energy-efficient materials, like halide perovskites. 
In this manuscript, we present a first-principles methodology for locally disordered (polymorphous) cubic 
inorganic and hybrid halide perovskites, rooted in the special displacement method, that offers a systematic 
and alternative approach to molecular dynamics for exploring finite-temperature properties. 
By enabling a unified and efficient treatment of anharmonic lattice dynamics, electron-phonon coupling, and positional polymorphism, 
our approach generates essential data to predict temperature-dependent phonon properties, free energies, band gaps, 
and effective masses. Designed with a high-throughput spirit, this framework has been
applied across a range of inorganic and hybrid halide perovskites: 
CsPbI$_3$, CsPbBr$_3$, CsSnI$_3$, CsPbCl$_3$, MAPbI$_3$, MAPbBr$_3$, MASnI$_3$, MAPbCl$_3$,
FAPbI$_3$, FAPbBr$_3$, FASnI$_3$, and FAPbCl$_3$. We provide a comprehensive comparison between theoretical 
and experimental results and we systematically uncover trends and insights into their electronic 
and thermal behavior. For all compounds, we demonstrate strong and consistent correlations between local structural disorder, band gap openings, 
and effective mass enhancements. We present anharmonic phonon quasiparticle dispersions and show that both electron-phonon coupling 
and thermal expansion contributions to the band gap are significantly affected by local disorder, improving agreement with experiment. 
Furthermore, we discuss deviations from these trends in Sn-based compounds, which we attribute to the enhanced stereochemical activity 
of the Sn lone pair. We also present phonon spectral functions of polymorphous structures 
that illustrate the breakdown of the phonon quasiparticle picture in these soft materials. For hybrid halide perovskites, we introduce the concept 
of reference structures to address the complex interplay of disordered orientations of MA and FA molecules 
with the inorganic sublattice distortions, affecting 
both anharmonicity, transverse optical and longitudinal optical phonon frequencies, and electron-phonon coupling. 
Our work underscores the potential of integrated, high-throughput computational frameworks to
transform the discovery and optimization of halide perovskite photovoltaic devices at finite temperatures.
\end{abstract}

\maketitle

\section{Introduction}
Halide perovskites, either layered, bulk, nanocrystals, or heterostructures are attracting immense interest due to 
their impressive power conversion efficiencies in optoelectronics, with new candidates emerging 
constantly~\cite{Kojima2009,Snaith2013,Filip2014,Protesescu2015,Volonakis2016,Eperon2016,Tsai2016,Brenner2016,Volonakis2017,Mao2018,Hoffman2019, Liu2020, Marchenko2020, Dey2021, Ahmadi2021, Jiang2022, Jacobsson2021, Unger2022, Sidhik2022, Metcalf2023, Sidhik2024, Filip2024}.
Several key challenges in the field of halide perovskites remain partially unexplored and continue to be active areas of research. 
These include gaining a deeper understanding of (i) the complex anharmonic potential energy surface 
responsible for the formation of local 
structure/disorder~\cite{Even2016,Beecher2016,Marronnier2018,Zhao2020,Fabini2020,Hehlen2022,Dirin2023,Zacharias2023npj,Weadock2023,DiezCabanes2023,Balvanz2024,CaicedoDvila2024,Dubajic2025}, 
anharmonic lattice vibrations~\cite{Quarti2015,Carignano2017,Yaffe2017,Katan2018,Marronnier2018,Fu2018,Ferreira2020,Zacharias2023npj,Zacharias2023}, and 
phase stability~\cite{Marronnier2017,Sutton2018,Marronnier2018,Doherty2021,He2024}, and 
(ii) anharmonic carrier-phonon coupling and its role to the electronic structure 
renormalization~\cite{Patrick2015,Quarti2016,Ning2022,Zacharias2023npj,Seidl2023}, temperature-dependent optical absorption, photoluminescence (PL)~\cite{Maculan2015,Milot2015,Wright2016,Kontos2018,Hsu2019,Mannino2020,Kahmann2020,Chen2020,Peters2022}, carrier mobilities~\cite{Stoumpos2013,Milot2015,Parrott2016,Ponc2019,Irvine2021,Cucco2024}, 
polaron formation~\cite{Zhu2015,Frost2017,Miyata2017,Bretschneider2018,Cinquanta2019,Ghosh2020,sterbacka2020,SrimathKandada2020,Puppin2020,Cannelli2021,Guzelturk2021,Buizza2021,Schilcher2021,Zhang2021,Yue2023,LafuenteBartolome2024,Baranowski2024}, 
and ultrafast non-equillibrium dynamics~\cite{Cinquanta2019,Zhang2023,Seiler2023,Biswas2024}. 

\begin{figure}[htb!]
 \begin{center}
\includegraphics[width=0.47\textwidth]{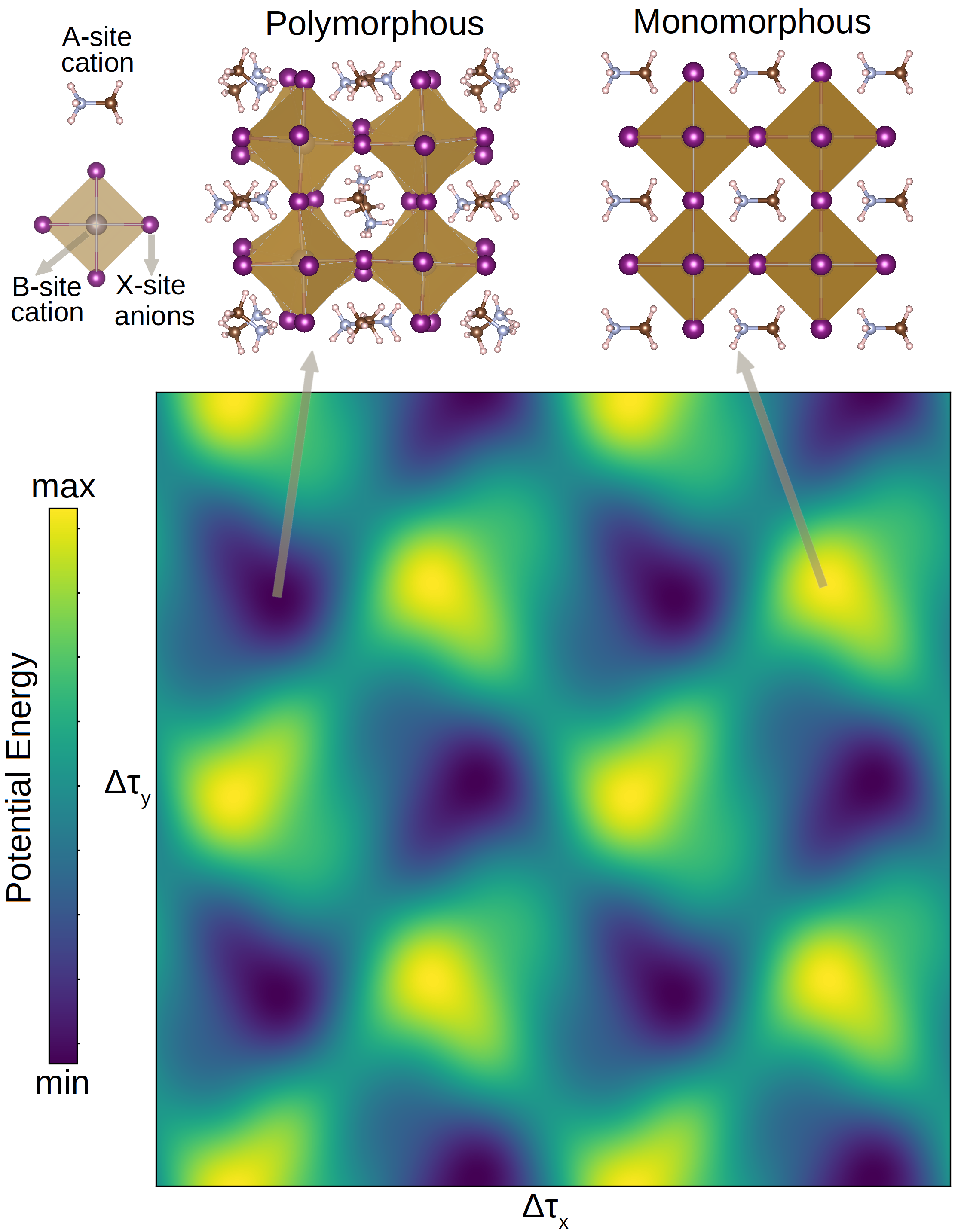}
 \end{center} 
\caption{ Top view of the schematic representation of an anharmonic PES showing 
minima and maxima defined by the polymorphous (locally disordered) and monomorphous halide perovskite 
structures, respectively. $\Delta \tau$ indicates atomic displacements away from static-equilibrium 
positions of the monomorphous structure along the Cartesian directions $x$ and $y$. 
The example ABX$_3$ structure used at the top is for MASnI$_3$ and consists of the A-site 
cation (MA), the B-site metal (Sn), and the X-site halide anions (I).
In this study we explore cases where A = methylammonium (MA), formamidinium (FA), or Cs, 
B = Pb or Sn, and X = I, Br, or Cl. 
\label{fig1} 
}
\end{figure}

The seminal work by Stoumpos {\it et al.}~\cite{Stoumpos2013} has brought to light the strong dependence of structural, optical, and electronic properties of 
hybrid (organic-inorganic) metal iodide perovskites on composition and synthesis conditions. 
One key element for understanding the peculiar properties of these materials is the description of correlated local disorder 
(positional polymorphism)~\cite{Zhao2020,Zhao2021,WangZhi2021,Zacharias2023npj,Quarti2024} in the tetragonal and cubic 
high-symmetry perovskite structures, not distinct in standard diffraction experiments~\cite{Comes1968,Mashiyama1998}. In particular, 
positional polymorphism represents multiple domains of local disorder at static-equilibrium that define a 
minimum in the system’s anharmonic potential energy surface (PES), as shown in Fig.~\ref{fig1}. These domains 
can be identified experimentally by local probes such as pair distribution functions~\cite{Beecher2016,Senn2016,Balvanz2024}
or nuclear magnetic resonance spectroscopy~\cite{Sabisch2025} and are present in crystals not exclusive 
to perovskites~\cite{Wang2020_Z,Wang2025,Zhao2024}.
In halide perovskites, local disorder manifests through B-site (metal) off-centering, which induces the 
formation of stereochemically active lone pairs~\cite{Fabini2020}, 
octahedral tilting, and a distribution of A-site cations within the octahedral voids.
Despite the presence of positional polymorphism in tetragonal and cubic phases, the overall structure, on a
{\it macroscopic level}, retains the average high-symmetry crystallographic form, as reported by traditional diffraction techniques.
The quasistatic network of local structures also mimics the very slow structural relaxations~\cite{Zacharias2023npj,Wang2025}, 
where the system does not settle into a single minimum but instead fluctuates between 
multiple low-energy configurations.

{\it Ab initio} electronic structure and electron-phonon calculations for halide perovskites typically utilize a 
simple monomorphous network, which can be described by a minimal unit 
cell~\cite{Huang2013,Even2013,Even2013b,Amat2014,Umari2014,Mosconi2016,Bokdam2016,Saidi2016,Whalley2017,Mao2018,Boubacar2019,Ponc2019,Ghaithan2020,Wang2020,Ning2022,Muhammad2022}. 
However, this structure is dynamically unstable, corresponding to a local 
maximum on the PES [Fig.~\ref{fig1}]. As a result, density-functional theory (DFT) calculations in halide perovskites 
based on monomorphous cells fail 
to capture the physics related to the more stable polymorphous networks, which are known to significantly influence 
the electronic structure~\cite{Zhao2020}.
In Ref.~[\onlinecite{Zacharias2023npj}], relying on the anharmonic special 
displacement method (ASDM)~\cite{Zacharias2016,Zacharias2020,Zacharias2023}, we have introduced a bottom-up {\it ab initio} 
approach for the exploration of polymorphous structures and a unified treatment of the complex interplay between local disorder, 
anharmonicity, and electron-phonon coupling. 
We have demonstrated using the case of inorganic Cs-based halide perovskites that local disorder plays a key role to accurately 
describe strongly overdamped vibrational dynamics, anharmonic electron-phonon coupling, and temperature-dependent band gaps. 
In Ref.~[\onlinecite{short}], we have generalized and applied our approach in a high-throughput framework to the computationally more 
demanding and technologically more significant case of cubic hybrid halide perovskites. 

In this paper, we provide a detailed description of our approach and conduct a comprehensive 
comparative analysis of 12 materials (4 FA-, 4 MA-, and 4 Cs-based compounds) through 
large-scale calculations for 120 polymorphous structures 
(i.e. 10 structures each) using supercells. We obtain the following results: 
(i) We introduce the concept of reference structures for hybrid halide perovskites, different than the monomorphous 
structures, where disorder is accounted initially only for the molecules. 
The reference structures enable the separation of the effects arising from the distinct dynamics of organic molecules and 
facilitate systematic comparisons with the electronic structure of the corresponding fully polymorphous states.
(ii) We quantify local disorder in terms of B-X bond length and B-X-B bond angle distortions and demonstrate strong correlations 
with the increase in band gap and effective masses induced by positional polymorphism. 
(iii) We demonstrate that the extent of local disorder is influenced by the size of the A-site cation, 
with FA-based compounds exhibiting the least positional polymorphism, followed by MA-based compounds, and 
Cs-based compounds displaying the highest degree of polymorphism. 
(iv) We compute anharmonic phonons at finite temperatures within the self-consistent phonon theory based on reference structures, 
and present phonon quasiparticle dispersions for all compounds. Our results compare well with diffuse scattering maps and 
phonon frequencies measured by various experimental techniques. We also discuss the role of A-site cations and X-site anions 
to transverse optical (TO) and longitudinal optical (LO) phonon splitting.
(v) We calculate the free energies and Boltzmann-weighted averages of temperature-dependent band gaps of all cubic compounds 
using several reference and polymorphous structures, to account for configurational entropy effects.
Our results highlight the critical role of local disorder in achieving accurate 
anharmonic electron-phonon coupling calculations, which result in unprecedented agreement with 
experiments~\cite{Milot2015,Kontos2018,Parrott2016,Sutton2018,Hsu2019,Mannino2020,Chen2020,Kahmann2020,Peters2022,Lpez2025}.
(vi) We distinguish the contributions of thermal expansion and electron-phonon coupling to band gap renormalization, 
demonstrating that they do not contribute equally, contrary to common assumptions~\cite{FranciscoLpez2019,Rubino2020}.
Overall, electron-phonon coupling dominates for Cs-based and MA-based compounds, reaching as high as 97\% while for 
FA-based compounds thermal expansion plays more important role. For locally disordered Sn-based compounds, we find an enhanced 
thermal expansion contribution to the band gap renormalization due to the enhanced lone pair activity, promoting
Sn-I-Sn bond angle bending under lattice expansion.
(vii) We calculate temperature-dependent hole and electron effective masses of all cubic compounds using both 
reference and polymorphous structures. The effective mass enhancement due to polymorphism and 
electron-phonon coupling reach as high as one order of magnitude, improving agreement between calculated and measured reduced effective masses. 
(viii) We show that, starting from maximally symmetrized reference structures, even one polymorphous configuration  
can be enough to calculate anharmonic electron-phonon properties in some systems. 
(ix) We show that positional polymorphism leads to a continuum of strongly overdamped vibrations and the breakdown of
the phonon quasiparticle picture obtained by standard approaches to anharmonicity applied to
monomorphous or reference cells~\cite{Patrick2015,Ning2022,Zacharias2023}. 
(x) We demonstrate the role of local disorder to accurately describe the band gap variations across different phases.

We organize the manuscript as follows: In Sec.~\ref{sec.Theory} we introduce the theory underpinning our 
electron-phonon coupling, temperature-dependent electron energies, polymorphous structure, and 
anharmonic phonon calculations via the ASDM. Section~\ref{sec.main_method_results} described 
our main methodology for computing anharmonic electron-phonon properties
in locally disordered halide perovskites, starting from reference structures. 
In the same section, we show the total energy lowering obtained for locally disordered structures 
with respect to reference structures and present strong correlations of the band gap widening and effective masses enhancements 
with bond length and angle variations. We also demonstrate our methodology for computing anharmonic phonon quasiparticle dispersions, 
and free energies, as well as Boltzmann-weighted averages of temperature-dependent band gaps and effective masses of MAPbBr$_3$ and FAPbI$_3$.
In Sec.~\ref{sec.case_by_case}, we provide a case-by-case analysis of all 12 halide perovskites, showing 
band gaps at the DFT and hybrid functional levels, DFT effective masses, analyzing anharmonic phonon dispersions and TO,LO phonon energies, 
and temperature-dependent band gaps and effective masses calculated within the ASDM. Where available we compare
our results with state-of-the-art $GW$ calculations. In the same section, we analyze the 
electron-phonon and thermal expansion contributions to the band gap renormalization for each material. 
In Sec.~\ref{sec.general_comp}, we present a general comparison and discussion of our results focusing 
on electronic structure modifications due to local disorder, the band gap across different polymorphs, LO-TO splitting, 
the breakdown of the phonon quasiparticle picture, and thermal induced renormalization.  
Section~\ref{sec.Conclusions} summarizes our key results and conclusions as well as provides avenues of future work.

\section{Theory} \label{sec.Theory}

In this work, we employ the ASDM~\cite{Zacharias2016,Zacharias2020,Zacharias2023} for 
the simulation of polymorphous structures, phonon anharmonicity at finite temperatures, and anharmonic electron-phonon coupling. 
The ASDM is classified as an efficient nonperturbative supercell approach. In this method, atoms are displaced according to 
a special linear combination of the computed phonon eigenvectors with amplitudes determined by the associated root mean squared 
displacements. It allows the generation of a single optimal structure that best represents the system at thermal equilibrium 
within the supercell under study. In other words, the ASDM generates, essentially, deterministic high ``quality'' locally
disordered and thermal configurations, replacing the need for stochastic samplings or molecular dynamics 
simulations used in nonperturbative simulations of phonon anharmonicity and/or electron-phonon coupling.

\subsection{Electron-phonon coupling}
Compared to perturbative methods for electron-phonon 
coupling~\cite{Baroni2001,Giustino2017,Sangalli_2019,Gonze2020,Engel2020,Zhou2021,Yang2022,Lee2023,Li2024}, nonperturbative 
approaches offer some advantages~\cite{Franceschetti2007,Zacharias2016,Monserrat2018,Zacharias2020}. 
For instance, nonperturbative methods inherently capture electron-phonon coupling through the adiabatic response 
of electrons to nuclear displacements in supercells, 
eliminating the need for explicit calculation of electron-phonon matrix elements. More, these approaches account for  
higher-order electron-multiphonon interactions~\cite{Saidi2016} and low frequency structural relaxations~\cite{Zacharias2023npj}, 
which are known to be important in halide perovskites. 
Another strength of nonperturbative methods lies in their ability to unify the treatment of anharmonicity and electron-phonon coupling, 
allowing to explore the complex potential energy surfaces of anharmonic systems~\cite{Franceschetti2007,Wiktor2017,Zacharias2020_PRB,Zacharias2023npj,Seidl2023}. 
This capability is central in this work to seamlessly accommodate positional polymorphism in our first principles calculations. 
Nonetheless, understanding the physics captured by nonperturbative approaches requires establishing 
a direct link to perturbative expressions.
The standard electron-phonon matrix element is defined as~\cite{Zacharias2016,Giustino2017}: 
\begin{eqnarray}\label{eq1}
  g_{nm\nu} = \sqrt{\frac{\hbar}{2M_0 \w_\nu}} \braket{\psi_n| \, \D U^{\{\tau\}}_{\rm KS} / \D x_\nu \,| \psi_m}
\end{eqnarray}
where $\hbar$ is the reduced Planck constant, $M_0$ is the proton mass, and $\psi_n$, $\psi_m$ represent two electron states coupled
by a phonon of normal coordinate $x_\nu$ and frequency $\omega_\nu$. $U_{\rm KS}$ is the density functional theory (DFT) 
Kohn-Sham (KS) potential determined by the electron density with the system at static-equilibrium and 
depends parametrically on the nuclei positions ${\{\tau\}}$. The KS energy calculated in a fixed nuclear configuration corresponds 
to a point on the potential energy surface (PES).

\begin{figure*}[p]
 \begin{center}
\includegraphics[width=0.95\textwidth]{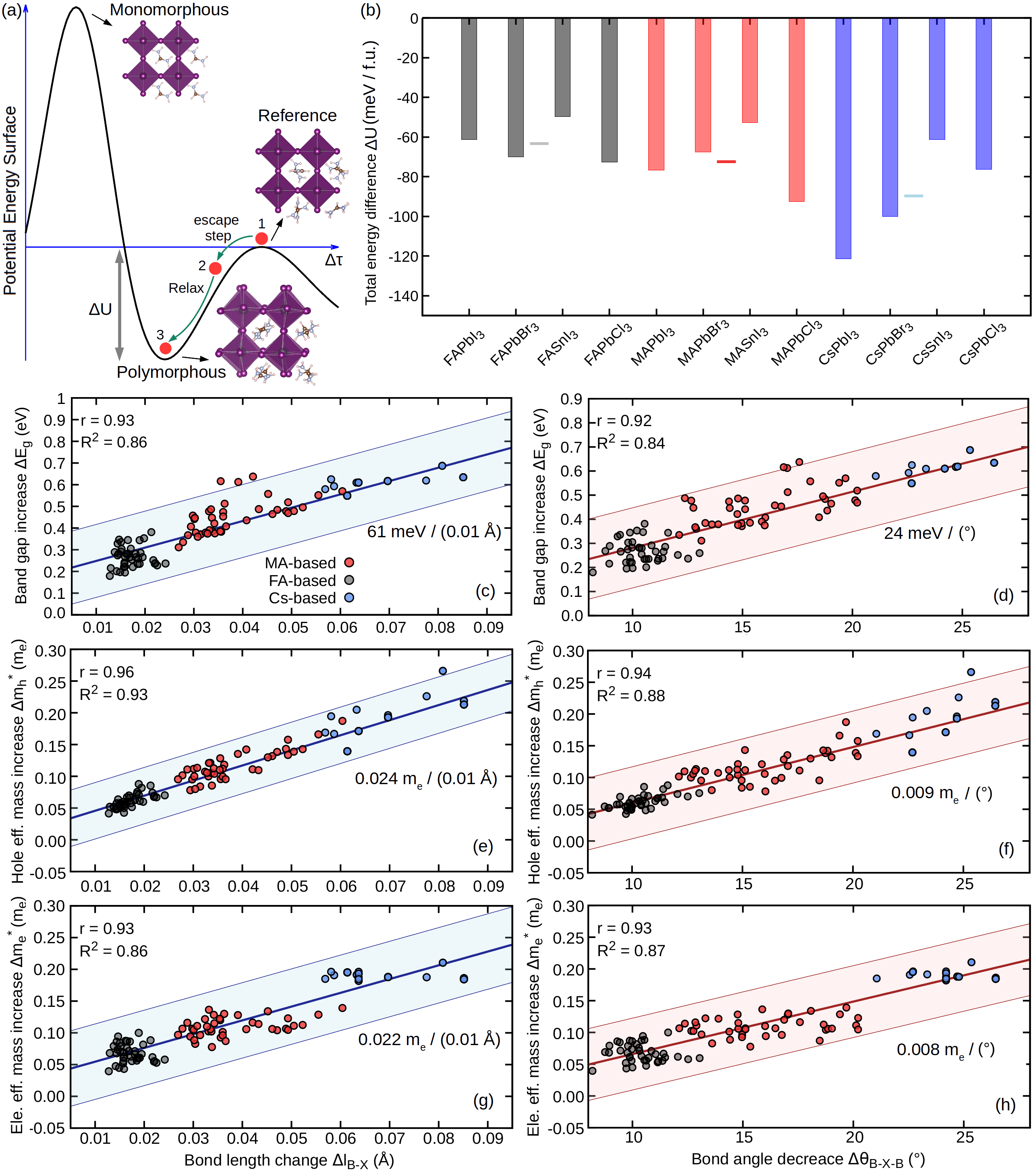}
 \end{center} 
\caption{ (a) Schematic representation of the transitions of the PES from the monomorphous to 
the polymorphous, and from the polymorphous to the reference structure, indicated by ball and stick models
of cubic FAPbI$_3$.
(b) DFT total energy lowering per formula unit ($\Delta U$) of polymorphous cubic halide perovskites 
relative to their reference structures, evaluated using Eq.~\eqref{eq11}, assuming $T=300$~K, and 10 configurations 
in $2 \times 2 \times 2$ supercells for each system. One formula unit (f. u.) refers to the unit cell containing 5 atoms.
Horizontal gray, red, and light-blue lines indicate the mean $\Delta U$ of FA-, MA-, and Cs-based compounds, respectively.
(c-h) Scatter plots showing the dependence of the band gap and effective mass increase due to
polymorphism on the average B-X bond length and B-X-B bond angle variations. The 120 data points correspond to 10 polymorphous 
configurations generated for 4 FA (black), 4 MA (red), and 4 Cs (blue) -based compounds. The straight thick lines represent
the least squares regression fit to the data, with the slopes indicated on each plot. 
The shaded regions represent three standard deviations on either side of the lines. 
All calculations refer to the DFT-PBEsol approximation. Our calculations for the 
band gaps and effective masses include spin-orbit coupling (SOC) effects.
\label{fig2} 
}
\end{figure*}

\subsection{Temperature-dependent electron energies}
It is apparent from Fig.~\ref{fig2}(a) that whenever the system is in static 
equilibrium (i.e. when the forces acting on the atoms are zero), 
it can reside at a local maximum, a local minimum or a saddle point on the PES. 
In view of Eq.~\eqref{eq1}, using either a monomorphous, a reference, or a polymorphous structure is 
expected to have a direct impact on first-principles 
electron-phonon calculations. Not only a different point on the PES is entering Eq.~\eqref{eq1}, but also the phonon frequencies $\w_\nu$ and 
electron wavefunctions $\psi_n$ are renormalized by choosing a different structure.  

In this work, we focus on electron-phonon renormalized KS energies $\ve_n$ 
which can be written in terms of normal coordinates as~\cite{Zacharias2016}:
\begin{eqnarray}\label{eq2}
\Delta \ve_n(T) &=& \frac{1}{2} \sum_\nu \frac{\D^2 \ve_n}{\D x^2_\nu} \sigma^2_{\nu,T} + 
            \frac{3}{4!} \sum_{\nu,\mu} \frac{\D^4 \ve_n}{\D x^2_\nu \D x^2_\mu} \sigma^2_{\nu,T}\sigma^2_{\mu,T} 
\nonumber
\\ &+& \mathcal{O}(\sigma^6) \cdots. 
\end{eqnarray}
Here, $\Delta \ve_n(T) = \ve_n(T) - \ve_n$, where $\ve_n(T)$ is the electron energy at temperature $T$ and $\ve_n$ is the electron energy 
with the system at static equilibrium. $\sigma^2_{\nu,T}$ represents the mean square displacements of the atoms along a phonon mode 
given by  
\begin{eqnarray}\label{eq3}
 \sigma^2_{\nu,T} = \frac{\hbar}{2M_0 \w_\nu}(2 n_{\nu,T} + 1), 
\end{eqnarray}
where the $n_{\nu,T}= [\exp(\hbar\omega_\nu / k_{\rm B}T)-1]^{-1}$ is the Bose-Einstein occupation of the mode and $k_{\rm B}$ is 
the Boltzmann constant. 
Eq.~\eqref{eq2} is obtained by taking a Taylor expansion of $\ve_n$ with respect to the normal coordinates $x_\nu$ and applying the 
standard thermal average as a multivariate Gaussian integral~\cite{Zacharias2016}.

To make a connection with perturbation theory, we adopt the adiabatic Allen-Heine (AH) theory~\cite{Allen1976,Allen1981}  
of temperature-dependent band structures and express $\ve_n(T)$ in terms of electron-phonon 
matrix elements, considering terms up to second order in normal coordinates:
\begin{eqnarray}\label{eq4}
\Delta \ve^{\rm AH}_n(T) = \bigg[\sum_{\beta\neq n} \frac{|g_{n \beta\nu}|^2}{\ve_n - \ve_\beta} + 
   h_{n\nu\nu}\bigg] (2n_{\nu,T} + 1)
\end{eqnarray}
where the first term inside the square brackets represents the Fan-Migdal contribution and 
$h_{n\nu\nu} =  \hbar/2M_0 \w_\nu \braket{\psi_n | \, \partial^2 U^{\{\tau\}}_{\rm KS}/ {\D x_\nu^2} \, |\psi_n} $ represents the 
Debye-Waller contribution to electron-phonon renormalized band structures~\cite{Giustino2017}.
We note that the electron-phonon matrix elements [Eq.~\eqref{eq1}] entering Eq.~\eqref{eq4}, when computed 
within the harmonic approximation, may be ill-defined if the system exhibits imaginary phonon modes. This is usually 
an artifact of applying the harmonic approximation to an inherently anharmonic system. This underscores the need to 
explicitly consider the system's anharmonicity, as discussed in Secs.~\ref{sec.Poly_explore} and~\ref{ASDM_theory}.
Another source of anharmonicity affecting the electronic structure at finite temperatures is thermal lattice 
expansion, which is treated separately in this work, as discussed in Ref.~[\onlinecite{short}] and Sec.~\ref{Sec_TE_comp_details}. 

\subsection{Polymorphous structure exploration} \label{sec.Poly_explore}
At a local maximum (or a saddle point), unstable phonons arise due to the negative curvature of the 
potential energy surface in certain directions, 
resulting in imaginary phonon frequencies ($\omega^2_\nu < 0$) that imply a dynamic instability. 
Hence, the monomorphous and reference structures, shown in Fig.~\ref{fig2}(a), are dynamically unstable. 
Displacing the atoms of the monomorphous or reference structures along phonon eigenvectors corresponding to imaginary phonons 
indicates a direction in which the system can lower its energy and move toward a more stable configuration. 
However, achieving the polymorphous structure (i.e., a local minimum on the PES) through a linear combination of such 
eigenvectors is a challenging task.
Instead, we first test displacing the atoms of the reference structure in a supercell along the phonon eigenvectors 
associated with imaginary phonons using displacements given by:
\begin{eqnarray}\label{eq5}
    \Delta \tau_{p \k \a} = (M_0/M_\k)^\frac{1}{2}\sum_{\nu,\w^2_\nu < 0}  e_{p \k \a,\nu} \, x_{\nu}.
\end{eqnarray}
Here $p$ is an index for the unit cells comprising the supercell, $\a$ denotes a Cartesian direction, and $x_{\nu}$ 
represents the normal coordinates that determine the magnitude of displacement. 
This initial step allows the system to escape from a saddle point [point 1 in Fig.~\ref{fig2}(a)] and move
toward a more stable configuration [point 2 in Fig.~\ref{fig2}(a)], depending on the value of $x_\nu$. 
As a final step, the system is driven toward a local minimum [point 3 in Fig.~\ref{fig2}(a)] by 
relaxing the atomic positions while keeping the lattice constants fixed. 

Using the example of CsPbI$_3$ for a $2 \times 2 \times 2$ supercell, we tested the values 
$x_{\nu} = \{0.05, 0.5, 1, 2, 3, 4, 5\}$\,\AA \, to initially 
displace the atoms of the reference structure using Eq.~\eqref{eq5}, followed by structural relaxation. For 
 $x_{\nu} = \{0.5, 1, 2\}$~\AA, the system reached a polymorphous structure, resulting in a total energy lowering of 121~meV relative 
to the reference structure. For $x_{\nu} = \{3, 4, 5\}$~\AA, the system converged to a saddle point with a total energy reduction of 
108 meV, whereas for $x_{\nu} = 0.05$~\AA, the system returned back to the reference structure.

As a more consistent escape step from the unstable stationary point, we employed Zacharias-Giustino (ZG) displacements, the key element of
the ASDM, which are defined as~\cite{Zacharias2016}
\begin{eqnarray}\label{eq6}
    \Delta \tau^{\rm ZG}_{p \k \a} (T) = (M_0/M_\k)^\frac{1}{2}\sum_{\nu,\w^2_\nu > 0} S_\nu \, e_{p \k \a,\nu} \, \sigma_{\nu,T}, 
\end{eqnarray}
Here, the summation is performed over stable phonons modes with $\w^2_\nu > 0$ and $S_\nu$ represents signs associated with each 
eigenvector. These signs are assigned combinatorially by {\tt ZG.x} of EPW~\cite{Lee2023} to ensure that the anisotropic 
displacement parameters are closely reproduced~\cite{Zacharias2020}. Using CsPbI$_3$, we tested ZG displacements for $T=0$~K. While 
the system after the escape step moves toward a less stable configuration than the reference structure, 
the final relaxation step leads to a polymorphous structure with the same energy lowering of 121~meV, as 
achieved using Eq.~\eqref{eq5} for $x_{\nu} = \{ 0.5, 1, 2 \}$~\AA. 
We note that for generating polymorphous structures with distinct atomic coordinates, one can employ different 
initial sets of ZG displacements by using various permutations of the set of signs $S_\nu$ or different temperatures $T$.

In this work, we employed ZG displacements [Eq.~\eqref{eq6}] for $T=0$~K as the escape step to explore effectively polymorphous structures, 
rather than using Eq.~\eqref{eq5}. 
The ZG method avoids biasing the system toward specific soft-mode distortions, which may otherwise lead to 
saddle points on the PES after relaxation. In addition, Eq.~\eqref{eq6} relies on mode amplitudes 
determined by phonon occupations via Boltzmann statistics, ensuring that the system is sampled in a manner consistent with the underlying PES. 
In contrast, Eq.~\eqref{eq5} requires an arbitrary choice of displacement amplitude, which may not correspond to a physically meaningful point 
on the PES.

\subsection{Anharmonic phonons via the ASDM} \label{ASDM_theory}
The degree of polymorphism, reflected by the total energy lowering relative to the reference structure, 
can also serve as a quantitative measure of the system's anharmonicity.
To address anharmonicity within the phonon quasiparticle picture, we employ the self-consistent phonon (SCP) theory~\cite{Hooton1955,Souvatzis2008,Brown2013} 
using the ASDM~\cite{Zacharias2023}. 
The SCP theory has been implemented successfully in several state-of-the-art approaches, including those 
described in Refs.~[\onlinecite{Hellman_2011,Hellman_2013,Errea2014,Tadano2014,Tadano_2015,Monacelli2021,Mingo_2021}]

In the SCP theory, an effective harmonic matrix of interatomic force constants (IFCs) at temperature $T$ 
is evaluated iteratively until convergence, a process that effectively corresponds to minimizing 
a reference system's trial free energy $F_0$ corresponding to a PES $U_0$. This can be understood 
in the context of the Gibbs-Bogoliubov inequality~\cite{Isihara1968,Bianco2017}, 
\begin{eqnarray}\label{eq7}
 F(T) \leq F_0(T) &=&  \braket{U - U_0}_T + F_{\rm vib}(T) 
\end{eqnarray}
where $F$ is the true free energy, $F_{\rm vib}$ is the vibrational free energy of the reference system, and 
$U$ is the PES of the actual system with the thermal average $\braket{.}_T$ taken with respect to the 
eigenstates of the reference system. 
The inequality in Eq.~\eqref{eq7} suggests that variational minimization of the trial free energy $F_0 (T)$ 
of the simpler reference system provides an upper bound on the true free energy $F(T)$ of the more complex anharmonic system. 

In Eq.~\eqref{eq7} the thermal average $\braket{U - U_0}_T$ is decoupled as $\braket{U}_T - \braket{U_0}_T$ due to the linearity 
of expectation under a normalized probability distribution, with both terms evaluated over the same reference ensemble.
Hence, the trial free energy can be expressed with respect to the phonon frequencies of the reference system as follows~\cite{Zacharias2023}:
\begin{eqnarray}\label{eq8}
F_0(T) &=&  \braket{U}_T -  \frac{M_0}{2} \sum_{\nu} \omega^2_{ \nu} \, \sigma^2_{ \nu,T} \nonumber \\
&+& \sum_{ \nu} \bigg[ \frac{\hbar \omega_{\nu}}{2} - k_{\rm B} T\, {\rm ln} [1+ n_{\nu}(T)] \bigg].
\end{eqnarray}
To minimize the reference system's free energy, it suffices to calculate iteratively the 
matrix of effective IFCs until self-consistency is achieved, i.e.:
\begin{eqnarray}  \label{eq9}
C_{p \k \a, p' \k' \a'}(T) = \Braket{ \frac{\partial^2 U^{\{\tau\}}_{\rm KS}} { \partial \tau_{p \k \a}  \partial \tau_{p' \k' \a'}}}_T.
\end{eqnarray}
To reach the above result one needs to perform the derivative of Eq.~\eqref{eq8} with respect to the IFCs, $C_{p \k \a, p' \k' \a'}$, and 
find the solution for $\D F_0(T) / \D C_{p \k \a, p' \k' \a'} = 0$ as derived in Refs.~[\onlinecite{Zacharias2023}] and~[\onlinecite{Bianco2017}].

In the ASDM for polymorphous systems~\cite{Zacharias2023}, obtaining an initial set (zeroth iteration)  
of effective IFCs involves computing the IFCs by finite differences 
of the polymorphous structure and applying the symmetries of the 
reference system, including both rotations and translations. Subsequent iterations involve the calculation 
of $C_{p \k \a, p' \k' \a'}(T)$, which is determined at each iteration by 
finite differences applied to a single thermally displaced configuration generated using Eq.~\eqref{eq6}. 
At each iterative step, the symmetry operations of the reference system are applied to $C_{p \k \a, p' \k' \a'}(T)$. 
Furthermore, to speed up convergence, linear mixing~\cite{Zacharias2023} between the IFCs is applied at each iteration.  

After convergence is achieved, the final set of IFCs, is used to generate 
effective harmonic phonons to simulate lattice anharmonicity, and thus 
anharmonic ZG displacements via Eq.~\eqref{eq6} which are applied on the polymorphous 
or reference structures. A DFT supercell calculation on the resulting thermally displaced structure enables the evaluation of a temperature-dependent charge density, 
opening the way for computing anharmonic electron-phonon properties~\cite{Zacharias2023npj}.
For example, the ASDM can be used for computing anharmonic electron-phonon renormalized electron energies given by Eq.~\eqref{eq2} using only a single configuration
without relying on the explicit computation of the electron-phonon matrix elements entering Eq.~\eqref{eq4}. 
Another significant advantage of the ASDM is its ability to incorporate higher-order contributions to electron-phonon coupling without additional computational cost. 
Furthermore, it offers the flexibility to initiate calculations from different points on the PES, thereby being able to straightforwardly
 account for the influence of polymorphism on the electronic structure, without the use of molecular dynamics simulations.



\begin{figure}[htb!]
 \begin{center}
\includegraphics[width=0.45\textwidth]{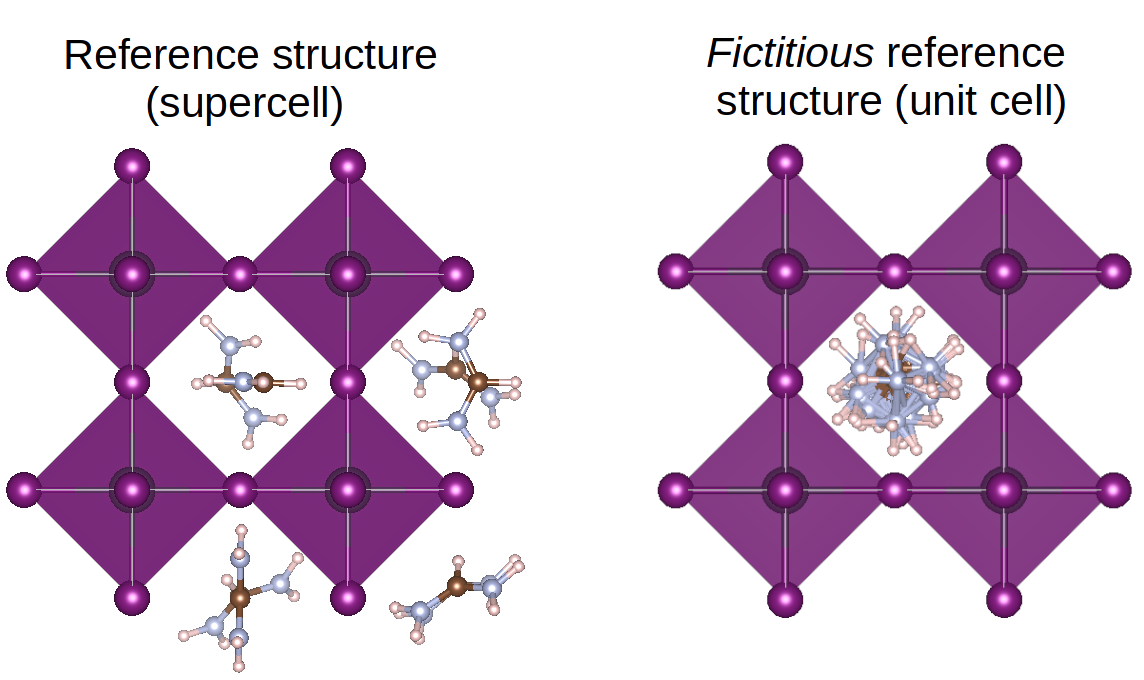}
 \end{center} 
\caption{ (a) Example reference structure of FAPbI$_3$ with relaxed random molecular orientations 
in a $2 \times 2 \times 2$ sypercell. 
(b) Example fictitious reference structure of FAPbI$_3$ obtained by applying unit cell translations 
to the molecules in the reference structure, placing all back to a single unit cell. 
This fictitious structure serves as the reference point for calculating the interatomic force constants within 
the SCP theory.
\label{fig3} 
}
\end{figure}

\section{Methodology and main results} \label{sec.main_method_results}

\subsection{Methodology} \label{sec.main_methodol}
To compute electron-phonon renormalized quantities in locally disordered hybrid halide perovskites, we employ the ASDM  
as described in Sec.~\ref{ASDM_theory} and Ref.~[\onlinecite{Zacharias2023}], but now incorporating a separate treatment 
for the organic molecules. The methodology is outlined below:
\begin{enumerate}
  \item Compute the dynamically unstable harmonic phonons of the monomorphous structure. 
  \item Construct a $2m \times 2n \times 2p$ supercell of the monomorphous structure, where $m$, $n$, $p$ are integers $\geq$ 1.
        Place the FA or MA molecules in different orientations, which can either be random or determined by a subgroup of 
        the structure's space group. Choosing a subgroup allows to explore the effect of symmetries on the IFCs in step 5 below. 
  \item Apply special displacements at temperature $T$ [Eq.~\eqref{eq6}] to the molecules and perform a geometry optimization 
        with the atoms of the inorganic network and lattice constants fixed. 
        The resulting geometry is the {\it reference structure} [Fig.~\ref{fig3}(a)].
  \item Starting from the reference structure, apply special displacements at $T$ [Eq.~\eqref{eq6}]
        to the inorganic network and perform a full geometry optimization 
        with the lattice constants fixed. The resulting geometry is the {\it polymorphous structure} [e.g. Fig.~\ref{fig1}].
  \item Compute the IFCs of the polymorphous structure and enforce translational invariance with respect to 
        a {\it fictitious reference structure} [Fig.~\ref{fig3}(b)], so that 
       if $\tau_{PK} = \tau_{p\k} + R_m$ and $\tau_{P'K'} = \tau_{p'\k'} + R_m$ then 
       \begin{equation}\label{eq10}
        \tilde{C}_{0\k\a, p'\k'\a'} = \frac{1}{N_p} \sum_{m} \tilde{C}_{P K \a, P'K'\a'},
       \end{equation}
       where $\tau_{PK}$ and $\tau_{p\k}$ represent atomic positions within the reference and fictitious structures, respectively,
       $\tilde{C}_{P K \a, P'K'\a'}$ are the IFCs computed for the polymorphous structure, and $N_p$ is the number of 
       unit cells contained in the reference structure. The summation
       over $m$ runs over $N_p$ pairs of atoms in the reference structure connected by lattice translations $R_m$ of the 
       fictitious unit cell. Depending on the size of the supercell and the orientation of the molecules, 
       if the reference system respects specific rotational symmetries (whether proper or improper), these symmetries should be 
       applied to $\tilde{C}_{P K \a, P'K'\a'}$, as described in Ref.~[\onlinecite{Zacharias2023}].
  \item Generate the effective dynamical matrix and obtain the phonons by diagonalization. Using the phonons, generate $\Delta\tau^{\rm ZG}$ 
        for $T$ [Eq.~\eqref{eq6}] and displace the nuclei of the reference structure. 
        The resulting geometry is the {\it ZG reference structure}. 
  \item Repeat steps 5 and 6 with linear mixing of the IFCs until self-consistency. 
        For step 5, {\it use the ZG reference instead of the polymorphous structure}.
  \item Use the self-consistent phonons to generate a ZG $2M \times 2N \times 2P$ supercell by displacing atoms of 
        the reference or polymorphous structures, where $M \geq m$, $N \geq n$, $P \geq p$, 
        for computing electron-phonon properties.  The resulting geometries are  
        the {\it ZG reference structure} or {\it ZG polymorphous structure}.
  \item When configurational entropy is accounted for, i.e. multiple ZG reference or polymorphous structures 
        are used to compute a property $O$,
        perform a Boltzmann-weighted average, defined as:
\begin{eqnarray}\label{eq11}
    \braket{O}_{j} = \frac{1}{Z} \sum_{j} e^{-F_j / k_{\rm B} T } O_j,
\end{eqnarray}
where $\braket{.}_{j}$ is a configurational average, $F_j$ is the free energy of the structure $j$ [Eq.~\eqref{eq7}], and 
$Z=\sum_j e^{-F_j / k_{\rm B} T }$ is the partition function. When no vibrational contributions are 
considered, the free energy is replaced by the DFT KS total energy.
\end{enumerate}

The choice to first obtain a reference structure, i.e. relaxing the organic molecules before relaxing the entire halide perovskite
system, rather than using the monomorphous structure, helps prevent artifacts arising from the net dipole moments of the molecules, 
particularly in the case of MA-based compounds. We note that for inorganic halide perovskites the reference structure is identical 
to the monomorphous structure. 
The generation of a reference structure in hybrid systems is also rationalized by the mostly distinct time 
scales and dynamics of the organic and inorganic components. 
The organic molecules exhibit, on majority, faster internal vibrations (energy scales $>$ 100 meV)  
and slower rotational and librational motions (energy scales $<$ 4~meV)  
compared to the inorganic lattice, whose phonon dynamics typically occur at intermediate energy scales of 4 -- 35 meV~\cite{Zacharias2023}. 
Starting from random molecular orientations allows the 
molecules to explore their local potential energy landscape, where their dipoles adjust to the electrostatic environment created 
by the interplay with the fixed inorganic lattice. Relaxing the molecules first ensures that these slower rotational and librational 
modes settle into stable local minima without interference from the lattice dynamics.
Additionally, it allows to mimic the effects of experimentally observed slow and stochastic molecular librational relaxations~\cite{Even2016}, which can not be 
deduced from the unstable lattice modes of the monomorphous structure using Eq.~\eqref{eq6}.
In the subsequent step, the whole structure 
is relaxed together, enabling the inorganic lattice to adjust and couple with the pre-optimized molecular orientations. This two-step 
approach captures the polymorphous nature of the cubic phase, where local distortions and dynamic interactions between the organic 
and inorganic sublattices coexist, in a quasistatic manner. Additionally, it provides a reference structure  
that serves as a well-defined baseline for subsequent phonon quasiparticle calculations, such as those performed within the framework of the SCP theory.

\begin{figure}[htb!]
 \begin{center}
\includegraphics[width=0.49\textwidth]{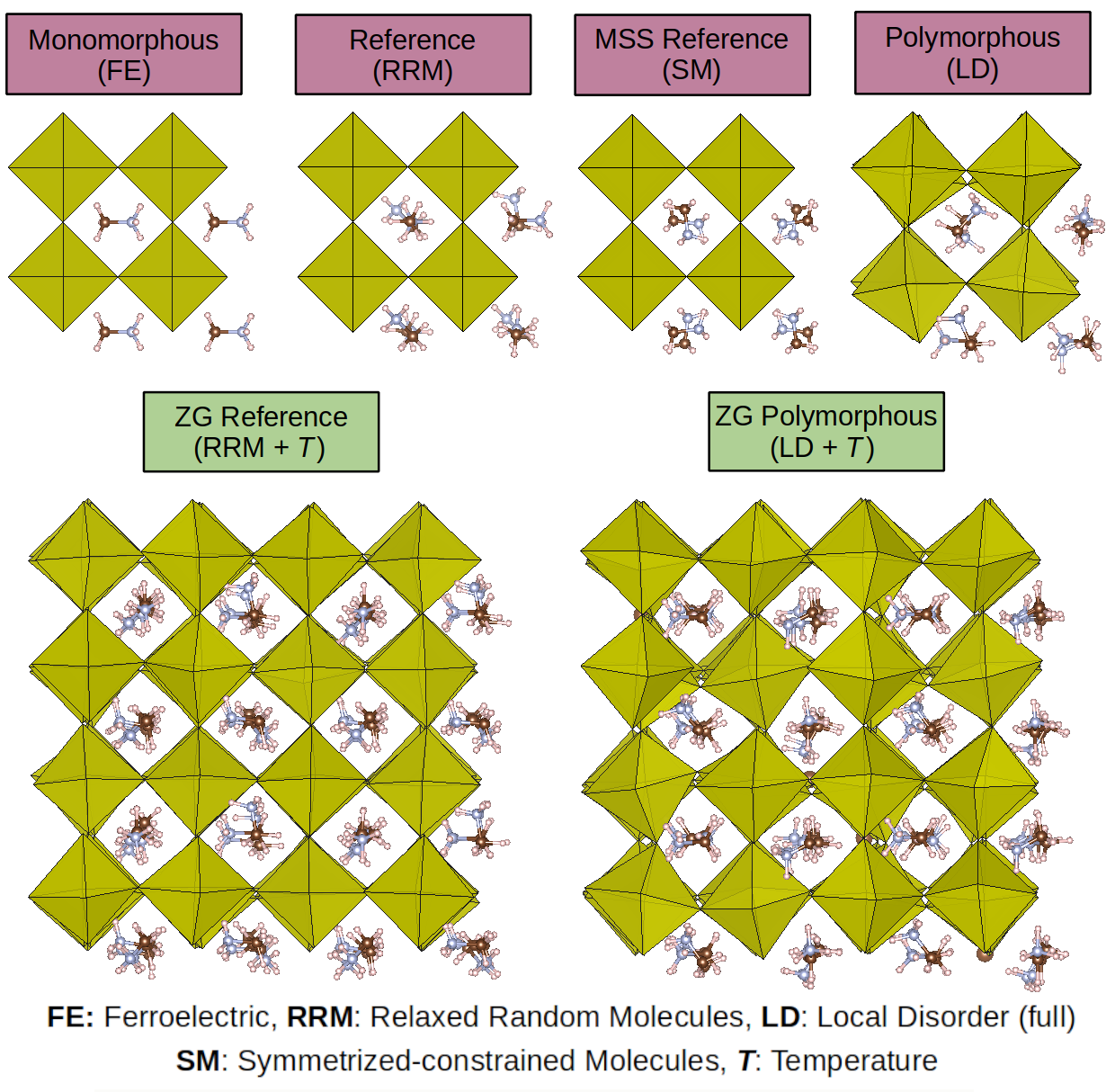}
 \end{center} 
\caption{Structures considered throughout this study using the example of MAPbBr$_3$. 
Reference structures can contain symmetrized-constrained or relaxed random molecular orientations. These structures lead 
to distinct polymorphous structures after geometry optimization. The ZG structures, labelled as green, incorporate 
electron-phonon coupling effects at finite temperatures ($T$). For the ZG structures, we considered a  
4$\times$4$\times$4 supercell (768 atoms) to accommodate a broader sampling of vibrational modes.
\label{fig4} 
}
\end{figure}

Overall, we consider the following cases, as shown in Fig.~\ref{fig4}: 
(i) {\it Monomorphous structures} that exhibit a ferroelectric arrangement of the molecules. 
(i) {\it Reference structures} 
that account for correlated local disorder in the organic sublattice by relaxing molecules from random initial orientations.
(ii) {maximally symmetrized supercells} (MSS) that are used as reference structures 
which adopt specific space group symmetries, while remaining constrained by the supercell size.
(iii) {\it Polymorphous structures} 
that account for correlated local disorder in both the organic and inorganic sublattices.
(iv) {\it ZG reference structures} which incorporate thermal vibrations 
throughout the network, with static-equilibrium atomic positions of the reference structure.
(v) {\it ZG polymorphous structures} which incorporate thermal vibrations 
throughout the network, with static-equilibrium atomic positions of the polymorphous structure.

\subsection{Polymorphous structures, total energy lowering, and electronic structure correlations}

We first apply steps 1-4 in the main methodology for generating 10 different reference and 
their corresponding polymorphous structures to account for configurational entropy effects in 12 materials.
All calculations refer to $2 \times 2 \times 2$  supercells using the PBEsol functional;  
more computational details are available in Sec.~\ref{computdet}. 
 Figure~\ref{fig2}(b) presents the Boltzmann-weighted average [Eq.~\eqref{eq11}] of the total energy lowering 
($\Delta U$) calculated for the polymorphous structures relative to the reference structures, 
assuming a temperature $T=300$~K. We note that testing 
different temperatures from 30 to 1000 K affects the values by no more than 4~meV per formula unit. 
On average, we find that the potential well depth, $\Delta U$, is shallower in FA-based compounds compared to MA-based compounds, 
which in turn exhibit a shallower potential well than Cs-based compounds. 
The observed trend arises from differences in correlated local disorder in the inorganic network, where, 
FA-based compounds exhibit the least structural distortions, 
MA-based compounds show moderate disorder, and Cs-based compounds experience the highest degree of octahedral tilting and 
bond length fluctuations, leading to a deeper potential well. 
This behavior is mainly explained by the size of the A-site cation. 
Larger organic molecules, such as FA, create a more sterically constrained environment that limits structural flexibility, 
whereas smaller cations like MA and Cs permit greater lattice distortions and allow local disorder to emerge more readily. 
This trend is consistent with the tolerance factor~\cite{short,Kieslich2014,Filip2018,Li2022}, where lower values, associated 
with smaller A-site cations, favor increased positional polymorphism.

We quantify the degree of local disorder in the inorganic network by calculating the average B-X bond length ($\Delta l_{\rm B-X}$)
and B-X-B bond angle variations ($\Delta \theta_{\rm B-X-B}$) relative to the idealized reference structures 
and demonstrate their correlations with the induced band gap increase ($\Delta E_{\rm g}$) and hole/electron effective 
mass enhancements ($\Delta m^*_{\rm h}$/$\Delta m^*_{\rm e}$), as shown in Figs.~\ref{fig2}(c)-(h). 
In all cases we obtain strong 
correlations indicated by the Pearson correlation, r, and R$^2$ coefficients, ranging from 0.84 to 0.96. 
Consistent with our findings on the potential well depth [Fig.~\ref{fig2}(a)], the data categorize into three distinct groups. 
FA-based compounds, lying on the left side of the plot, exhibit $\Delta l_{\rm B-X}$ between 0.013 -- 0.024~\AA\, and 
$\Delta \theta_{\rm B-X-B}$ ranging from 8.2 -- 13.0$^\circ$. MA-based compounds, located in the middle, show 
$\Delta l_{\rm B-X}$ from 0.027 -- 0.060~\AA\, and $\Delta \theta_{\rm B-X-B}$ between 12.1 -- 20.2$^\circ$. On the right, 
Cs-based compounds display the largest structural deviations, with $\Delta l_{\rm B-X}$ spanning 0.057 -- 0.085~\AA\, 
and $\Delta \theta_{\rm B-X-B}$ ranging from 21.1 -- 26.5$^\circ$. 
Notably, MA-based compounds cover the widest range of structural variations, which we attribute to the larger 
dipole moments of the MA cations that can significantly influence the electronic structure~\cite{Quarti2015,Motta2015}.

\begin{figure}[t!]
 \begin{center}
\includegraphics[width=0.49\textwidth]{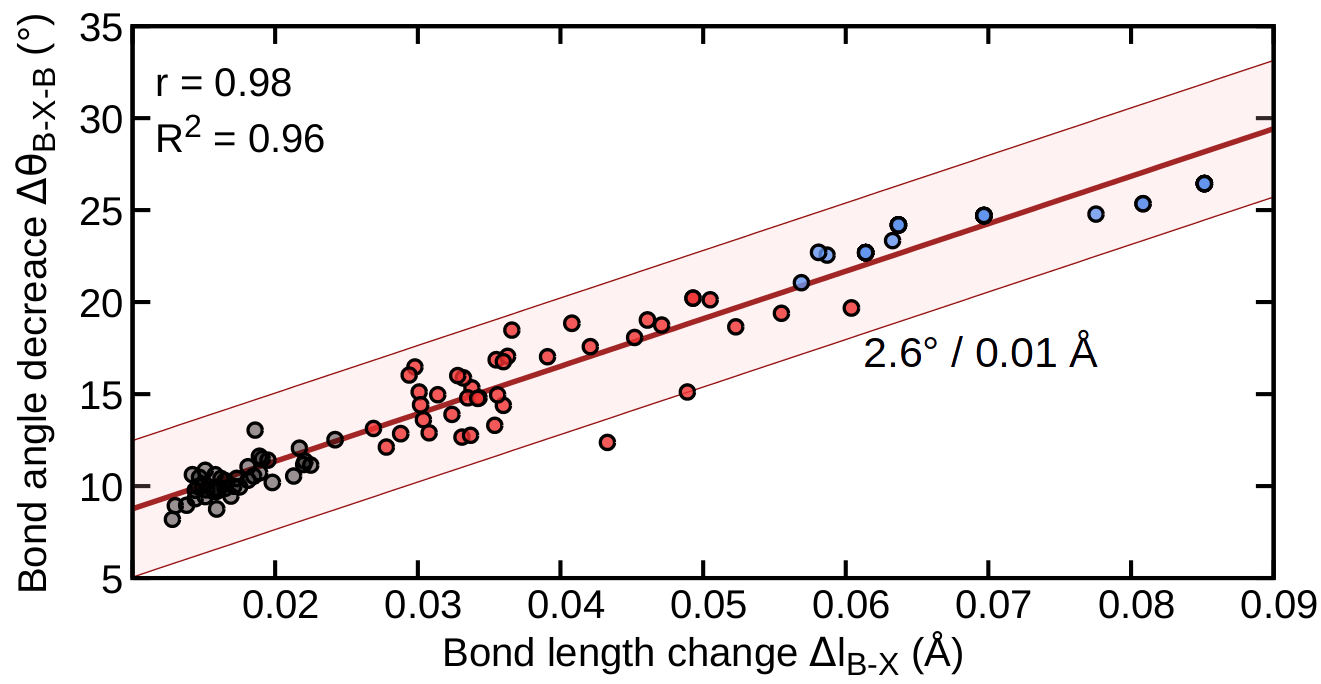}
 \end{center} 
\caption{Dependence of the average B-X-B bond angle decrease on the average B-X bond length variation. 
The 120 data points correspond to 10 polymorphous configurations generated for 4 FA (black), 4 MA (red), and 
4 Cs (blue) -based compounds. The straight thick line represents
the least squares regression fit to the data, with the slope indicated on the plot. The shaded region represents three
standard deviations on either side of the lines. 
\label{fig5} 
}
\end{figure}

In view of Figs.~\ref{fig2}(c)-(h), it is clear that both bond length and bond angle variations correlate closely
with the band gap and effective mass enhancements, indicating that these structural distortions are mutually related. 
Indeed, Fig.~\ref{fig5} shows an almost perfect correlation ($r=0.98$) between the two distortion 
measures, suggesting that only one of these two features is sufficient to explain changes in the electronic 
properties induced by local disorder.

The strong correlation between local disorder and electronic properties arises from orbital 
hybridization effects, where distortions in B-X-B angles modify the degree of overlap between the 
B-site cation and X-site halide orbitals~\cite{Filip2014}. In the ideal cubic reference structure, the B-X-B connectivity 
facilitates strong metal-halide sp-p hybridization, leading to smaller band gaps, highly dispersive electronic 
bands with very low effective masses. However, when local disorder is present, such as octahedral tilting 
and bond-length fluctuations, the overlap between B-p/s orbitals and X-p orbitals is reduced, decreasing the 
bandwidth and thus increasing both the band gap and effective masses~\cite{short}. 
The stereochemical activity of the ns$^2$ lone pair emerges as a consequence of this disorder, with 
symmetry-breaking distortions inducing an asymmetrical electron density around the B-site metal, which is more 
prominent in Sn-based compounds~\cite{Fabini2020,Balvanz2024}.

\begin{figure*}[p]
 \begin{center}
\includegraphics[width=0.98\textwidth]{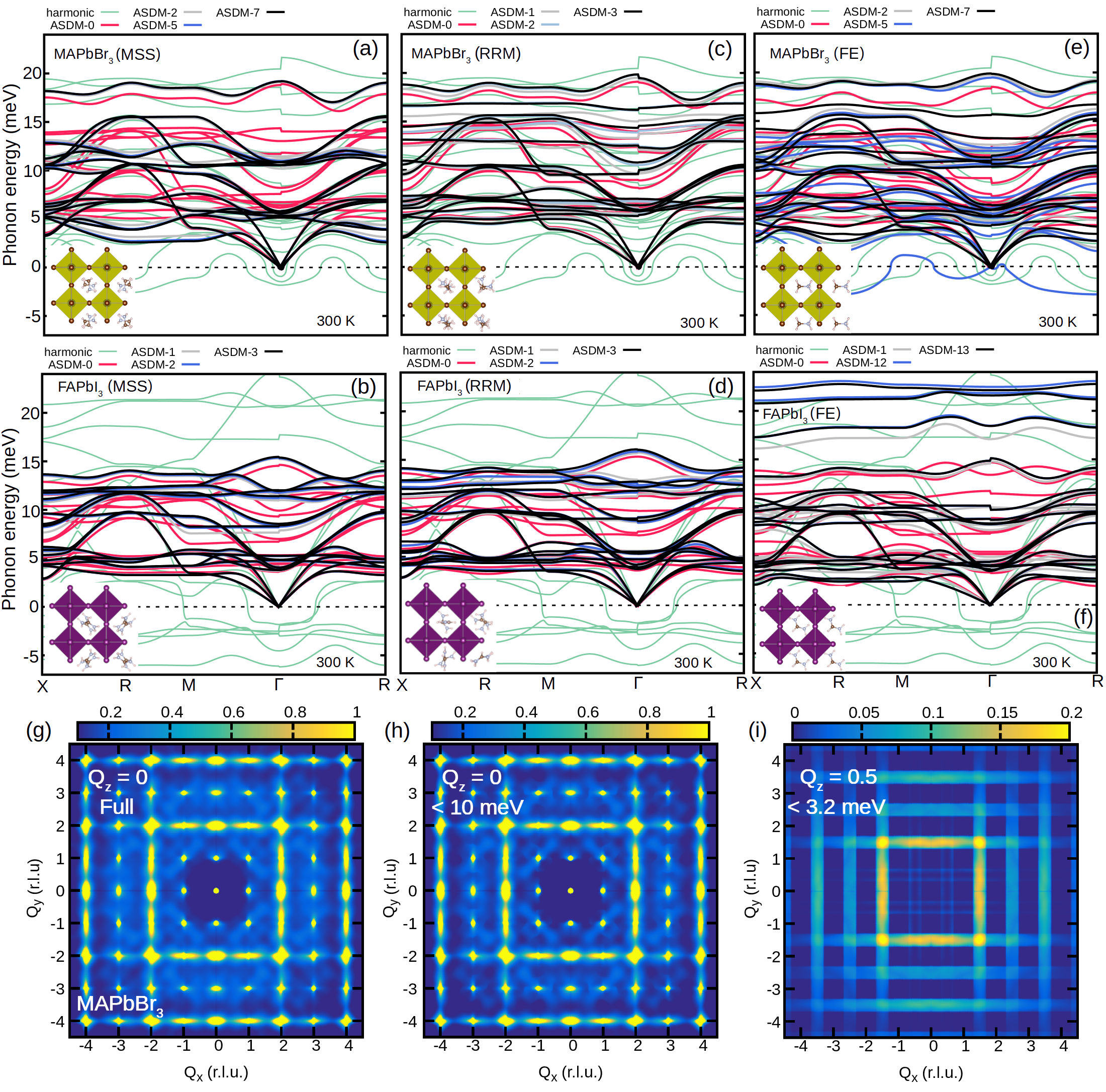}
 \end{center} 
\caption{ (a-f) Convergence of the ASDM anharmonic phonon dispersions at 300~K calculated for MAPbBr$_3$ and FAPbI$_3$ 
using  $2 \times 2 \times 2$  supercells and the PBEsol approximation. 
In (a) and (b), we use maximally symmetrized structures (MSS) with space group P422 as the reference. 
In (c) and (d), we use reference structures with relaxed random molecular (RRM) orientations.
In (e) and (f), the molecules adopt a single ferroelectric (FE) orientation. 
The structures are displayed in the bottom left corner of each plot. The green and red curves represent 
harmonic phonon dispersions calculated for the monomorphous and polymorphous (ASDM-0) structures, respectively. 
Subsequent ASDM-$i$ steps, where $i$ represents an integer index for the iteration, are shown as gray and blue.
The converged ASDM phonon dispersion is shown as black. (g,h) One-phonon diffuse scattering intensity of MAPbBr$_3$ at 
300~K calculated for [$Q_x$$Q_y$0] plane using phonons across the full energy range 
(g) and phonons with energies below $10$~meV (h). (f) One-phonon diffuse scattering intensity of MAPbBr$_3$ at
300~K calculated for [$Q_x$$Q_y$0.5] plane using phonons with energies below $3.2$~meV. 
The components of the scattering wavevectors are given in reciprocal lattice units (r.l.u) 
of the minimal cubic unit-cell. 
\label{fig6} 
}
\end{figure*}

\subsection{Phonon anharmonicity and diffuse scattering}

Figures~\ref{fig6}(a)-(f) show the convergence of the phonon dispersion of MAPbBr$_3$ and FAPbI$_3$ calculated for three different 
reference structures using the ASDM through the steps 1-7 in the main methodology [Sec.~\ref{sec.main_methodol}]. 
All ASDM calculations are for $2 \times 2 \times 2$  supercells. 
In all plots, we focus in the energy range 0--25~meV related mostly to lattice vibrations  
of the inorganic network; high energy flat phonon bands representing molecular vibrations are not shown for clarity. 
In Sec.~\ref{computdet}, we further show the percentage contribution of molecular vibrations in the 
energy range 0--25~meV. 
In all plots, we also include the phonon dispersion of the monomorphous structure calculated within the harmonic 
approximation (green curve), which reveals instabilities across the reciprocal space.  

In Figs.~\ref{fig6}(a,b), we present symmetrized anharmonic phonon dispersions, obtained by arranging the 
MA or FA molecules in the reference structure to respect specific space group symmetries, while remaining 
constrained by the supercell size, i.e. the MSS structure. 
The symmetrized reference structure is constructed such that the molecules are positioned to preserve 8 symmetries 
of the Pm$\bar{3}$m (No. 221) space group. 
Those 8 symmetries represent the tetragonal subgroup P422 (No. 89) which is a subgroup of Pm$\bar{3}$m. 
All ASDM steps, yield stable anharmonic phonons across the entire reciprocal space, with reasonable convergence achieved 
as early as the second iteration (ASDM-2) for both MAPbBr$_3$ and FAPbI$_3$. Notably, the symmetrized phonons calculated 
for the polymorphous structure (ASDM-0) successfully capture the key features 
of the converged anharmonic phonon dispersion. However, the ASDM-0 fail to capture the low-frequency flat band
of MAPbBr$_3$ along the R-M path, which undergoes a significant softening upon inclusion of phonon self-energy corrections, 
reaching a minimum energy as low as 2.5 meV [Fig.~\ref{fig6}(a)]. 
This branch is the most sensitive feature of the phonon dispersion due to the ultraslow dynamical hopping of the system between 
different minima in the PES which cannot be captured by the quasistatic polymorphous framework. 
As shown below, the starting orientation of the molecules in the reference structure can impact
profoundly these low energy phonons. 

In Figs.~\ref{fig6}(c,d), we present the ASDM convergence performance when starting random orientations of the molecules are chosen, 
effectively eliminating all rotational symmetries in the MAPbBr$_3$ and FAPbI$_3$ reference structures. 
We readily observe that all ASDM iterations produce dynamically stable phonons, with the translationally 
symmetrized phonons for the polymorphous structure at ASDM-0 closely approximating those 
obtained from the fully converged phonon dispersion at ASDM-3.
In contrast to the results shown in Figs.~\ref{fig6}(a,b), where certain rotational symmetries are preserved in the MSS 
structure, the phonon dispersion shown in Figs.~\ref{fig6}(c,d) reveals degeneracy splittings. 
This confirms that the molecular orientations influence the lattice vibrations of the inorganic network.

As shown in Figs.~\ref{fig6}(e,f), the effect of molecular orientation on the phonon dispersion becomes even more 
pronounced when the reference structure corresponds to the monomorphous structure, characterized by a single 
ferroelectric orientation of the MA or FA molecules. 
Unlike the symmetric and random reference structures of MAPbBr$_3$, the ASDM procedure fails to achieve convergence 
in the ferroelectric case, exhibiting persistent phonon instabilities [Fig.~\ref{fig6}(e)], indicated by negative 
phonon energies, at various iterations (e.g., ASDM-5). Notably, as the ASDM procedure progresses, certain flat phonon 
bands associated with the orientations of the organic molecules become imaginary (not shown), 
highlighting the inherent instability of the ferroelectric MA-based configurations.
The phonon instabilities observed in the ferroelectric orientation arise because the MA molecule’s intrinsic dipole 
conflicts with the lattice's tendency to minimize electrostatic energy and maintain mechanical stability. 
The phonon instabilities also suggest that MA-based hybrid halide perovskites do not support a stable ferroelectric 
state for the molecules, favoring alternative configurations (i.e. antiferroelectric or random molecular orientations) 
that better accommodate the coupling between the lattice dynamics of the organic and inorganic framework.
Our results for ferroelectric FAPbI$_3$ up to ASDM-13 indicate that a ferroelectric fluctuation of the molecules 
may be dynamically stable in FA-based compounds. This stability is likely due to the minimal net dipole moment 
of the FA molecule, which reduces internal electric fields that could destabilize the ferroelectric phase. 
This result is also consistent with studies on solution-processed FAPbI$_3$ layers, suggesting that substituting MA with FA cations 
enhances lattice robustness and contribute to material durability~\cite{Smecca2016}. 
In Sec.~\ref{computdet}, we show that the phonon modes with energies above $15$~meV are primarily associated with vibrations 
of the organic sublattice, demonstrating that the ferroelectric state leads to further decoupling between 
the vibrational dynamics of the inorganic and organic sublattices.

\begin{figure*}[p]
 \begin{center}
\includegraphics[width=0.92\textwidth]{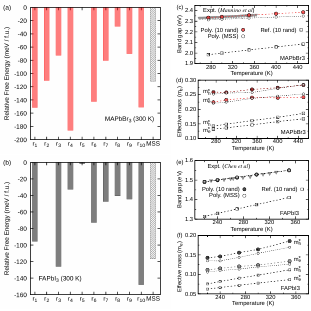}
 \end{center} 
\caption{ (a,b) Relative free energy at 300~K for 10 ZG polymorphous configurations (r$_1$-r$_{10}$) of 
cubic MAPbBr$_3$ (a) and FAPbBr$_3$ (b), referenced to the least stable configuration, 
calculated using Eq.~\eqref{eq8} and 2$\times$2$\times$2 supercells.
The ZG polymorphous configurations are obtained starting from geometry optimization of the
reference structures with random molecular orientations and then through ASDM. 
ASDM converged phonons calculated for each structure [e.g. black lines in Figs.~\ref{fig6}(c,d)] are used to 
compute the free energy. The relative free energy of the ZG polymorphous configuration starting from the
 maximally symmetrized supercell (MSS) is shown for comparison. 
(c-f) Band gaps and electron/hole effective masses (m$_{\rm e}^*$/m$_{\rm h}^*$) calculated 
for MAPbBr$_3$ and FAPbI$_3$ using ZG polymorphous (filled discs) and ZG reference (open squares) 
configurations in 4$\times$4$\times$4 supercells accounting for both electron-phonon coupling and thermal lattice expansion. 
The effect of spin-orbit coupling is also included. Boltzmann-weighted averages [Eq.~\eqref{eq11}] are obtained for 10 structures. 
Open circles represent calculations using the ZG polymorphous configuration of the MSS.
All calculated temperature-dependent band gaps are shifted by 1.44 and 1.10 eV to match experimental data 
of MAPbBr$_3$ and FAPbI$_3$ at 300~K, respectively. These corrections are consistent with 
the PBE0 values for the band gaps reported in Table~\ref{table.1}. 
Experimental gaps (gray triangles) are from Refs.~[\onlinecite{Mannino2020}] and~[\onlinecite{Chen2020}].
\label{fig7} 
}
\end{figure*}

To validate our approach for calculating the phonon dispersions of hybrid halide perovskites we report 
diffuse scattering maps of MAPbBr$_3$ at 300~K.
In Fig.~\ref{fig6}(g), we show the phonon-induced diffuse scattering map of MAPbBr$_3$ calculated 
for the [$Q_x$$Q_y$0] plane, where $Q_\a$ represent Cartesian components of the scattering wavevector $\bQ$. 
The scattering map is computed using the phonons calculated for the MSS structure [Fig.~\ref{fig6}(a)], 
neglecting multi-phonon contributions~\cite{Zacharias2021PRB}.
Our computed scattering map exhibits excellent qualitative agreement with the experimental maps reported in
Ref.~[\onlinecite{Dubajic2025}], reproducing all main features, including the vertical and horizontal diffuse 
scattering rods that connect high intensity Bragg peaks, formed when $Q_x$ and $Q_y$ are integers. 
In Fig.~\ref{fig6}(h), we present the phonon-induced diffuse scattering map of MAPbBr$_3$ in the same plane, 
considering now only phonons with energies below 10 meV. A comparison of the maps in Figs.~\ref{fig6}(g) and (h) clearly 
reveals that all major features are reproduced, demonstrating that low-energy phonons dominate diffuse 
scattering in MAPbBr$_3$. In Fig.~\ref{fig6}(i), we show the phonon-induced scattering map of MAPbBr$_3$ 
calculated for the [$Q_x$$Q_y$0.5] plane, focusing now on ultralow energy phonons with energies below 3.2~meV driving 
dynamical octahedral tilting and hopping between different minima in the PES. 
These structural fluctuations are primarily associated with the low-energy flat phonon band along the R-M path in Fig.~\ref{fig6}(a) and 
give rise to diffuse scattering rods connecting equivalent wavevectors at the M and R points, consistent with calculations 
for the high-symmetry cubic phase of CsPbBr$_3$~\cite{Zacharias2023npj,LaniganAtkins2021}. We note that we exclude any 
quasielastic contributions to diffuse scattering arising from polymorphism in the inorganic sublattice, as this is beyond 
the scope of the present work. However, in Sec.~\ref{Sec.Overdamped_ph}, we examine the influence of local disorder on phonons, 
leading to strongly overdamped vibrational dynamics.

\subsection{Free energies, temperature-dependent band gaps, and effective masses}

We remark that the 10 polymorphous structures explored for each cubic hybrid halide perovskite in 
2$\times$2$\times$2 supercells yield distinct DFT total energy lowerings and phonon spectra. 
These variations arise from the sensitivity of local disorder to molecular orientations as 
discussed previously. 
To account for the influence of local disorder, configurational entropy, and thermal effects on the electronic 
structure across various ZG polymorphous configurations, we compute the free energy of each configuration and use it to 
obtain Boltzmann-weighted averages [Eq.~\eqref{eq11}]. 
Figures~\ref{fig7}(a)-(f) show calculations for the relative free energies of each ZG polymorphous configuration
and Boltzmann averaged band gaps and effective masses of cubic MAPbBr$_3$ and FAPbI$_3$. The free 
energies in Figs.~\ref{fig7}(a,b) are evaluated with respect to the least stable configurations, r2 and r5. 
We also include the relative free energy calculated for the ZG polymorphous configuration obtained by combining
the ASDM with geometry optimization for the MSS structure.
Although starting from the MSS structure yields one of the most stable configurations, 
it does not lead to the absolute minimum in stability. Other random molecular orientations facilitate 
local disorder within 2$\times$2$\times$2 supercells more effectively, leading to lower energy ZG polymorphous structures.

In Figs.~\ref{fig7}(c) and (e), we compare our calculated temperature-dependent band gaps using 
ZG polymorphous (filled discs) and ZG reference (open squares) structures of cubic MAPbBr$_3$ and FAPbI$_3$ 
with experimental data from Refs.~[\onlinecite{Mannino2020}] and~[\onlinecite{Chen2020}]. 
Accounting for local disorder in our calculations significantly improves agreement with experiment. 
This improvement arises from two key effects: (i) positional polymorphism influences the electronic structure, 
leading to a substantial band gap increase of 0.44 eV for MAPbBr$_3$ and 0.24 eV for FAPbI$_3$ 
relative to the reference structures (see Table~\ref{table.1}), and (ii) it reduces the temperature coefficient of the band gap,
 quantified by a decrease in $dE_{\rm g}/dT$ of 51\% for MAPbBr$_3$ and 62\% for FAPbI$_3$. 
A quantitative comparison of $dE_{\rm g}/dT$ extracted from experiments and calculations, along with the 
respective contributions of electron-phonon coupling and thermal lattice expansion, is presented in Sec.~\ref{sec.case_by_case}.
Importantly, we demonstrate in Figs.~\ref{fig7}(c-f) that a single ZG polymorphous configuration obtained from the MSS 
structure can yield accurate results for MAPbBr$_3$ and FAPbI$_3$, comparable to those obtained using 10 ZG polymorphous configurations.

In Figs.~\ref{fig7}(d) and (f), we present the temperature dependence of the  
hole and electron effective masses in cubic MAPbBr$_3$ and FAPbI$_3$, exhibiting a linear relationship with temperature. 
Compared to calculations for the reference structures, the effective masses obtained using ZG polymorphous configurations 
are almost twice as large. 
In fact, for the ZG reference structure of cubic MAPbBr$_3$ we obtain $m^*_{\rm h, ref} = 0.15$, 
$m^*_{\rm e, ref} = 0.14$,  and a reduced effective masses of $\mu_{\rm ref} = 0.07$ at $T=300$~K
in units of the free electron mass $m_0$.
In contrast, for the ZG polymorphous structure of cubic MAPbBr$_3$, these values increase substantially
to $m^*_{\rm h, poly} = 0.26$,  $m^*_{\rm e, poly} = 0.23$, and $\mu_{\rm poly} = 0.12$  at the same temperature. 
The corresponding values at $T=300$~K for cubic FAPbI$_3$ are 
$m^*_{\rm h, ref} = 0.10$, $m^*_{\rm e, ref} = 0.08$, $\mu_{\rm ref} = 0.04$,  
$m^*_{\rm h, poly} = 0.16$, $m^*_{\rm e, poly} = 0.13$, and $\mu_{\rm poly} = 0.07$.  
We note that our calculations are at the DFT-level and can be further improved by considering quasiparticle corrections within the
$GW$ approximation, which are known to enhance the effective 
masses~\cite{Huang2013,Umari2014,Brivio2014,Bokdam2016,Mosconi2016,Davies2018,Wang2020,Filip2021,Sajedi2022} 
In Sec.~\ref{sec.case_by_case}, we provide a more detailed discussion how the reduced effective masses of polymorphous structures 
compare to experimental measurements and $GW$ calculations.

\begin{table*}[t!]
\caption{Lattice constants ($L$), DFT band gaps ($E_{\rm g}$), PBE0 band gaps ($E^{\rm PBE0}_{\rm g}$), DFT hole ($m^*_{\rm h}$) 
and electron ($m^*_{\rm e}$) effective masses, and reduced effective masses ($\mu$) of the reference (r) and polymorphous (p) structures of inorganic and hybrid halide perovskites. 
Band gaps and effective masses are calculated using $2\times2\times2$ supercells (40 and 96 atoms), $3\times3\times3$ uniform ${\bf k}$-grids, 
and a plane wave cutoff of 120~Ry, including the effect of spin-orbit coupling. DFT corrections to the band gaps [$\Delta E_{\rm g}$($T$)] and effective masses 
[$\Delta m^*_{\rm h}$($T$)] due to electron-phonon coupling and thermal expansion at temperatures $T$ are calculated with the ASDM 
using $4\times4\times4$ ZG supercells (320 and 768 atoms), $1\times1\times1$ uniform ${\bf k}$-grids, and a plane wave 
cutoff of 60~Ry, including the effect of spin-orbit coupling. Values were obtained as averages over 10 distinct 
configurations of the reference and polymorphous structures. 
Experimental band gaps of the cubic phases, $E^{\rm expt.}_{\rm g}$, and experimental reduced effective masses of the 
low-temperature phases ($\mu^{\rm expt.}$) are reported for comparison.  
$\Delta_s$ represents an empirical correction to the calculated band gaps to match $E^{\rm expt.}_{\rm g}$, which 
is consistent with the PBE0 correction $E^{\rm PBE0}_{\rm g} - E_{\rm g}$.
}
\setlength{\arrayrulewidth}{0.6pt} 
\begin{tabular*}{\textwidth}{@{\extracolsep{\fill}}l c c c c c c c c c c c c c}
\hline \hline
       & $L$  & $E_{\rm g}$  & $E^{\rm PBE0}_{\rm g}$ & $\Delta E_{\rm g}$($T$) & $E^{\rm expt.}_{\rm g}$ & $\Delta_s$ 
       &  $m^*_{\rm h}$  & $\Delta m^*_{\rm h}$($T$)  &   $m^*_{\rm e}$ & $\Delta m^*_{\rm e}$($T$) & $\mu$ & $\mu^{\rm expt.}$  \\
       &  \AA & (eV)  & (eV)  &  (eV) &  (eV)  & (eV)  & ($m_0$) &  ($m_0$) & ($m_0$) & ($m_0$) & ($m_0$) & ($m_0$) \\  
\hline 
r-CsPbI$_3$ & 6.25  & -0.05  & 1.15 & 0.39 (650 K) & -        & 1.02 & -0.02  & -             & -0.02 & 0.22 (650 K) & -0.010 & - \\
p-CsPbI$_3$ & 6.25  &  0.60  & 1.84 & 0.13 (650 K) & 1.78$^a$ & 1.02 &  0.21  & 0.16 (650 K) &  0.17 & 0.08 (650 K) & 0.094 & 0.114 $\pm$ 0.01$^k$ \\ [3pt]
r-CsPbBr$_3$ & 5.87 &  0.24  & 1.52 & 0.41 (430 K) & -        & 1.38 &  0.04  & 0.17 (430 K) &  0.04 & 0.14 (430 K) & 0.020 & - \\
p-CsPbBr$_3$ & 5.87 &  0.85  & 2.30 & 0.16 (430 K) & 2.39$^b$ & 1.38 &  0.23  & 0.11 (430 K) &  0.23 & 0.10 (430 K) & 0.115 & 0.126 $\pm$ 0.01$^k$ \\ [3pt]
r-CsSnI$_3$ & 6.14  & -0.27  & 0.67 & 0.52 (500 K) & -        & 0.86 & -0.07  & 0.20 (500 K) & -0.10 & 0.18 (500 K) & -0.041 & - \\ 
p-CsSnI$_3$ & 6.14  &  0.28  & 1.30 & 0.20 (500 K) & 1.35$^c$ & 0.86 &  0.07  & 0.05 (500 K) &  0.09 & 0.08 (500 K) & 0.039 & - \\ [3pt]
r-CsPbCl$_3$ & 5.60 &  0.59  & 2.16 & 0.33 (330 K) & -        & 1.77 &  0.12  & 0.16 (330 K) &  0.13 & 0.14 (330 K) & 0.062 & - \\ 
p-CsPbCl$_3$ & 5.60 &  1.19  & 2.79 & 0.10 (330 K) & 3.02$^d$ & 1.77 &  0.29  & 0.07 (330 K) &  0.31 & 0.08 (330 K) & 0.150 & 0.202 $\pm$ 0.01$^l$ \\ [3pt]
r-MAPbI$_3$  & 6.31 & 0.03   & 1.23 & 0.26 (315 K) & -        & 1.00 &  0.02  & 0.09 (315 K) &  0.02 & 0.07 (315 K) & 0.010 & - \\ 
p-MAPbI$_3$  & 6.31 & 0.52   & 1.75 & 0.09 (315 K) & 1.61$^e$ & 1.00 &  0.16  & 0.06 (315 K) &  0.13 & 0.04 (315 K) & 0.072 & 0.104$^m$ \\ [3pt]
r-MAPbBr$_3$ & 5.97 & 0.39   & 1.82 & 0.17 (300 K) & -        & 1.44 &  0.09  & 0.06 (300 K) &  0.09 & 0.05 (300 K) & 0.045 & - \\ 
p-MAPbBr$_3$ & 5.97 & 0.83   & 2.27 & 0.08 (300 K) & 2.34$^b$ & 1.44 &  0.19  & 0.07 (300 K) &  0.18 & 0.05 (300 K) & 0.092 & 0.106$^n$, 0.117$^m$ \\ [3pt]
r-MASnI$_3$  & 6.23 & -0.09  & 0.88 & 0.22 (300 K) & -        & 1.02 & -0.05  & 0.09 (300 K) & -0.03 & 0.09 (300 K) & -0.019 & - \\ 
p-MASnI$_3$  & 6.23 & 0.28   & 1.30 & 0.13 (300 K) & 1.31$^f$ & 1.02 &  0.06  & 0.03 (300 K) &  0.09 & 0.04 (300 K) & 0.036 & - \\ [3pt]
r-MAPbCl$_3$  & 5.68 & 0.72  & 2.34 & 0.13 (300 K) & -        & 1.73 &  0.15  & 0.05 (300 K) &  0.16 & 0.04 (300 K) & 0.077 & - \\ 
p-MAPbCl$_3$  & 5.68 & 1.24  & 2.88 & 0.12 (300 K) & 3.07$^g$ & 1.73 &  0.26  & 0.08 (300 K) &  0.27 & 0.08 (300 K) & 0.132 & - \\ [3pt]
r-FAPbI$_3$  & 6.36 & 0.10   & 1.33 & 0.18 (300 K) & -        & 1.10 &  0.03  & 0.07 (300 K) &  0.03 & 0.05 (300 K) & 0.015 & \\ 
p-FAPbI$_3$  & 6.36 & 0.34   & 1.59 & 0.09 (300 K) & 1.53$^h$ & 1.10 &  0.10  & 0.06 (300 K) &  0.09 & 0.04 (300 K) & 0.047 & 0.09, 0.095$^m$ \\ [3pt]
r-FAPbBr$_3$ & 5.99 & 0.46   & 1.89 & 0.13 (300 K) & -        & 1.50 &  0.10  & 0.04 (300 K) &  0.10 & 0.03 (300 K) & 0.050 & -\\ 
p-FAPbBr$_3$ & 5.99 & 0.70   & 2.13 & 0.10 (300 K) & 2.29$^b$ & 1.50 &  0.16  & 0.05 (300 K) &  0.15 & 0.04 (300 K) & 0.077 & 0.115, 0.13$^m$ \\ [3pt]
r-FASnI$_3$  & 6.33 & 0.11   & 1.13 & 0.18 (300 K) & -        & 0.85 &  0.03  & 0.04 (300 K) &  0.03 & 0.05 (300 K) & 0.015 & - \\ 
p-FASnI$_3$  & 6.33 & 0.42   & 1.48 & 0.08 (300 K) & 1.38$^i$ & 0.85 &  0.08  & 0.03 (300 K) &  0.11 & 0.04 (300 K) & 0.046 & - \\ [3pt]
r-FAPbCl$_3$  & 5.74 & 0.89  & 2.51 & 0.09 (300 K) & -        & 1.71 &  0.17  & 0.04 (300 K) &  0.18 & 0.04 (300 K) & 0.087 & - \\ 
p-FAPbCl$_3$  & 5.74 & 1.17  & 2.79 & 0.03 (300 K) & 2.91$^j$ & 1.71 &  0.23  & 0.05 (300 K) &  0.25 & 0.05 (300 K) & 0.120 & - \\ 
\hline
\hline
\end{tabular*}

\vspace*{0.15cm}
\raggedright
{\footnotesize $^{a}$ Ref.~[\onlinecite{Sutton2018}], $^{b}$ Ref.~[\onlinecite{Mannino2020}], $^{c}$ Ref.~[\onlinecite{Kontos2018}], 
$^{d}$ Ref.~[\onlinecite{Peters2022}], $^{e}$ Ref.~[\onlinecite{Milot2015}], $^{f}$ Ref.~[\onlinecite{Parrott2016}], $^{g}$ Ref.~[\onlinecite{Hsu2019}],
$^{h}$ Ref.~[\onlinecite{Chen2020}],  $^{i}$ Ref.~[\onlinecite{Kahmann2020}], $^{j}$ Ref.~[\onlinecite{Lpez2025}], $^k$ Ref.~[\onlinecite{Yang2017}], \\ 
$^l$ Ref.~[\onlinecite{Baranowski2020}], $^m$ Ref.~[\onlinecite{Galkowski2016}], $^n$  Ref.~[\onlinecite{Baranowski2024}]
}
\label{table.1}
\end{table*}

\section{Case-by-Case Analysis of 12 Materials} \label{sec.case_by_case}

\begin{figure*}[htb!]
 \begin{center}
\includegraphics[width=0.95\textwidth]{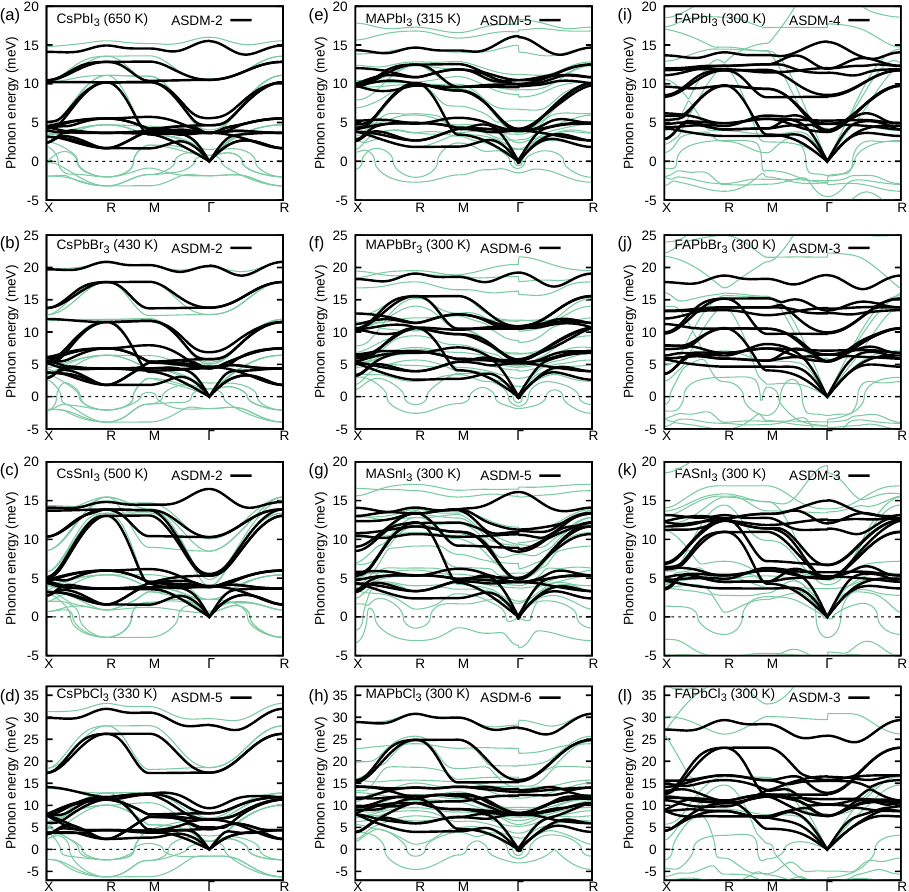}
 \end{center} 
\caption{ Anharmonic phonon dispersions at finite temperatures (black curves) calculated for the cubic phases of each material 
with the ASDM. On each plot we indicate the ASDM-$i$ iteration for which convergence is achieved. The green  
curve represents the harmonic phonon dispersion obtained for the monomorphous structure. 
All phonon dispersions include long-range dipole-dipole interaction corrections, leading to LO-TO splitting.
\label{fig8} 
}
\end{figure*}

\begin{figure*}[p]
 \begin{center}
\includegraphics[width=0.84\textwidth]{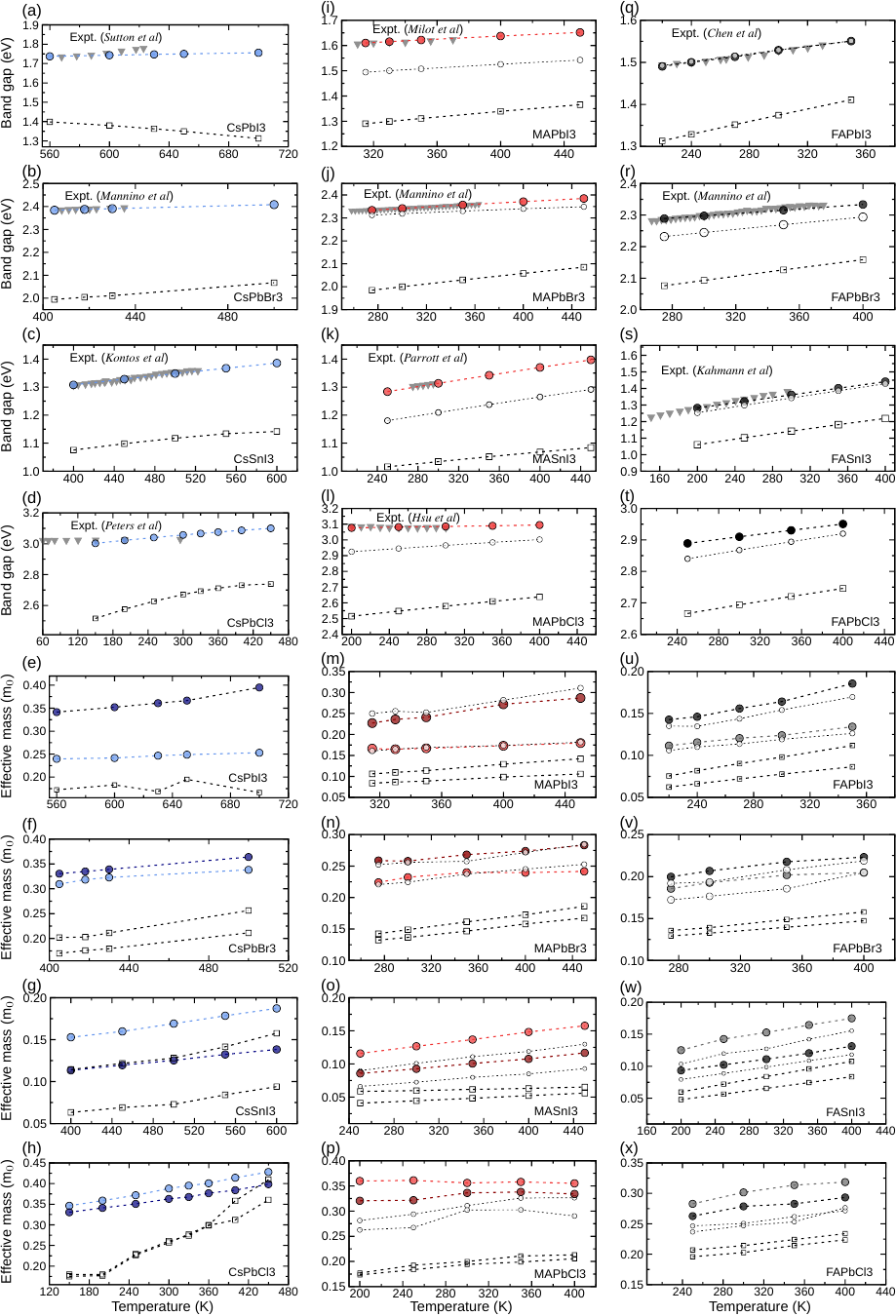}
 \end{center} 
\vspace*{-0.5cm}
\caption{Temperature-dependent band gaps and carrier effective masses calculated with the ASDM and Boltzmann averages [Eq.~\eqref{eq11}]
over 10 ZG polymorphous (coloured) and 10 ZG reference (open squares) structures. For hybrid compounds, we show the result
using one ZG polymorphous configuration derived from the MSS structure (open circles). Experimental band gaps (gray triangles)
are from Refs.~[\onlinecite{Milot2015,Kontos2018,Parrott2016,Sutton2018,Hsu2019,Mannino2020,Chen2020,Kahmann2020,Peters2022,Lpez2025}].
Dark blue, dark red, and black in (e-h), (m-p), and (u-x) indicate the hole effective masses. Light blue, light red, and gray 
in (e-h), (m-p), and (u-x) indicate the electron effective masses.
\label{fig9} 
}
\end{figure*}

In this section, we discuss our results for each cubic halide perovskite considered in this study: 
CsPbI$_3$, CsPbBr$_3$, CsSnI$_3$, CsPbCl$_3$, MAPbI$_3$, MAPbBr$_3$, MASnI$_3$, MAPbCl$_3$,
FAPbI$_3$, FAPbBr$_3$, FASnI$_3$, and FAPbCl$_3$. 
We follow this specific order to maintain consistency with the companion paper~\cite{short}, which focuses 
on APbI$_3$, APbBr$_3$, and ASnI$_3$. In this work, we provide further analysis and results on these systems 
and also include findings for Cl-based compounds.
We present first-principles calculations of the band gaps and effective masses at the DFT level for both 
reference and polymorphous structures (Table~\ref{table.1}). For the band gaps, we additionally report results 
obtained using the PBE0 hybrid functional~\cite{Perdew1996}, which has been found to reliably describe the band gaps of 
three-dimensional halide perovskites~\cite{Zacharias2023npj,Garba2025}.
All reported effective masses and reduced effective masses are in units of the free electron mass $m_0$. 
Temperature-dependent anharmonic phonon dispersions are shown in 
Fig.~\ref{fig8}. TO and LO phonon frequencies are reported in Table~\ref{table.2}, which are 
compared  against experimental results from various techniques.
Furthermore, in Fig.~\ref{fig9}, we report temperature-dependent band gaps and effective masses computed using the ASDM. A comparative analysis 
between the reference and polymorphous structures is provided in Table~\ref{table.1} as well as in Figs.~\ref{fig9} and~\ref{fig10}, highlighting the impact of 
local disorder on electron-phonon coupling and thermal lattice expansion contributions to band gap renormalization. 
In Sec.~\ref{Sec.Overdamped_ph}, we also investigate the influence 
of positional polymorphism on vibrational dynamics through calculations of phonon spectral functions, shown in Fig.~\ref{fig11}. 
We validate our computational methodology and findings by comparing our results with experimental measurements.

\subsection{CsPbI$_3$}

We first describe our results for cubic CsPbI$_3$ without considering thermal effects. 
Our DFT-PBEsol calculations for the reference (monomorphous) structure reveal a semimetallic behavior, 
characterized by a spurious exchange in orbital character between the conduction and valence band extrema 
[Fig.~\ref{fig12}(a)].
We thus assign a negative value to the band gap of $-0.05$~eV, as reported in Table~\ref{table.1}. 
Accounting for positional polymorphism resolves the orbital exchange problem [Fig.~\ref{fig12}(b)] and yields a
band gap widening of 0.65~eV relative to the monomorphous structure.
This band gap increase is the largest among all compounds studied in this work, consistent 
with the largest energy lowering reported for CsPbI$_3$ in Fig.~\ref{fig2}(b). 
Going beyond DFT and using the PBE0 functional to improve the description of electronic correlations, 
we obtain a band gap of 1.15~eV for the monomorphous structure. 
This value is notably lower than the experimental value of 1.78~eV at 650 K~\cite{Sutton2018}. However, for the locally 
disordered network, the theoretical value improves significantly, yielding 1.84~eV, showing agreement 
with the experimental value.

At the DFT-PBEsol level, the sign of the effective masses 
is also reversed due to the orbital exchange character in the monomorphous structure, leading to extremely 
small hole and electron effective masses of $m^*_{\rm h, ref}=-0.02$ and
$m^*_{\rm e, ref}=-0.02$. As with the band gap, incorporating
local disorder corrects this issue leading to $m^*_{\rm h, poly}=0.21$ 
and $m^*_{\rm e, poly}=0.17$.
These corrections correspond to hole and electron effective mass enhancement factors 
due to polymorphism of $\lambda_{\rm h} = 11.5$ and $\lambda_{\rm e} = 9.5$, respectively; for the 
definition of $\lambda$ refer to Sec.~\ref{sec.el-ph_calc}. 
We note that a more meaningful comparison is with the effective masses 
reported for the orthorhombic structure at the DFT level, which are
$m^*_{\rm h, ortho}=0.12$ and $m^*_{\rm e, ortho}=0.14$~\cite{Sutton2018}, as this structure accommodates 
octahedral tilting, which is absent in the monomorphous cubic phase. 
In this case, the effective mass enhancements captured by polymorphism 
are $\lambda_{\rm h} = 0.75$ and $\lambda_{\rm e} = 0.21$. Furthermore, our calculated effective masses $m^*_{\rm h, poly}=0.21$
and $m^*_{\rm e, poly}=0.17$ agree well with $GW$ values reported for the orthorhombic CsPbI$_3$ of 
$m^*_{{\rm h}, GW}=0.24$ and $m^*_{{\rm e}, GW}=0.23$~\cite{Sutton2018}. 
Our value for the reduced effective mass of polymorphous CsPbI$_3$ is $\mu_{\rm poly}= 0.094$ which is close to
the experimental value~\cite{Yang2017} of $\mu = 0.114 \pm 0.01$ at $T = 2$~K.
Our value also agrees with previous $GW$ calculations for orthorhombic 
CsPbI$_3$ reporting a reduced effective mass of 0.093~\cite{Filip2021} and 0.12~\cite{Sutton2018}.

We now examine the anharmonic phonon dispersion.
The cubic phase of CsPbI$_3$ stabilizes above 539~K~\cite{Marronnier2018,Sutton2018} 
and thus we focus on temperatures around this value. 
In Fig.~\ref{fig8}(a), we present the calculated ASDM phonon dispersion at 650 K, which converges as early as 
the second iteration, fully alleviating the instabilities present in the harmonic approximation for the monomorphous structure. 
Our calculations for the TO modes yield $\w_{{\rm \, TO}_1} = 3.68$, $\w_{{\rm \, TO}_3} = 4.7$, and  $\w_{{\rm \, TO}_4} = 10.7$~meV, 
each of which is doubly degenerate, while $\w_{{\rm \, TO}_2}=3.72$~meV remains triply degenerate corresponding to the silent mode. 
For the LO modes, we obtain  $\w_{{\rm \, LO}_1} = 3.71$, $\w_{{\rm \, LO}_2} = 5.6$, and  $\w_{{\rm \, LO}_3} = 15.6$~meV.
Our values compare well with THz-TDS measurements on thin films containing cubic shaped CsPbI$_3$ nanocrystals of about 
15~nm~\cite{Andrianov2019} reported in Table~\ref{table.1}. 
Our values for $\w_{{\rm \, LO}_1} $ and $\w_{{\rm \, LO}_2}$ also agree well 
with the experimental values of 3.3 and 5.3~meV obtained from exciton-phonon sidebands in the 
PL spectra of CsPbI$_3$ nanocrystals of size 9.31~nm at 3~K~\cite{Lv2021}. 
Deviations between our calculations, which are based on single-crystal reference structures, and experimental 
results on nanocrystals are expected due to size confinement effects, surface ligand interactions, and structural disorder. 
The latter is addressed in Sec.~\ref{Sec.Overdamped_ph}. These factors can alter the lattice dynamics, leading to shifts 
or broadening of phonon modes.

Now we consider the impact of thermal effects, arising from electron-phonon coupling and 
thermal expansion, on the band gap and effective masses. 
Accounting for the extra renormalization on the electronic structure due to thermal effects  
we obtain a band gap opening of $\Delta E_{\rm g, ref} (650 {\rm ~K})=0.39$~eV and 
$\Delta E_{\rm g, poly} (650 {\rm ~K})=0.13$~eV for the ZG reference and ZG polymorphous structures, 
respectively (see Table~\ref{table.1}). The corresponding electron and hole thermal effective mass enhancements are 
$\lambda_{\rm e, ref} (650 {\rm ~K}) = 11$, $\lambda_{\rm h, poly} (650 {\rm ~K})= 0.76$, 
and $\lambda_{\rm e, poly} (650 {\rm ~K})= 0.47$.
We note that we were not able to extract a hole effective mass enhancement for monomorphous cubic 
CsPbI$_3$ due to numerical instabilities related to the orbital exchange character at the DFT level. 
The large reductions due to positional polymorphism in $\Delta E_{\rm g}$ and $\lambda_{\rm e}$ at 650~K show effectively that the 
thermal induced renormalization is strongly reduced at the presence of local disorder.

In Fig.~\ref{fig9}(a) we compare our calculations for the band gap of cubic CsPbI$_3$ as a function of 
temperature using ZG polymorphous (blue discs) and ZG reference (open squares) networks with experimental data from 
Ref.~[\onlinecite{Sutton2018}] (gray triangles). To match the experimental data, we shift our 
DFT values by $\Delta_s=1.02$~eV close to our PBE0 correction of 1.24~eV (Table~\ref{table.1}); 
notably, the shift of 1.02~eV is also consistent with the $GW$ correction to DFT of 1.08~eV for orthorhombic CsPbI$_3$~\cite{Sutton2018}.
From Fig. \ref{fig9}(a), we observe that using the ZG reference structure not only underestimates the experimental values 
but also results in a band gap closure with increasing temperature, yielding a slope 
of $dE_{\rm g} / dT = -6.1 \times 10^{-4}$~eV/K. 
This incorrect trend is also reflected in previous 
electron-phonon calculations for cubic CsPbI$_3$ in Refs.~[\onlinecite{Saidi2016}] and~[\onlinecite{Ning2022}].
As shown in Fig.~\ref{fig10}, the negative variation of the band gap with temperature when 
employing a ZG reference structure is attributed to both electron-phonon coupling and thermal lattice expansion. In fact,
both $dE_{\rm g} / dT|_{\rm EPC}$ and $dE_{\rm g} / dT|_{\rm TE}$ indicate a closing of the band gap with temperature.  
Importantly, employing ZG polymorphous networks recovers the correct trend and reverses the signs 
of $dE_{\rm g} / dT|_{\rm EPC}$ and $dE_{\rm g} / dT|_{\rm TE}$, leading to the band gap widening 
with temperature, consistent with experiments. Our analysis suggests that the electron-phonon coupling
dominates thermal induced band gap renormalization, contributing 97\% as shown in Figs.~\ref{fig10}(a) and (b).
The total $dE_{\rm g} / dT = 1.3 \times 10^{-4}$ eV/K, however, still underestimates the extracted slope 
from experimental data, which is $8.3 \times 10^{-4}$ eV/K. We attribute this discrepancy to (i) the DFT treatment 
of electron-phonon coupling~\cite{Li2024}, (ii) the neglect of thermal lattice expansion in phonon calculations, and (iii) 
insufficient experimental data above 540~K, which limits the accuracy of the extracted slope.

In Fig.~\ref{fig9}(e), we also report the temperature variation of the hole and electron effective masses in the 
range 560 -- 700~K. Using a ZG reference network, we are unable to extract hole effective masses, while the electron
effective masses exhibit numerical instabilities, fluctuating around $0.18$~$m_0$ without displaying a clear trend. 
Using ZG polymorphous networks instead, yields an almost linear variation for both electron (light blue) and 
hole (dark blue) effective masses with slopes 
$\lambda^T_{{\rm h, poly}} =  3.7 \times 10^{-4}$~$m_0 / {\rm K}$ and $\lambda^T_{{\rm e, poly}} =  1.0 \times 10^{-4}$~$m_0 / {\rm K}$. 
We note that as for the band gap, electron-phonon coupling dominates the thermal induced effective masses enhancements, 
contributing 95\% to $\lambda^T_{{\rm h, poly}}$ and 89\% to $\lambda^T_{{\rm e, poly}}$.
The reduced effective mass computed for the ZG polymorphous structures varies linearly from $\mu_{\rm poly} (560~{\rm K}) = 0.141$ 
to $\mu_{\rm poly} (700~{\rm K}) = 0.154$. 

\subsection{CsPbBr$_3$}

Here, we describe our calculations for cubic CsPbBr$_3$. 
The DFT-PBEsol calculations for the monomorphous structure yield a band gap of $0.24$~eV, as reported in Table~\ref{table.1}. 
Similar to CsPbI$_3$, polymorphism leads to a large band gap opening of 0.59~eV relative to the monomorphous structure, 
compatible with the relative large energy lowering shown in Fig.~\ref{fig2}(b). 
Combining the PBE0 functional for exchange-correlation and the monomorphous structure, we obtain a band gap
of 1.52~eV.
This value is significantly below the experimental value of 2.39~eV at 430~K~\cite{Mannino2020}.
The theoretical value improves significantly when using the locally disordered network, 
yielding 2.30 eV, which aligns more closely with the experimental value.

At the DFT-PBEsol level, the effective masses are quite small, 
with $m^*_{\rm h, ref}=0.04$ and
$m^*_{\rm e, ref}=0.04$. Incorporating
local disorder through our polymorphous networks increase 
the effective masses, resulting to $m^*_{\rm h, poly}=0.23$
and $m^*_{\rm e, poly}=0.23$ which compare well with the hole effective masses of 0.226 and $0.203 \pm 0.016$ obtained 
from $GW$ calculations for the orthorhombic phase and angle-resolved photoelectron spectroscopy measurements~\cite{Sajedi2022}. 
The hole and electron effective mass enhancement factors due to 
corrections arising from local disorder are $\lambda_{\rm h} = 4.75$ and $\lambda_{\rm e} = 4.75$. 
These large enhancement factors yield excellent agreement between the 
reduced effective mass of polymorphous CsPbBr$_3$ of $\mu_{\rm poly}= 0.115$ and the experimental 
value~\cite{Yang2017} of $0.126 \pm 0.01$ at $T = 2$~K. Furthermore, our calculated value 
compares well with the $GW$ value of 0.102~$m_0$ reported for orthorhombic 
CsPbBr$_3$ in Ref.~[\onlinecite{Filip2021}].

Now we study the anharmonic phonon dispersion and subsequently the impact on the band gap 
and effective masses arising from electron-phonon coupling and thermal expansion.
We focus on temperatures above 403~K, where the cubic phase of CsPbBr$_3$ becomes stable~\cite{He2018,Mannino2020}. 
In Fig.~\ref{fig8}(b), we show the converged ASDM phonon dispersion at 430 K (second iteration) 
in excellent agreement with those reported in Refs.~[\onlinecite{Zacharias2023npj,LaniganAtkins2021}].
For the TO modes, our calculations yield $\w_{{\rm \, TO}_1} = 4.4$, $\w_{{\rm \, TO}_3} = 5.8$, and  $\w_{{\rm \, TO}_4} = 13.8$~meV 
which are doubly degenerate, and $\w_{{\rm \, TO}_2}=4.6$~meV, which is triply degenerate. For the LO modes, 
we obtain  $\w_{{\rm \, LO}_1} = 4.5$, $\w_{{\rm \, LO}_2} = 6.9$, and  $\w_{{\rm \, LO}_3} = 20.3$~meV.
These values compare well with THz Kerr measurements~\cite{Frenzel2023} for orthorhombic CsPbBr$_3$ single crystals at 80~K, 
which reveal strongly coupled TO modes of $3.7$ and $5.4$~meV.
We note that the overestimation in theoretical values is largely attributed to the different structural phases considered, 
as the phonon dispersions of the cubic and orthorhombic phases are known to differ~\cite{Zacharias2023}.
Our calculated value of $\w_{{\rm \, LO}_3}$ agrees with the 20~meV obtained from fits 
of the one-phonon Fr\"ohlich coupling model to PL linewidth measured for CsPbBr$_3$ 
nanocrystals~\cite{Cherrette2023} with cubic symmetry and size 9.3~nm in the 15--295~K range. Corresponding fits 
to PL measurements for orthorhombic CsPbBr$_3$ single crystals in the 10--295~K range yield a value 
of 16.0~meV~\cite{Sebastian2015}. Our computed $\w_{{\rm \, LO}_3}$ also compares well with 18.9~meV obtained by 
THz-TDS measurements for a CsPbBr$_3$ pellet at room temperature.

As show in Table~\ref{table.1}, the thermal induced renormalization on the electronic structure of 
reference and polymorphous networks yields a band gap widening of $\Delta E_{\rm g, ref} (430 {\rm ~K})=0.41$ and 
$\Delta E_{\rm g, poly} (430 {\rm ~K})=0.16$~eV. The corresponding hole and electron effective mass enhancements are 
$\lambda_{\rm h, ref} (430 {\rm ~K}) = 4.25$, $\lambda_{\rm e, ref} (430 {\rm ~K})=3.5$, 
$\lambda_{\rm h, poly} (430 {\rm ~K})= 0.48$ and $\lambda_{\rm e, poly} (430 {\rm ~K})= 0.43$. 
The large reductions due to positional polymorphism in $\Delta E_{\rm g}$ and $\lambda$ at 430~K can be 
explained by the strong reduction in electron-phonon self-energy corrections arising from the weaker response of 
electrons to lattice vibrations in a disordered environment~\cite{short}.

In Fig.~\ref{fig9}(b) we compare our temperature-dependent band gap calculations using 
ZG polymorphous (blue discs) and ZG reference (open squares) cubic CsPbBr$_3$ 
with experiments from Ref.~[\onlinecite{Mannino2020}] (gray triangles). Our DFT calculations are shifted
 by $\Delta_s = 1.38$~eV to match the experimental data; $\Delta_s$ compares well with our PBE0 correction of 1.45~eV 
obtained for the polymorphous structure (see Table~\ref{table.1}).
Our calculations reveal that using the ZG reference structure yields 
a band gap opening with temperature but underestimates the experimental values by more than 0.35~eV, which 
cannot be explained by more accurate treatments to exchange-correlation or electron-phonon corrections. 
Furthermore, our calculated slope of $dE_{\rm g} / dT = 7.7 \times 10^{-4}$~eV/K
overestimates the experimental value~\cite{Mannino2020} of $3.4 \times 10^{-4}$~eV/K by approximately 125~\%. 
As illustrated in Fig.~\ref{fig10}, both $dE_{\rm g} / dT|_{\rm EPC}$ and $dE_{\rm g} / dT|_{\rm TE}$ 
computed for the ZG reference structure contribute to the band gap widening with temperature, 
with the electron-phonon contribution accounting for 82\%, as shown in Figs.~\ref{fig10}(a) and (b).
In contrast, the use of ZG polymorphous networks significantly suppresses both the electron-phonon coupling 
and thermal expansion contributions, with electron-phonon coupling dominating eventually by 94~\%.
The total $dE_{\rm g} / dT = 2.6 \times 10^{-4}$ eV/K calculated using ZG polymorphous structures
is in good agreement with the experimental slope of $3.4 \times 10^{-4}$~eV/K, as illustrated in Fig.~\ref{fig10}(c).

Figure~\ref{fig9}(f) shows the variation of the hole and electron effective masses with temperature in the 
range 400 -- 500~K calculated for the ZG reference (open squares) and ZG polymorphous (blue discs) networks. 
Our analysis for the slopes yields
$\lambda^T_{{\rm h, ref}}  =  6.0 \times 10^{-4}$ 
and $\lambda^T_{{\rm e, ref}}  =  4.3 \times 10^{-4}$, 
$\lambda^T_{{\rm h, poly}} = 3.5  \times 10^{-4}$,
 and $\lambda^T_{{\rm e, poly}} = 2.7  \times 10^{-4}$~$m_0 / {\rm K}$, showing
that the presence of local disorder decreases the slopes by approximately 40~\%. 
We also note here that electron-phonon coupling contributes to more than 90\% to the thermal induced 
effective masses enhancements for both the monomorphous and polymorphous structures. 
The reduced effective mass obtained for ZG polymorphous cubic CsPbBr$_3$ varies from 
$\mu_{\rm poly}(405~{\rm K}) = 0.159$ to $\mu_{\rm poly}(500~{\rm K}) = 0.175$. 

\subsection{CsSnI$_3$}

In this section, we discuss our results for cubic CsSnI$_3$. 
Similar to CsPbI$_3$, our DFT-PBEsol calculations yield a semimetallic behavior for monomorphous CsSnI$_3$, 
due to the spurious exchange in orbital character between the band extrema.
We thus assign a negative value to the band gap of $-0.27$~eV, as reported in Table~\ref{table.1}. 
Taking into account positional polymorphism addresses the band inversion issue and results in a 0.56~eV 
widening of the band gap compared to the monomorphous structure.
By using the PBE0 functional for improved treatment of electronic correlations, we obtain a band gap
of 0.67 eV for the monomorphous structure.
This value is significantly lower than the experimental value of 1.35~eV at 500~K~\cite{Kontos2018}. 
Accounting for local disorder yields a PBE0 value of 1.30~eV, which matches the experimental result. 

The DFT-PBEsol effective masses exhibit a reversal in sign due to the orbital exchange characteristics of 
the monomorphous structure, resulting in hole and electron effective masses 
of $m^*_{\rm h, ref} = -0.07$ and $m^*_{\rm e, ref} = -0.10$. As with the band gap, 
the incorporation of local disorder corrects this problem, yielding hole and electron effective masses 
of $m^*_{\rm h, poly} = 0.07$ and $m^*_{\rm e, poly} = 0.09$. 
These values compare well with quasiparticle self-consistent $GW$ calculations for the monomorphous cubic
structure, yielding $m^*_{{\rm h}, GW} = 0.069$ and light-electron effective mass $m^*_{{\rm e}, GW} = 0.068$~\cite{Huang2013}.
However, this agreement is fortuitous, as the $GW$ results do not include SOC, whose omission introduces a significant error 
together with the neglect of local disorder. In general, SOC effects significantly influence the effective masses in both the 
monomorphous and polymorphous phases of cubic halide perovskites~\cite{Zacharias2023npj}. 
Our corrections due to local disorder in cubic CsSnI$_3$ correspond 
to effective mass enhancement factors of $\lambda_{\rm h} = 1$ and $\lambda_{\rm e} = 0.9$, respectively, 
and result in a reduced effective mass of $\mu_{\rm poly} = 0.04$, the smallest among 
all inorganic compounds studied here.

Here, we examine the ASDM phonon dispersion and how electron-phonon coupling and thermal expansion affect 
the band gap and effective masses of cubic 
CsSnI$_3$. The cubic phase of CsSnI$_3$ is observed above 440~K~\cite{Kontos2018} 
and as such, we perform calculations using temperatures around this value. 
Figure~\ref{fig8}(c) shows the anharmonic phonon dispersion at 500 K, which converges at 
the second ASDM iteration and agrees well with previous calculations within the SCP theory~\cite{Patrick2015,Zacharias2023}. 
The zone-centered modes lie at $\w_{{\rm \, TO}_1} = 3.9$, $\w_{{\rm \, TO}_2}=4.0$, $\w_{{\rm \, TO}_3} = 5.3$, and  $\w_{{\rm \, TO}_4} = 10.3$~meV, 
$\w_{{\rm \, LO}_1} = 4.0$, $\w_{{\rm \, LO}_2} = 5.6$, and  $\w_{{\rm \, LO}_3} = 16.5$~meV, close to 
those reported for CsPbI$_3$ at 650~K. Moreover, our values of $\w_{{\rm \, LO}_1} $, $\w_{{\rm \, LO}_2}$ compare well 
with the PL phonon sidebands of 3.3 and 5.3~meV obtained for CsPbI$_3$ nanocrystals at 3~K~\cite{Lv2021}. 

The electronic structure renormalization due to electron-phonon coupling and thermal expansion
leads to an overall band gap opening of $\Delta E_{\rm g, ref} (500 {\rm ~K})=0.52$~eV and 
$\Delta E_{\rm g, poly} (500 {\rm ~K})=0.20$~eV as reported for the reference and polymorphous structures 
in Table~\ref{table.1}. The effective mass enhancements are 
$\lambda_{\rm h, ref} (500 {\rm ~K}) = 2.86$, $\lambda_{\rm e, ref} (500 {\rm ~K}) = 1.8$ 
$\lambda_{\rm h, poly} (500 {\rm ~K})= 0.71$, and $\lambda_{\rm e, poly} (500 {\rm ~K})= 0.89$, showing again 
a large impact of positional polymorphism on thermal induced corrections.

\begin{figure*}[p]
 \begin{center}
\includegraphics[width=0.75\textwidth]{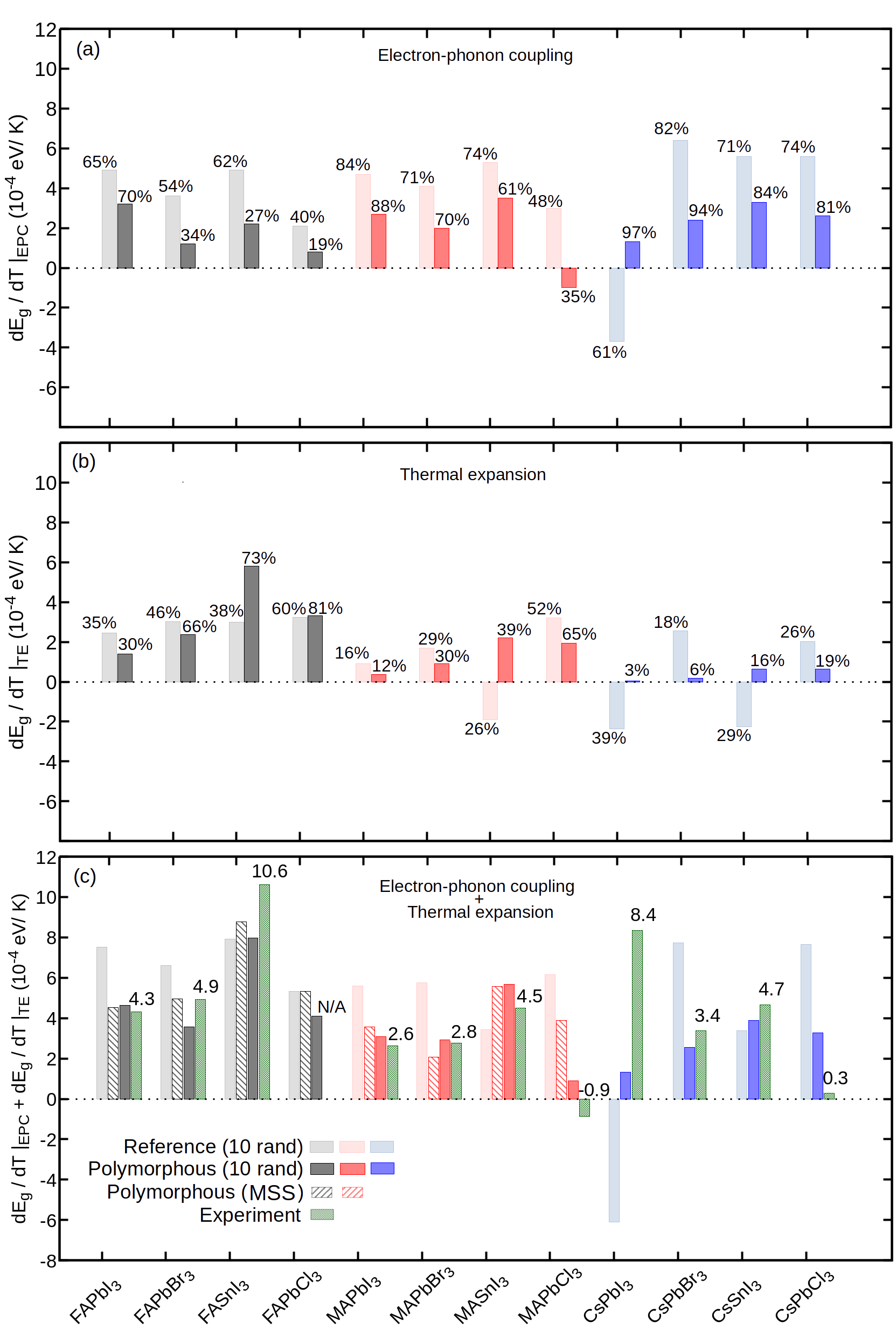}
 \end{center} 
\caption{ (a,b,c) Contributions of electron-phonon coupling $dE_{\rm g} / dT|_{\rm EPC}$ (a) and thermal lattice 
expansion $dE_{\rm g} / dT|_{\rm TE}$ (b) to the total temperature coefficient of the band gap $dE_{\rm g} / dT$ (c). 
The total $dE_{\rm g} / dT$ values correspond to slopes obtained from linear fits to the calculated band gaps shown 
in Fig.~\ref{fig9}. Light red, light blue, and light gray bars represent average data obtained for 10 ZG reference structures. 
Red, blue, and gray bars represent average data obtained for 10 ZG polymorphous structures. In (c), we also include data obtained 
for a single ZG polymorphous configuration of the MSS structure (patterned bar). Experimental data (green bars) are extracted by linear fits 
to the measurements~\cite{Milot2015,Kontos2018,Parrott2016,Sutton2018,Hsu2019,Mannino2020,Chen2020,Kahmann2020,Peters2022,Lpez2025} reported in Fig.~\ref{fig9}. 
The percentages in (a) and (b) indicate the contributions of $dE_{\rm g} / dT|_{\rm EPC}$ and $dE_{\rm g} / dT|_{\rm TE}$ to their total.
In (c), the numbers indicate the experimental values in units of $10^{-4}$ eV/K. 
\label{fig10} 
}
\end{figure*}

In Fig.~\ref{fig9}(c), we report the band gap of cubic CsSnI$_3$ as a function of temperature 
computed using ZG polymorphous (blue discs) and ZG reference (open squares) structures. We compare our calculations
 with measurements from Ref.~[\onlinecite{Kontos2018}] (gray triangles). 
Our DFT values are shifted by $\Delta_s = 0.86$~eV, which compares well with the PBE0 correction of 
1.02~eV obtained using the polymorphous structure (Table~\ref{table.1}). 
The temperature-dependent band gaps calculated for the reference structure exhibit a band gap 
widening with increasing temperature, but underestimate the experimental values by over 0.2~eV.
The calculated band gap temperature coefficient is $dE_{\rm g} / dT = 3.4 \times 10^{-4}$~eV/K
for the reference structure which compares well with the corresponding coefficient extracted from experimental data of 
$4.7 \times 10^{-4}$~eV/K. However, this agreement is fortuitous, 
as electron-phonon coupling and thermal expansion contribute with opposite
sign, as shown in Figs.~\ref{fig10}(a) and (b). In fact, the computed band gap reduction with lattice expansion 
is due to the artificial orbital exchange character in reference cubic CsSnI$_3$. Reversing the sign of this 
contribution yields $dE_{\rm g} / dT = 7.9 \times 10^{-4}$~eV/K which overestimates
the experimental value by 68\%.
Importantly, the use of polymorphous networks results in a positive $dE_{\rm g} / dT_{\rm TE}$ 
and an overall $dE_{\rm g} / dT = 3.9 \times 10^{-4}$ eV/K, which is in close agreement with 
experiments. Similar to the other Cs-based compounds, our analysis shows a dominant
electron-phonon coupling contribution of 84\%. 

In Fig.~\ref{fig9}(g), we show the temperature dependence of the hole and electron effective masses of 
cubic CsSnI$_3$ in the range 400 -- 600~K. 
The slopes obtained for the data of the reference (open squares) and polymorphous (blue discs)
structures are $\lambda^T_{{\rm h, ref}}  =  2.1 \times 10^{-4}$,
and $\lambda^T_{{\rm e, ref}}  =  1.5 \times 10^{-4}$,
$\lambda^T_{{\rm h, poly}} = 1.2  \times 10^{-4}$,
 and $\lambda^T_{{\rm e, poly}} = 1.7  \times 10^{-4}$~$m_0 / {\rm K}$. 
Similar to the other polymorphous compounds, electron-phonon coupling 
is the dominant effect contributing to more than 90\% to the thermal induced 
effective masses enhancements. For polymorphous cubic CsSnI$_3$, we evaluate a 
reduced effective mass of $\mu_{\rm poly}(400~{\rm K}) = 0.065$ and $\mu_{\rm poly}(600~{\rm K}) = 0.080$. 

\subsection{CsPbCl$_3$}

Here, we present our findings for cubic CsPbCl$_3$. The DFT-PBEsol calculations for the monomorphous structure 
yield a band gap of 0.59~eV, as shown in Table~\ref{table.1}. Similar to other inorganic Pb-based compounds, 
polymorphism induces a substantial band gap widening of 0.60~eV compared to the monomorphous structure.
By applying the PBE0 functional for exchange-correlation to the monomorphous structure, we obtain a band gap
of 2.16~eV which underestimates the experimental value of $3.02$~eV at 295~K, obtained from PL measurements of the 
free exciton peak~\cite{Peters2022}. However, when local disorder is taken into account, the PBE0 band 
gap opens to 2.79~eV, improving agreement with the experimental result.

The effective masses calculated for the monomorphous structure
are $m^*_{\rm h, ref}=0.12$ and
$m^*_{\rm e, ref}=0.13$. Using locally disordered structures,
the effective masses increase to $m^*_{\rm h, poly}=0.29$ 
and $m^*_{\rm e, poly}=0.31$ which correspond to effective mass enhancement factors of
$\lambda_{\rm h} = 1.42$ and $\lambda_{\rm e} = 1.38$. 
The corrections arizing from polymorphism improve considerably the agreement with 
the experimental reduced effective mass of $0.202\,\pm\,0.01$ at $T = 2$~K~\cite{Baranowski2020}, 
increasing the calculated value for CsPbCl$_3$ from $\mu_{\rm ref}=0.062$ to $\mu_{\rm poly}= 0.150$. 
Despite this important improvement, our computed $\mu_{\rm poly}$ still underestimates experiment by 25\%, 
which we attribute to neglecting quasiparticle and excitonic effects~\cite{Filip2021,Lang2015} from our calculations. 

We now examine the anharmonic phonon dispersion and subsequently the thermally induced renormalization of the 
band gap and effective masses. 
The cubic phase of CsPbCl$_3$ stabilizes above 325~K~\cite{He2021}. 
In Fig. \ref{fig8}(d), we present the converged ASDM phonon dispersion at 330~K, achieved at the fifth iteration, 
where phonon energies extend up to 30 meV. This range is broader compared to other inorganic halide perovskites, 
owing to the small mass of Cl.
The energies of the degenerate TO modes at the $\Gamma$ point are $\w_{{\rm \, TO}_1} = 4.6$, $\w_{{\rm \, TO}_2}=6.8$,
$\w_{{\rm \, TO}_3} = 8.2$, and  $\w_{{\rm \, TO}_4} = 17.4$~meV and of the LO modes are
$\w_{{\rm \, LO}_1} = 5.0$, $\w_{{\rm \, LO}_2} = 9.4$, and  $\w_{{\rm \, LO}_3} = 28.2$~meV.
The value of $\w_{{\rm \, LO}_1}$ compares well with the LO phonon frequency of 4.3~meV, obtained from an one-phonon model fit to 
the PL linewidths of single crystals in the 185-295~K range, as reported in Ref.~[\onlinecite{Sebastian2015}].
Our data is also compared with Raman measurements~\cite{Guilloux2023} of TO modes at 170~K and 
for LO modes at 2~K for cubic-shaped CsPbCl$_3$ nanocrystals of 8~nm, as shown in Table~\ref{table.2}. 
Deviations between theory and experiment can be attributed to various factors, including quantum confinement, surface effects, 
the dielectric screening environment, and variations in locally disordered domains, as well as the computational setup, such as the choice 
of functional and lattice parameters.
Notably, our calculated phonon energies compare better with data obtained from infrared spectroscopy data on
thin films and single crystals of MAPbCl$_3$ at 300 K~\cite{Sendner2016}, which report TO phonon energies of 8.2 and 11.0~meV and LO phonon
energies of 8.7 and 27.9~meV.  

As reported in Table~\ref{table.1}, the thermal induced renormalization on the electronic structure of
 monomorphous and polymorphous networks yields a band gap widening of $\Delta E_{\rm g, ref} (330 {\rm ~K})=0.33$~eV and 
$\Delta E_{\rm g, poly} (330 {\rm ~K})=0.10$~eV. The corresponding hole and electron effective mass enhancements are 
$\lambda_{\rm h, ref} (330 {\rm ~K}) = 1.33$, $\lambda_{\rm e, ref} (330 {\rm ~K})=1.08$, 
$\lambda_{\rm h, poly} (330 {\rm ~K})= 0.24$ and $\lambda_{\rm e, poly} (330 {\rm ~K})= 0.26$, showing the local disorder 
surpresses thermal induced corrections at 330~K.  

Figure~\ref{fig9}(d) shows the band gap of cubic CsPbCl$_3$ in the temperature range 150--450~K
obtained using the ZG polymorphous (blue discs) and ZG reference (open squares) structures.
While the cubic phase of CsPbCl$_3$ stabilizes above 325~K~\cite{He2021}, we also conduct calculations at temperatures 
as low as 150~K, where temperature-dependent PL measurements of the free exciton peak (gray triangles) 
are available~\cite{Peters2022}.
DFT calculations are shifted by $\Delta_s = 1.77$~eV to facilitate comparison with experiment, close to the PBE0 correction 
of 1.60~eV obtained for the polymorphous structure (see Table~\ref{table.1}).
Our data for the ZG reference structure show a band gap widening with temperature, where the rate of 
change gradually decreases.
Fitting a straight line to our data within the temperature range 
of 150--450~K yields a slope of $dE_{\rm g} / dT = 7.6 \times 10^{-4}$~eV/K, 
which significantly overestimates $0.3 \times 10^{-4}$~eV/K extracted from experimental data in the range 60--295~K.
As illustrated in Fig.~\ref{fig10}, both $dE_{\rm g} / dT|_{\rm EPC}$ and $dE_{\rm g} / dT|_{\rm TE}$ 
computed using the ZG reference structure contribute to the band gap widening with temperature, 
with the electron-phonon contribution accounting for 74\%, as illustrated in Figs.~\ref{fig10}(a) and (b). 
When we employ ZG polymorphous networks both effects become less pronounced, 
with electron-phonon coupling contributing 81\% to the total $dE_{\rm g} / dT = 3.2 \times 10^{-4}$ eV/K.
Although this value shows better agreement, it still significantly overestimates 
the experimental slope of $0.3 \times 10^{-4}$~eV/K, unlike in the case of CsPbBr$_3$ 
where our calculated slope matches closely experiments. 
We attribute this different behavior to the fact that
excitonic effects, which are neglected from our calculations, play a more important role 
in CsPbCl$_3$, as evidenced by its higher binding energy and stronger exciton interactions~\cite{Filip2021,Fasahat2025}. 

In Fig.~\ref{fig9}(h), we present hole and electron effective masses as a function of temperature 
calculated for the ZG reference (open squares) and ZG polymorphous (blue discs) structures. 
We observe that the effective masses of the polymorphous structure follow a clear linear trend, 
which is more physically consistent, in contrast to the behavior observed for the reference structure.
Fitting a straight line to the data yields the slopes 
$\lambda^T_{{\rm h, ref}}  =  8.0 \times 10^{-4}$, $\lambda^T_{{\rm e, ref}}  =  6.2 \times 10^{-4}$, 
$\lambda^T_{{\rm h, poly}} = 2.2  \times 10^{-4}$,  and $\lambda^T_{{\rm e, poly}} = 2.7  \times 10^{-4}$~$m_0 / {\rm K}$, showing
that local disorder decreases significantly the variation of the effective masses with temperature. 
Similar to our analysis for the band gap, electron-phonon coupling contribution dominates 
thermal induced renormalization of the effective masses being more than 90\%. 
Finally, we report that the reduced effective mass calculated for polymorphous cubic CsPbCl$_3$ 
varies linearly from $\mu_{\rm poly}(150~{\rm K}) = 0.169$ to $\mu_{\rm poly}(450~{\rm K}) = 0.206$. 

\begin{table*}[p]
\caption{Optical phonon energies (in meV) at the $\Gamma$ point. In parenthesis, 
we give values of the LO phonons before LO-TO splitting. 
For Cs-based compounds, $\w_{{\rm \, TO}_2}$ is a silent triply degenerate mode, and the other TO modes are infrared-active and doubly degenerate.
Experimental data on single crystals (SCs), polycrystalline pellets (PCPs), thin films (TFs), nanocrystals (NCs), and NC films (NCFs) with 
temperatures are reported. Experiments refer to the following spectroscopy and scattering techniques: 
Photoluminescence (PL), THz Kerr, Raman, Infrared (IR), Inelastic Neutron Scattering (INS), and
Terahertz Time-Domain Spectroscopy (THz-TDS). The temperature range used for fitting the PL linewidth (LW) is also indicated. 
}
\label{table.2}
\centering
\begin{tabular*}{\textwidth}{@{\extracolsep{\fill}}l c c c c c c c c | c c c c}
\hline \hline
\textbf{} & $\w_{{\rm \, TO}_1}$ & $\w_{{\rm \, TO}_2}$ & $\w_{{\rm \, TO}_3}$ & $\w_{{\rm \, TO}_4}$ & $\w_{{\rm \, TO}_5}$ & $\w_{{\rm \, TO}_6}$ & $\w_{{\rm \, TO}_7}$ & $\w_{{\rm \, TO}_8}$ & $\w_{{\rm \, LO}_1}$ & $\w_{{\rm \, LO}_2}$ & $\w_{{\rm \, LO}_3}$ & $\w_{{\rm \, LO}_4}$ \\
\hline
CsPbI$_3$, 650~K    & 3.68  & 3.72  & 4.7 & 10.7 & - & - & - & - & 3.71 (3.68)  & 5.6 (4.7)  & 15.6 (10.7)  & - \\
THz-TDS, NCF, 300~K~\cite{Andrianov2019} & 4.1  & 5.9 & 8.8 & 11.2 & - & - & - & - &  4.8  & 6.9 & 9.3  & 15.2 \\
PL replica, NC, 3~K~\cite{Lv2021} &  \multicolumn{8}{c|}{\,} & 3.3 & 5.3 &  &  \\  [4pt] 
CsPbBr$_3$, 430~K   & 4.4  & 4.6  & 5.8 & 13.8 & - & - & - & - & 4.5 (4.4)  & 6.9 (5.8)  & 20.3 (13.8)  & - \\
THz Kerr, SC, 80 K~\cite{Frenzel2023} & \multicolumn{2}{c}{3.7} & 5.4 & \multicolumn{5}{c|}{\,} &  & &  &  \\  
PL LW, NC, 15-295 K~\cite{Cherrette2023} &  \multicolumn{8}{c|}{\,} &  &  & 20.0 &  \\  
PL LW, SC, 10-295 K~\cite{Sebastian2015} &  \multicolumn{8}{c|}{\,} &  &  & 16.0 & \\  
THz-TDS, PCP, 300 K \cite{Nagai2018}      &  \multicolumn{8}{c|}{\,} &  &  & 18.9  & \\  [4pt]  
CsSnI$_3$, 500~K  & 3.9  & 4.0  & 5.3 & 10.3 & - & - & - & - & 4.0 (3.9) & 5.6 (5.3)  & 16.5 (10.3)  & - \\ [4pt]
CsPbCl$_3$, 330~K   & 4.6  & 6.8  & 8.2 & 17.4 & - & - & - & - & 5.0 (4.6)  & 9.4 (8.2)  & 28.2 (17.4)  & - \\
PL LW, SC, 185-295 K \cite{Sebastian2015} &  \multicolumn{8}{c|}{\,} & 4.3 & &  &   \\ 
Raman, NC, 170 \& 2 K \cite{Guilloux2023} & 3.1  & 6.0  & 8.4 & 9.8 & 13.5 & 14.5 &  &  &   & 14.2  & 25.0  & \\ [4pt]
MAPbI$_3$, 315~K & 3.9 & 3.94 & 3.94 & 4.0 & 9.6 & 9.98 & 10.0 & 10.4 & 3.93 (3.9) & 4.2 (3.94) & 10.4 (10.0) & 16.2 (10.4) \\
IR, TF/SC, 300 K \cite{Sendner2016} &  \multicolumn{4}{c}{4.0} & \multicolumn{4}{c|}{7.8} & \multicolumn{2}{c}{5.0} & & 16.5  \\ 
IR, SC, 300 K \cite{Boldyrev2020} &  \multicolumn{4}{c}{4.2} & \multicolumn{4}{c|}{8.3} & \multicolumn{2}{c}{4.6} & & 16.2  \\  
THz-TDS, SC, 300 K \cite{Nagai2018} &  \multicolumn{8}{c|}{\,} & \multicolumn{2}{c}{3.8} &  & 16.2    \\  
INS, SC, 5 K \cite{Ferreira2020} & 2.3 & \multicolumn{7}{c|}{\,} & \multicolumn{2}{c}{3.8} & 11.7 & 15.5  \\  
PL LW, TF, 150-370 K~\cite{Wright2016} &  \multicolumn{8}{c|}{\,} & & & 11.5 &   \\ [4pt] 
MAPbBr$_3$, 300 K   & 5.1 & 5.3 & 5.6 & 5.7 & 10.5 & 10.6 & 10.8 & 10.9 & 5.2 (5.1) & 5.8 (5.5)  & 10.9 (10.6) & 19.2 (10.9) \\
IR, TF/SC, 300 K \cite{Sendner2016} &  \multicolumn{4}{c}{5.6} & \multicolumn{4}{c|}{9.1} & \multicolumn{2}{c}{6.3} &  & 20.1  \\ 
IR, SC, 300 K \cite{Anikeeva2023} &  \multicolumn{4}{c}{5.9} & \multicolumn{4}{c|}{9.7} & \multicolumn{2}{c}{6.0} & & 20.7  \\  
IR, SC, 300 K \cite{Wang2019} &  \multicolumn{4}{c}{5.5} & \multicolumn{2}{c}{8.3} & \multicolumn{2}{c|}{11.9} & \multicolumn{2}{c}{6.3} & 12.0 & 21.0  \\ 
THz-TDS, SC, 300 K \cite{Nagai2018} &  \multicolumn{8}{c|}{\,} & \multicolumn{2}{c}{5.7} & 9.1 & 20.1  \\ 
INS, SC, 5~K \cite{Ferreira2020} &  \multicolumn{8}{c|}{\,} & \multicolumn{2}{c}{5.1} & 11.6 & 16.9  \\  
Raman, SC, 20 K \cite{Ferreira2020} &  \multicolumn{8}{c|}{\,} & 4.1 & 5.9 & 11.9 & 16.3  \\  
THz Kerr, SC, 80 K~\cite{Frenzel2023} & \multicolumn{4}{c}{4.7} & \multicolumn{4}{c|}{} & & & &   \\ 
PL LW, TF, 100-370 K \cite{Wright2016} &  \multicolumn{8}{c|}{\,} & & & \multicolumn{2}{c}{15.3}   \\ [4pt] 
MASnI$_3$, 300~K    & 4.47 & 4.54 & 4.85 & 5.0 & 8.4 & 8.9 & 10.7 & 11.3 & 4.48 (4.47)  & 4.92 (4.85) & 11.0 (8.4) & 16.1 (11.3) \\ [4pt]
MAPbCl$_3$, 300~K   & 7.81 & 7.99 & 8.05 & 8.3 & 12.2 & 13.0 & 15.3 & 15.7 & 7.82 (7.81) & 8.3 (8.05) & 13.6 (13.0) & 27.7 (15.7) \\
IR, TF/SC, 300 K \cite{Sendner2016} &  \multicolumn{4}{c}{8.2} & \multicolumn{4}{c|}{11.0} & \multicolumn{2}{c}{8.7} & & 27.9  \\ 
THz-TDS, SC, 300 K \cite{Nagai2018} &  \multicolumn{8}{c|}{\,} & \multicolumn{2}{c}{8.2} & 12.4 &  \\ [4pt] 
FAPbI$_3$, 300~K    & 3.79 & 3.80 & 4.8 & 5.3 & 8.3 & 8.6 & 11.9 & 12.0 & 4.0 (3.8) & 4.9 (4.8) & 11.4 (8.6) & 15.5 (11.9) \\
PL replica, NC, 3.6 K~\cite{Fu2018} &  \multicolumn{8}{c|}{\,} &  \multicolumn{2}{c}{3.2} & 11.1 & 15.4   \\  
INS, SC, 5~K~\cite{Ferreira2020} &  \multicolumn{8}{c|}{\,} & \multicolumn{2}{c}{3.9} & 9.8 & 15.1  \\  
PL LW, TF, 10-370 K \cite{Wright2016} &  \multicolumn{8}{c|}{\,} &  & & $11.5$ &   \\ [4pt] 
FAPbBr$_3$, 300~K   & 5.4 & 5.5 & 6.5 & 7.1 & 9.7 & 10.0 & 13.3 & 13.7 & 5.5 (5.4) & 6.53 (6.50) & 12.9 (10.0) & 18.8 (13.3) \\
INS, SC, 5~K~\cite{Ferreira2020} &  \multicolumn{8}{c|}{\,} & \multicolumn{2}{c}{5.5} & 13.1 & 22.2  \\  
PL replica, NC, 5.5~K~\cite{Cho2021} &  \multicolumn{8}{c|}{\,} & \multicolumn{2}{c}{4.4} &  & 18.8  \\  
PL LW, TF, 10-370 K \cite{Wright2016} &  \multicolumn{8}{c|}{\,} & & & \multicolumn{2}{c}{15.3}   \\ [4pt] 
FASnI$_3$, 300~K    & 4.93 & 4.95 & 5.21 & 5.8 & 6.7 & 6.8 & 12.1 & 12.6 & 4.98 (4.95)  & 5.24 (5.21) & 11.3 (6.7) & 15.0 (12.1) \\ [4pt]
FAPbCl$_3$, 300~K   & 7.56 & 7.62 & 10.1 & 11.4 & 12.4 & 12.9 & 15.8 & 16.6 & 7.64 (7.56) & 10.2 (10.1) & 15.5 (12.9) & 25.8 (15.8) \\ [2pt]
\hline
\hline
\end{tabular*}
\end{table*}

\subsection{MAPbI$_3$}

We now describe our results for MA-based compounds, starting with cubic MAPbI$_3$. 
The DFT-PBEsol calculations for both the monomorphous and reference structures give a 
semimetallic behavior with a band gap of $0.06$ and $0.03$~eV. 
Unlike in the case of CsPbI$_3$, no electronic band edge inversion is observed [see Figs.~\ref{fig12}(c) and (d)]. 
However, using reference structures helps alleviating the problem of an asymmetric charge distribution
induced by the net dipole moment of the MA molecule in the monomorphous structure. 
Incorporating local disorder through polymorphous networks yields an average
band gap widening of 0.49~eV relative to the reference structures.
As reported in Table~\ref{table.1}, our PBE0 calculation yields a band gap
of 1.23~eV for the reference structure which 
underestimates the experimental value of 1.61~eV at 315~K~\cite{Milot2015}. 
Using polymorphous networks, the theoretical value improves significantly, 
as we find a band gap of 1.75~eV, close to previous 
$GW$ calculations on the orthorhombic~\cite{Davies2018} and tetragonal~\cite{Umari2014} phases, reporting 
1.57 and 1.67~eV, respectively. 
$GW$ band gaps of 1.67~eV for the cubic phase were reported in Refs.~[\onlinecite{Bokdam2016}] and [\onlinecite{Brivio2014}], 
based on a ferroelectrically distorted monomorphous structure. However, this symmetry-broken configuration does not reflect
the correct macroscopic Pm$\bar{3}$m symmetry of cubic MAPbI$_3$, leading to unphysical 
results such as artificial Rashba splitting.

Similar to monomorphous cubic CsPbI$_3$, the DFT-PBEsol hole and electron effective masses are extremely 
small: $m^*_{\rm h, ref}=0.02$ and $m^*_{\rm e, ref}=0.02$. Using
locally disordered structures increases the effective masses considerably to $m^*_{\rm h, poly}=0.16$
and $m^*_{\rm e, poly}=0.13$ amounting to hole and electron effective mass enhancement factors 
of $\lambda_{\rm h} = 7$ and $\lambda_{\rm e} = 5.5$, respectively. 
The large mass enhancements lead to the increase of the reduced effective mass 
from $\mu_{\rm ref}=0.01$ to $\mu_{\rm poly}= 0.072$ improving significantly the comparison 
with the experimental value~\cite{Galkowski2016} of $\mu = 0.104$ at $T = 2$~K and 
the calculated value for orthorhombic MAPbI$_3$ of 0.11 obtained within the $GW$ 
approximation~\cite{Davies2018}.  

We now focus on the anharmonic phonon dispersion and then on the impact of thermal effects on the band gap and effective masses. 
Since the cubic phase of MAPbI$_3$ is stable only above 315~K~\cite{Kawamura2002,Milot2015}, our analysis 
is performed above this temperature regime.
In Fig.~\ref{fig8}(e), we present the ASDM phonon dispersion of the MAPbI$_3$ reference structure at 315~K, 
focusing on the energy range associated with the phonon dynamics of the inorganic network. The ASDM converges 
by the fifth iteration, effectively mitigating the instabilities across the reciprocal space obtained in the harmonic 
approximation for the ferroelectric monomorphous structure. 
In addition, the harmonic phonon dispersion exhibits discontinuities at the $\Gamma$ point, such as a splitting of 1~meV at an 
energy of approximately 15~meV. These discontinuities originate from anisotropies in the Born effective charge and high-frequency dielectric 
tensors, which are calculated for the ferroelectric monomorphous structure. These anisotropies, in turn, arise from the net dipole moment 
of the MA molecule, which breaks the macroscopic symmetry of the lattice and alters the long-range dipole-dipole interactions. As a result, 
the non-analytical contribution to the dynamical matrix is affected, leading to direction-dependent LO-TO splitting and 
discontinuities in the dispersion~\cite{Zacharias2023}.
The discontinuities at the $\Gamma$ point are alleviated in the ASDM phonon dispersion, which correctly reflects 
the isotropic cubic lattice. This is achieved by using average Born effective charge and high-frequency dielectric tensors 
calculated for the polymorphous structures. This demonstrates that, although local disorder is 
present in polymorphous structures, on average, those structures reflect the isotropy of the lattice.
The ASDM phonon dispersion of MAPbI$_3$ spans a similar 
energy range (0--15~meV) as that of CsPbI$_3$ shown in Fig.~\ref{fig8}(a), and exhibits comparable features, such 
as the ultrasoft flat band along the R-M direction and the phonon frequencies at the $\Gamma$ point. 
The differences between the two dispersions are primarily attributed to the contributions of the Cs atom 
and the MA molecule and secondarily to the different temperatures used.
Our calculated transverse optical and longitudinal optical phonon frequencies at the $\Gamma$ point for 315~K 
are $\w_{{\rm \, TO}_{1-4}} = 3.9-4.0$, $\w_{{\rm \, TO}_{5-8}} = 9.6-10.4$, 
$\w_{{\rm \, LO}_1} = 3.93$, $\w_{{\rm \, LO}_2} = 4.2$, $\w_{{\rm \, LO}_3} = 10.4$, and $\w_{{\rm \, LO}_4} = 16.2$~meV, as 
reported in Table~\ref{table.2}.
These values compare well with infrared spectroscopy measurements on single crystals at room temperature, reporting TO modes of 4.0 and 7.8~meV and 
LO modes of 5.0 and 16.5~meV as in Ref.~[\onlinecite{Sendner2016}], and TO modes of 4.2 and 8.3~meV along with LO modes of 4.6 and 16.2~meV 
as in Ref.~[\onlinecite{Boldyrev2020}]. 
Similarly, our calculated LO phonon energies agree with THz time-domain spectroscopy measurements at 
room temperature~\cite{Nagai2018}, which find values of 3.8 and 16.2~meV, 
as well as neutron scattering measurements at 5 K~\cite{Ferreira2020} of 3.8, 11.7, and 15.5~meV. 
We note that our value of $\w_{{\rm \, LO}_3} = 10.4$~meV compares well with the effective LO phonon frequency of 11.5~meV 
used to fit temperature-dependent PL linewidth data in Ref.~[\onlinecite{Wright2016}]. 
Small deviations between theory and experiment are attributed to the functional used for the DFT calculations, different lattice constants,
as well as to the fact that, due to supercell size constraints, our reference structure preserves only a subgroup of the full Pm$\bar{3}$m symmetry.

Accounting for the renormalization on the electronic structure due to electron-phonon coupling and thermal expansion,
we obtain a band gap opening of $\Delta E_{\rm g, ref} (315 {\rm ~K})=0.26$~eV and 
$\Delta E_{\rm g, poly} (315 {\rm ~K})=0.09$~eV for the reference and polymorphous structures, 
respectively (see Table~\ref{table.1}). The corresponding effective mass enhancements are 
$\lambda_{\rm h, ref} (315 {\rm ~K}) = 4.5$, $\lambda_{\rm e, ref} (315 {\rm ~K}) = 3.5$, 
$\lambda_{\rm h, poly} (315 {\rm ~K})= 0.38$, and $\lambda_{\rm e, poly} (315 {\rm ~K})= 0.31$.
The pronounced reductions in $\Delta E_{\rm g}$, $\lambda_{\rm h}$, and $\lambda_{\rm e}$ due to polymorphism demonstrate that 
thermally induced renormalization effects are significantly suppressed in the presence of local positional disorder.
Furthermore, our calculations at 315~K suggest an increase of 37\% and 31\% due to electron-phonon coupling of the hole and electron
effective masses, respectively, which compares well with the 28\% increase at 300~K calculated for both hole and 
electron effective masses in the orthorhombic phase using many-body perturbation theory~\cite{Schlipf2018}. 

In Fig.~\ref{fig9}(i), we present the temperature-dependent band gaps of cubic MAPbI$_3$ calculated 
using Boltzmann-weighted averages over 10 ZG polymorphous configurations (red discs) and their corresponding 
ZG reference structures (open squares) at each temperature.
We also report temperature-dependent band gaps computed using only one ZG polymorphous network obtained by geometry
optimization of the MSS structure (open circles). 
We compare our results with the experimental data from Ref.~[\onlinecite{Milot2015}] (gray triangles). 
To match the experimental data, we shift our DFT values by $\Delta_s = 1.00$~eV close to 
our PBE0 correction of 1.23~eV (Table~\ref{table.1}); 
notably, the 1.00~eV shift aligns with the $GW$ correction to the DFT value reported for orthorhombic MAPbI$_3$~\cite{Davies2018}.
Our analysis on data obtained for the reference structures yields a slope of $dE_{\rm g} / dT = 5.6 \times 10^{-4}$~eV/K 
which overestimates the experimental value of $2.6 \times 10^{-4}$~eV/K by 115~\%.
This overestimation is also observed in electron-phonon calculations performed on ferroelectric monomorphous 
cubic MAPbI$_3$ in Ref.~[\onlinecite{Saidi2016}].
Accounting for positional polymorphism suppresses both electron-phonon and thermal expansion 
induced band gap renormalization leading to a slope of $dE_{\rm g} / dT = 3.1 \times 10^{-4}$~eV/K, which significantly 
improves the agreement with experiment. 
Importantly, even when a single ZG polymorphous configuration is used, starting from the MSS structure, the 
extracted slope of $3.6 \times 10^{-4}$~eV/K remains in excellent agreement with experiment, as illustrated 
also by the bar plot in Fig.~\ref{fig10}(c).
Our analysis shown in Figs.~\ref{fig10}(a) and (b) suggests that electron-phonon coupling in both reference and polymorphous structures
dominates thermal band gap renormalization, contributing to 84\% and 88\%, respectively. 

In Fig.~\ref{fig9}(m), we report the temperature variation of the hole and electron effective masses in the 
range 315 -- 450~K. Using both ZG reference and ZG polymorphous networks, we obtain a linear variation of
the effective masses with temperature.  
Our analysis for the slopes yields
$\lambda^T_{{\rm h, ref}}  =  2.7 \times 10^{-4}$
and $\lambda^T_{{\rm e, ref}}  =  1.6 \times 10^{-4}$,
$\lambda^T_{{\rm h, poly}} = 4.6  \times 10^{-4}$,
 and $\lambda^T_{{\rm e, poly}} = 1.0  \times 10^{-4}$~$m_0 / {\rm K}$, showing
that the presence of local disorder increase/decrease the 
slopes $\lambda^T_{{\rm h}}$/$\lambda^T_{{\rm e}}$ by 70\%/38\%.
As for the band gap renormalization, using a single ZG polymorphous configuration obtained from the 
MSS structure results in excellent agreement with Boltzmann-weighted averages over 10 polymorphous configurations.
We remark that in all cases electron-phonon 
coupling dominates the thermal induced effective masses enhancements contributing, for example, 
97\% to $\lambda^T_{{\rm h, poly}}$ and 99\% to $\lambda^T_{{\rm e, poly}}$.
The reduced effective mass computed for the ZG polymorphous structure varies linearly from 
 $\mu_{\rm poly}(315~{\rm K}) = 0.096$ to $\mu_{\rm poly}(450~{\rm K}) = 0.110$.

\subsection{MAPbBr$_3$}

Now we present our calculations for cubic MAPbBr$_3$. 
The DFT-PBEsol band gap calculated for the reference structure is $0.39$~eV, as reported in Table~\ref{table.1}. 
Compared to CsPbBr$_3$, polymorphism leads to a similar, but smaller, band gap opening of 0.44~eV 
relative to the reference structure. 
Using the PBE0 functional, we obtain an average band gap for the reference structures of 
1.82~eV. 
This value underestimates the experimental band gap of 2.34~eV at 300~K~\cite{Mannino2020}.
Using locally disordered networks, our theoretical value shows significant improvement,
yielding 2.27~eV which considerably enhances the agreement with the experiment.
Our value also compares well with the $GW$ band gap of 2.56~eV obtained 
for the tetragonal phase of MAPbBr$_3$~\cite{Mosconi2016}.
The $GW$ overestimation of the band gap by 0.22~eV is attributed to the use of a 
phase other than cubic and to the scalar relativistic treatment of the screened Coulomb interaction $W$.

At the DFT-PBEsol level, the hole and electron effective masses are
$m^*_{\rm h, ref}=0.09$ and $m^*_{\rm e, ref}=0.09$. Incorporating
local disorder through our polymorphous networks increase 
the effective masses to $m^*_{\rm h, poly}=0.19$ 
and $m^*_{\rm e, poly}=0.18$. The effective mass enhancement factors due to these corrections 
are $\lambda_{\rm h} = 1.11$ and $\lambda_{\rm e} = 1.00$. 
These mass enhancements lead to the increase of the reduced effective mass of cubic MAPbBr$_3$
from $\mu_{\rm ref}=0.045$ to $\mu_{\rm poly}= 0.092$ improving 
considerably the comparison with the experimental values~\cite{Galkowski2016,Baranowski2024} of $0.106$ and 0.117 at $T = 2$~K. 
As with the band gap, the $GW$ calculations of Ref.~[\onlinecite{Mosconi2016}] yield an overestimated value of 0.144 
for tetragonal MAPbBr$_3$, likely due to the scalar relativistic treatment of the screened Coulomb interaction $W$.

Now we examine ASDM phonon dispersion and subsequently the effect of electron-phonon coupling and 
thermal expansion on the band gap and effective masses. 
We study temperatures above 250~K, where the cubic phase of MAPbBr$_3$ is stable~\cite{Mashiyama2003,Mannino2020}. 
Figure~\ref{fig8}(f) shows the ASDM phonon dispersion of the MAPbBr$_3$ reference structure at 300~K, 
focusing on phonons energies related to the inorganic network. The ASDM converges 
by the sixth iteration, alleviating the harmonic phonon instabilities calculated
for the ferroelectric monomorphous structure. 
As for the case of MAPbI$_3$, the discontinuities at the $\Gamma$ point 
are alleviated in the ASDM phonon dispersion by using average Born effective charge and dielectric tensors computed for 
the polymorphous structures, reflecting the isotropy of the cubic lattice.
The anharmonic phonons related to the inorganic network of MAPbBr$_3$ lie within the
energy range of 0--20~meV, similar to that of CsPbBr$_3$ shown in Fig.~\ref{fig8}(b). Apart from the contributions 
related to the Cs atom and MA molecule [see Figs.~\ref{fig14}(a,b)] and the different temperature used, both phonon dispersions 
exhibit similar characteristics, including the ultrasoft flat band along the R-M direction.
However, the frequency of the high-energy TO mode at the $\Gamma$ point of MAPbBr$_3$ is about 3 meV lower 
in energy compared to that of CsPbBr$_3$. 
Particularly, our anharmonicity calculations for MAPbBr$_3$ at 300~K yield
$\w_{{\rm \, TO}_{1-4}} = 5.1-5.7$, $\w_{{\rm \, TO}_{5-8}} = 10.5-10.9$,
$\w_{{\rm \, LO}_1} = 5.2$, $\w_{{\rm \, LO}_2} = 5.8$, $\w_{{\rm \, LO}_3} = 10.9$, and $\w_{{\rm \, LO}_4} = 19.2$~meV, as
reported in Table~\ref{table.2}.
These values are in good agreement with PL, inelastic neutron scattering, infrared, Raman, or other THz spectroscopy measurements 
for MAPbBr$_3$~\cite{Sendner2016,Wright2016,Nagai2018,Wang2019,Ferreira2020,Anikeeva2023,Frenzel2023}.
In particular, focusing on room temperature measurements, the following values are reported: 
(i) TO modes of 5.6 and 9.1~meV and LO modes of 6.3 and 20.1~meV~\cite{Sendner2016}, 
(ii) TO modes of 5.9 and 9.7~meV and LO modes of 6.0 and 20.7~meV~\cite{Anikeeva2023}, 
(iii) TO modes of 5.5 and 8.3, and 11.9~meV and LO modes of 6.3, 12.0, and 21.0~meV~\cite{Wang2019}, and 
(iv) LO modes of 5.7, 9.1, and 20.1~meV~\cite{Nagai2018}.
We note that an effective LO phonon frequency of 15.3~meV was used in Ref.~[\onlinecite{Wright2016}] to fit an one-phonon Fr\"ohlich 
coupling model to PL linewidths in the 100--370~K range. This value lies nearly midway between our 
computed $\w_{\,{\rm LO}_3} = 10.9$~meV and $\w_{\,{\rm LO}_4} = 19.2$~meV.
Discrepancies between theory and experiment are related to the choice of functional used in the DFT calculations, differences 
between the lattice constants, and to the supercell size used, in which our reference structure preserves only a subgroup of the 
full Pm$\bar{3}$m symmetry.

As show in Table~\ref{table.1}, the thermal induced renormalization on the electronic structure of reference and polymorphous 
networks yields a band gap widening of $\Delta E_{\rm g, ref} (430 {\rm ~K})=0.17$ and 
$\Delta E_{\rm g, poly} (430 {\rm ~K})=0.08$~eV. The corresponding hole and electron effective mass enhancements are 
$\lambda_{\rm h, ref} (430 {\rm ~K}) = 0.66$, $\lambda_{\rm e, ref} (430 {\rm ~K})=2.8$, 
$\lambda_{\rm h, poly} (430 {\rm ~K})= 0.37$ and $\lambda_{\rm e, poly} (430 {\rm ~K})= 0.28$. 
It is clear that both  $\Delta E_{\rm g}$ and $\lambda$ are diminished due to positional polymorphism; 
this is attributed to the reduced coupling of electrons to lattice vibrations in a distorted octahedral network, 
which in turn decreases the Fan-Migdal electron-phonon self-energy correction~\cite{short}.

In Fig.~\ref{fig9}(j), we present the band gap of cubic MAPbBr$_3$ as function of temperature evaluated 
as Boltzmann-weighted average over 10 ZG polymorphous configurations (red discs) and their
ZG reference structures (open squares).
Temperature-dependent band gaps computed using a single ZG polymorphous network obtained through 
geometry optimization of the MSS structure are also reported (open circles).
We compare our results with the experimental data from Ref.~[\onlinecite{Mannino2020}] (gray triangles). 
Our DFT calculations are shifted by our PBE0 correction of 1.44~eV (see Table~\ref{table.1}) 
to match the experimental data. 
Calculations indicate that using reference structures results in a band gap widening with temperature, 
but underestimates the experiment by more than 0.30 eV, a discrepancy that can be resolved by 
employing ZG polymorphous structures.
In addition, our calculated slope of $dE_{\rm g} / dT = 5.8 \times 10^{-4}$~eV/K
for the ZG reference structures overestimates the experimental value~\cite{Mannino2020} of $2.8 \times 10^{-4}$~eV/K 
by more than a factor of 2. As shown in Figs.~\ref{fig10}(a) and (b) for the reference structures, both $dE_{\rm g} / dT|_{\rm EPC}$ and $dE_{\rm g} / dT|_{\rm TE}$ 
contribute to the temperature-induced band gap widening, with the electron-phonon contribution accounting for 71\%.
The use of polymorphous networks reduces both the electron-phonon coupling 
and thermal expansion contributions, with electron-phonon coupling remaining the dominant effect, accounting for approximately 70\%.
The total slope calculated using ZG polymorphous structures is $dE_{\rm g} / dT = 2.9 \times 10^{-4}$ eV/K which 
agrees with the experimental one of $2.8 \times 10^{-4}$~eV/K.  
Remarkably, even when a single ZG polymorphous configuration is used, starting from the MSS structure, maintains
excellent agreement with the experimental data, with the extracted slope being $2.1 \times 10^{-4}$~eV/K. 

Figure~\ref{fig9}(n) shows temperature-dependent hole and electron effective masses 
in the range 275 -- 450~K calculated using ZG reference (open squares) and ZG polymorphous (red discs) networks. 
Unlike the hole effective mass of 0.15~$m_0$ at 300 K obtained for the reference structures, 
our calculated value of 0.26~$m_0$ for the polymorphous structures at 300 K is in excellent 
agreement with angle-resolved photoemission measurements at room temperature, which report a value 
of $0.25 \pm 0.05$~$m_0$\cite{Zu2019}.
Importantly, even when using a single ZG polymorphous configuration obtained 
from the MSS structure [open circles in Fig.~\ref{fig9}(n)],
 the results compare well with our values obtained from Boltzmann-weighted 
averages over 10 polymorphous configurations, i.e. when including configurational entropy effects.
Linear fits to the data give slopes of $\lambda^T_{{\rm h, ref}}  =  2.5 \times 10^{-4}$ 
and $\lambda^T_{{\rm e, ref}}  =  2.0 \times 10^{-4}$, $\lambda^T_{{\rm h, poly}} = 1.5  \times 10^{-4}$,
 and $\lambda^T_{{\rm e, poly}} = 0.9  \times 10^{-4}$~$m_0 / {\rm K}$, showing
that local disorder decreases the slopes by 40~\% and 55~\%, close to the values obtained for CsPbBr$_3$. 
We note that in our calculations, electron-phonon coupling dominates by 85\%--92\% the thermal-induced 
effective masses enhancements for both the reference and polymorphous structures. 
The reduced effective mass obtained for polymorphous cubic MAPbBr$_3$ varies linearly 
from  $\mu_{\rm poly}(275~{\rm K}) = 0.120$ to $\mu_{\rm poly}(450~{\rm K}) = 0.130$.

\subsection{MASnI$_3$}

In this section, we discuss our results for cubic MASnI$_3$. 
Similar to CsPbI$_3$ and CsSnI$_3$, our DFT-PBEsol calculations yield a semimetallic behavior 
for the reference structures of MASnI$_3$
and a spurious exchange in orbital character between the band extrema.
We thus assign a negative value to the band gap of $-0.09$~eV, as reported in Table~\ref{table.1}. 
The band inversion issue is addressed when taking into account positional polymorphism which results 
in a band gap widening of 0.37~eV. 
Our PBE0 calculations for the reference structures give a band gap
of 0.88 eV, which underestimates the experimental value of 1.31~eV at 300~K~\cite{Parrott2016}.
Employing locally disordered networks results in a band gap opening to 1.30~eV, which is in 
agreement with experiment and compares well with the $GW$ value of 1.1~eV reported
for tetragonal MASnI$_3$~\cite{Umari2014}.

Similar to CsPbI$_3$ and CsSnI$_3$, the effective masses of the reference structures exhibit a reversal in sign 
due to orbital exchange at the DFT-PBEsol level. This results in hole and electron effective masses of
$m^*_{\rm h, ref} = -0.05$ and $m^*_{\rm e, ref} = -0.03$. 
As with the band gap, local disorder corrects this problem and DFT-PBEsol hole and electron effective masses 
become $m^*_{\rm h, poly} = 0.06$ and $m^*_{\rm e, poly} = 0.09$. These effective masses are the smallest among all
hybrid halide compounds studied here and expected to be at the origin of the high mobility reported for MASnI$_3$
in Ref.~[\onlinecite{Stoumpos2013}]. The adjustments due to local disorder correspond 
to effective mass enhancement factors of $\lambda_{\rm h} = 1.2$ and $\lambda_{\rm e} = 3$, respectively, 
The reduced effective mass of cubic MASnI$_3$ increase from $\mu_{\rm ref}=0.019$ to $\mu_{\rm poly}= 0.036$.

Now we examine phonon anharmonicity in cubic  
MASnI$_3$. The cubic phase of MASnI$_3$ is observed above 275~K~\cite{Takahashi2011,Parrott2016} 
and therefore we perform calculations using temperatures around this value. 
Figure~\ref{fig8}(g) shows the anharmonic phonon dispersion at 300~K in the energy range 0--16~meV, 
which converges at the fifth ASDM iteration. 
The discontinuities at the $\Gamma$ point present in the harmonic phonon dispersion arise
from the use of a ferroelectric monomorphous structure, which induces anisotropy in the 
Born effective charge and dielectric tensors, and thus in the long-range contribution to 
the dynamical matrix. 
As with the other MA-based compounds, this issue is resolved in the anharmonic phonon dispersion by using averaged 
Born effective charges and dielectric tensors calculated from the polymorphous structures.
The anharmonic phonon dispersion related to the inorganic network of MASnI$_3$ 
exhibits similarities with the one reported for CsSnI$_3$ [Fig.~\ref{fig8}(c)], 
including an identical energy range of 0--16~meV and the low energy flat band along the R-M direction. 
The main differences are attributed to the contributions of the A-site cation, the different temperatures considered, 
and the choice of reference structure, which preserves only a subgroup of the full Pm$\bar{3}$m symmetry.
For MASnI$_3$, the zone-center frequencies 
$\w_{{\rm \, TO}_1}  = 4.47$, $\w_{{\rm \, TO}_4}  = 8.4$, $\w_{{\rm \, LO}_2}  = 4.92$, and $\w_{{\rm \, LO}_4}  = 16.1$~meV,
compare well with the spectroscopy data reported for MAPbI$_3$ single crystals at room 
temperature: 4.0, 7.8, 5.0, and 16.5~meV~\cite{Sendner2016}, and 4.2, 8.3, 4.6, and 16.2~meV~\cite{Boldyrev2020}, respectively. 

The thermal-induced electronic structure renormalization leads to an overall band gap opening 
of $\Delta E_{\rm g, ref} (300 {\rm ~K})=0.22$~eV and 
$\Delta E_{\rm g, poly} (300 {\rm ~K})=0.13$~eV as reported for the ZG reference and ZG polymorphous structures
in Table~\ref{table.1}. The hole and electron effective mass enhancement factors are 
$\lambda_{\rm h, ref} (300 {\rm ~K}) = 1.8$, $\lambda_{\rm e, ref} (300 {\rm ~K}) = 3.0$ 
$\lambda_{\rm h, poly} (300 {\rm ~K})= 0.5$, and $\lambda_{\rm e, poly} (300 {\rm ~K})= 0.4$, showing 
that local disorder reduces thermal induced corrections.

Figure~\ref{fig9}(k) shows the band gap of cubic MASnI$_3$ as a function of temperature. 
Red discs and open squares represent Boltzmann-weighted averages over the band gaps computed for 10 ZG 
polymorphous configurations and their ZG reference structures, respectively.
Open circles represent temperature-dependent band gaps computed using a single ZG polymorphous network obtained from
the MSS structure.
Grey triangles are experimental data from Ref.~[\onlinecite{Parrott2016}]. 
Our DFT values are shifted by $\Delta_s =0.90$~eV, which aligns with our PBE0 correction of 
1.02~eV obtained for the polymorphous structures. This shift also aligns
with the $GW$ correction of 0.79~eV reported in Ref.~[\onlinecite{Umari2014}] for tetragonal MASnI$_3$.
The reference structures exhibit a band gap 
widening with increasing temperature, but they underestimate the experimental values by about 0.3~eV.
The calculated temperature coefficient of the band gap for the reference structures is $dE_{\rm g} / dT = 3.4 \times 10^{-4}$~eV/K,
which shows good agreement with the experimentally determined value of $4.5 \times 10^{-4}$~eV/K. 
Similar to the case of CsSnI$_3$, this agreement appears coincidental, since electron-phonon coupling and thermal 
expansion contribute with opposite signs, as illustrated in Figs.~\ref{fig10}(a) and (b). Specifically,
the negative thermal expansion contribution arises from the orbital exchange character 
in the reference structures of cubic MASnI$_3$. Adjusting the sign of this contribution results in
$dE_{\rm g} / dT = 7.2 \times 10^{-4}$~eV/K, which overestimates the experimental value by 60~\%.
Using ZG locally disordered networks gives a positive $dE_{\rm g} / dT_{\rm TE}$ 
and an overall $dE_{\rm g} / dT = 5.7 \times 10^{-4}$ eV/K which is closer to experiment. 
Notably, even a single ZG polymorphous configuration, derived from the MSS structure, reproduces a 
nearly identical temperature coefficient of $5.6 \times 10^{-4}$~eV/K.
Similar to Cs-based and MA-based compounds, our analysis shows that electron-phonon coupling 
is the leading contribution, accounting for 61\%.

Figure~\ref{fig9}(o) shows the temperature dependence of the hole and electron effective masses of 
cubic MASnI$_3$ in the range 250 -- 450~K. 
Linear fits to the data obtained for the ZG reference (open squares) and ZG polymorphous (red discs)
structures are $\lambda^T_{{\rm h, ref}}  =  0.8 \times 10^{-4}$,
and $\lambda^T_{{\rm e, ref}}  =  0.3 \times 10^{-4}$,
$\lambda^T_{{\rm h, poly}} = 1.5  \times 10^{-4}$,
  and $\lambda^T_{{\rm e, poly}} = 2.1 \times 10^{-4}$~$m_0 / {\rm K}$. 
A single ZG polymorphous configuration derived from the MSS structure [open circles in Fig.~\ref{fig9}(o)] yields 
similar slopes to those obtained from Boltzmann-weighted averages over 10 ZG polymorphous configurations but 
underestimates the values by about 0.02.
Electron-phonon coupling is the leading thermal effect, contributing to more 
than 75\% to the effective masses enhancements. For polymorphous 
cubic MASnI$_3$, the reduced effective mass exhibits a linear temperature dependence, increasing 
from  $\mu_{\rm poly}(250~{\rm K}) = 0.049$ to $\mu_{\rm poly}(450~{\rm K}) = 0.067$.

\subsection{MAPbCl$_3$}

Here, we present our findings for cubic MAPbCl$_3$. The DFT-PBEsol calculations for the reference structures
yield a band gap of 0.72~eV, as shown in Table~\ref{table.1}. 
Local disorder induces a substantial band gap widening of 0.52~eV, which is the largest among all 
MA-based compounds studied in this work, consistent 
with the largest energy lowering reported for MAPbCl$_3$ in Fig.~\ref{fig2}(b). 
Our PBE0 calculations for the reference structures give a band gap of 2.34~eV which underestimates
significantly the experimental values of 2.88~eV and $3.07$~eV at 300~K, obtained from PL measurements of the major 
emission peak~\cite{Maculan2015,Hsu2019}.
Significant improvements are observed when employing locally disordered networks in combination with the PBE0 
approximation, yielding a band gap of 2.88~eV which is in excellent agreement with experiments. 
A $GW$ band gap of 3.46~eV has been previously reported for tetragonal MAPbCl$_3$~\cite{Mosconi2016}, which overestimates experiment
by approximately 0.4--0.6~eV. This discrepancy has been primarily attributed to the use of a scalar relativistic approximation for 
the screened Coulomb interaction, neglecting spin-orbit coupling for $W$~\cite{Filip2024}.

The effective masses calculated for the reference structures
are $m^*_{\rm h, ref}=0.15$ and $m^*_{\rm e, ref}=0.16$. Using locally disordered structures,
the effective masses increase to $m^*_{\rm h, poly}=0.26$  and $m^*_{\rm e, poly}=0.27$. The associated hole and electron 
effective mass enhancement factors due to polymorphism are $\lambda_{\rm h} = 0.73$ and $\lambda_{\rm e} = 0.63$. 
Corrections arising from polymorphism increase the reduced effective mass from $\mu_{\rm ref} = 0.077$ to $\mu_{\rm poly} = 0.132$. 
The latter value lies within the inherent accuracy of the $GW$ value of 0.159 reported for tetragonal MAPbCl$_3$ 
in Ref.~[\onlinecite{Mosconi2016}] that neglects spin-orbit coupling in $W$.

The cubic phase of MAPbCl$_3$ stabilizes above 180~K~\cite{Hsu2019} 
and therefore we perform calculations using temperatures above this value. 
In Fig.~\ref{fig8}(h), we present the converged ASDM phonon dispersion at 300~K, achieved at the sixth iteration, 
in the energy range associated with the vibrations of the inorganic sublattice (0--31~meV). 
Similar to other hybrid halide perovskites, the harmonic phonon dispersion displays phonon instabilities, lifted degeneracies, 
and discontinuities at the $\Gamma$ point due to the fixed molecular orientation
in the ferroelectric monomorphous structure. As for other hybrid halide perovskites, in our anharmonic phonon calculations 
we resolve these issues by including phonon self-energy corrections, taking the MSS structure as a reference, 
and computing the non-analytic contribution to the dynamical matrix based on Born effective charge and high-frequency dielectric 
tensors of the polymorphous structures. 
The anharmonic phonon dispersions of MAPbCl$_3$ and CsPbCl$_3$ [Fig.~\ref{fig8}(d)] have some similar characteristics, 
spanning comparable energy ranges and both exhibiting an ultrasoft, flat phonon band along the R-M direction. 
The primary differences arise from the contributions of the A-site cations and the selection of the MSS structure that retains only a 
subgroup of the full Pm$\bar{3}$m symmetry.
Table~\ref{table.2} reports the frequencies at the $\Gamma$ point of MAPbCl$_3$: 
$\w_{{\rm \, TO}_{1-4}}  = 7.8-8.3$, $\w_{{\rm \, TO}_{5-6}}  = 12.2-13.0$, 
 $\w_{{\rm \, TO}_{7-8}} = 15.3-15.7$, 
$\w_{{\rm \, LO}_1} = 7.8$, $\w_{{\rm \, LO}_2} = 8.3$, $\w_{{\rm \, LO}_3} = 13.6$, and $\w_{{\rm \, LO}_4} = 27.7$~meV.
A good agreement is observed with experimental data from infrared and THz-TDS spectroscopy on MAPbCl$_3$ at room temperature, 
which identify TO modes at 8.2 and 11.0~meV and LO modes at 8.2, 8.7, 12.4, and 27.9~meV~\cite{Sendner2016, Nagai2018}.

We next investigate the renormalization of the band gap and effective masses 
due to electron-phonon coupling and thermal expansion. 
Figure~\ref{fig9}(l) shows the band gap of cubic MAPbCl$_3$ in the temperature range 200--400~K
obtained using 10 ZG polymorphous (red discs) and 10 ZG reference (open squares) structures for each temperature.
We also include temperature-dependent band gaps computed with the ZG polymorphous configuration of the MSS structure (open cirles).
Experimental data (gray triangles) are from Ref.~[\onlinecite{Hsu2019}].
To enable direct comparison with experiment, the DFT data are rigidly shifted by $\Delta_s =1.73$~eV. 
This value matches the PBE0 correction of 1.64~eV obtained for the polymorphous structures (see Table~\ref{table.1}).
Our data for the ZG reference structures show a temperature-induced band gap widening, with a temperature coefficient 
of $dE_{\rm g} / dT = 6.2 \times 10^{-4}$~eV/K. This value significantly overestimates the experimentally derived 
coefficient of $-0.9 \times 10^{-4}$~eV/K, which indicates that the band gap remains nearly constant with temperature in 
the measured range.
As illustrated in Figs.~\ref{fig10}(a) and (b), both $dE_{\rm g} / dT|_{\rm EPC}$ and $dE_{\rm g} / dT|_{\rm TE}$ 
computed for the reference structures lead to band gap widening, 
with electron-phonon coupling accounting for 48\%. Notably, this is the only case among Cs-based and MA-based 
compounds where thermal expansion is the leading contribution to the thermal induced band gap renormalization.  
When ZG polymorphous networks are employed, electron-phonon coupling results in band gap narrowing, uniquely among all compounds studied, 
while thermal expansion continues to induce a band gap widening, though its contribution is reduced. 
The total temperature coefficient of the band gap for the polymorphous structures 
is $dE_{\rm g} / dT = 0.9 \times 10^{-4}$~eV/K, which closely 
matches the experimental value, as shown in Fig.~\ref{fig10}(c). Overall, we attribute the nearly constant variation of the band gap 
with temperature to the opposing effects of electron-phonon coupling and thermal expansion. 
Unlike for the other MA-based compounds, a single ZG polymorphous configuration derived from the MSS structure 
does not provide a sufficiently accurate approximation for computing temperature-dependent band gaps, yielding a 
temperature coefficient of $3.8 \times 10^{-4}$~eV/K.

In Fig.~\ref{fig9}(p), we present the temperature dependence of the hole and electron effective masses calculated using 
the ASDM for both the reference (open squares) and polymorphous structures (red discs). The effective masses of the ZG 
reference structures exhibit a monotonic increase with temperature, whereas those in the ZG polymorphous structures remain 
nearly constant, mirroring the trend observed for the band gap. Fitting a straight line to the data yields the slopes 
$\lambda^T_{{\rm h, ref}}  =  1.6 \times 10^{-4}$, $\lambda^T_{{\rm e, ref}}  =  1.8 \times 10^{-4}$, 
$\lambda^T_{{\rm h, poly}} = 0.9  \times 10^{-4}$, and $\lambda^T_{{\rm e, poly}} = 0.2  \times 10^{-4}$~$m_0 / {\rm K}$, showing
that local disorder affects significantly the variation of the effective masses with temperature. 
We remark that now electron-phonon coupling contribution dominates 
thermal induced renormalization of the effective masses being more than 70\%. 
We also report that the reduced effective mass calculated for polymorphous cubic MAPbCl$_3$ 
remains linearly constant being 
$\mu_{\rm poly}(200~{\rm K}) = 0.170$ and $\mu_{\rm poly}(400~{\rm K}) = 0.172$.
Finally, we note that as for the band gap, a single ZG polymorphous configuration derived from the MSS structure 
[open circles in Fig.~\ref{fig9}(p)] is not a good approximation to explain the temperature variation of the 
effective masses. 

\subsection{FAPbI$_3$}

We next present our results for cubic FAPbI$_3$. 
DFT-PBEsol calculations for the reference structure reveal a semimetallic behavior with a band gap of 0.10~eV.
However, unlike CsPbI$_3$ [Figs.~\ref{fig12}(a) and (b)], our charge density plots [Figs.~\ref{fig12}(f) and (g)] 
do not display any spurious exchange in orbital character between the conduction and valence band extrema.
Accounting for positional polymorphism leads to a relatively small band gap widening of 0.24~eV, consistent
with the relatively small degree of polymorphism in FA-based compounds (Fig.~\ref{fig2}).
Going beyond DFT, PBE0 calculations give a band gap of 1.33~eV for the reference structures. 
Although this underestimates the experimental value of 1.53~eV at 300~K~\cite{Chen2020}, it remains 
in reasonable agreement. Employing polymorphous networks results in a band gap of 1.59~eV, 
aligning better with experiment and again highlighting the importance of local disorder in accurate 
band gap predictions. Our result also compares well with $GW$ calculations for tetragonal 
FAPbI$_3$, which report a band gap of 1.71~eV~\cite{Muhammad2022}. 
Previous $GW$ calculations, based on idealized ferroelectric monomorphous structures that are not representative 
of the actual physical system yielding an artificial Rashba splitting, report a band gap of 1.47~eV 
and 1.48~eV~\cite{Bokdam2016,Muhammad2020}.
We emphasize that band structures of polymorphous structures do not exhibit a Rashba band splitting of 
the doubly degenerate band extrema~\cite{short,Zacharias2023npj} as they reflect, on average, the system's 
centrosymmetricity. 

At the DFT-PBEsol level, the spurious semimetallic behavior leads to extremely 
small hole and electron effective masses of $m^*_{\rm h, ref}=0.03$ and
$m^*_{\rm e, ref}=0.03$. As with the band gap, incorporating
local disorder increases the effective masses to $m^*_{\rm h, poly}=0.10$ 
and $m^*_{\rm e, poly}=0.09$, correpondig to effective mass enhancement factors 
of $\lambda_{\rm h} = 2.3$ and $\lambda_{\rm e} = 2.0$, respectively. 
Our calculated reduced effective mass increases significantly from $\mu_{\rm ref} = 0.015$ 
to $\mu_{\rm poly} = 0.047$ for polymorphous FAPbI$_3$. However, this value still substantially 
underestimates the experimental measurements for orthorhombic and tetragonal FAPbI$_3$, 
reported as 0.09 and 0.095, respectively~\cite{Galkowski2016}.
This underestimation is attributed to the semilocal treatment of exchange-correlation effects within DFT and 
the relatively small degree of polymorphism exhibited by FA-based compounds. 
Previous $GW$ calculations~\cite{Wang2020} on a ferroelectrically distorted monomorphous configuration of cubic FAPbI$_3$, 
(though unphysical, exhibiting a relatively large $\mu_{ GW} = 0.121$) 
report an increase in the DFT hole and electron effective masses of 72\% and 93\%, respectively.
Applying these corrections to our DFT effective masses of polymorphous FAPbI$_3$ yields a 
renormalized value of $\mu_{{\rm poly} + GW} = 0.086$, in excellent agreement with the experimental values.  

In the following we present the anharmonic phonon dispersion. The cubic phase of FAPbI$_3$ stabilizes above 285~K~\cite{Fabini2016} 
and thus we perform calculations using temperatures around this value. 
In Fig.~\ref{fig8}(i), we present the calculated ASDM phonon dispersion at 300 K, which converges by the 
fourth iteration. We focus on the energy range 0--16~meV, mostly associated with vibrations of the 
inorganic atoms. Accounting for anharmonicity with the ASDM fully alleviates the strong instabilities present 
in the harmonic phonon dispersion calculated for the ferroelectric monomorphous structure. 
The differences between the anharmonic phonon dispersions of cubic MAPbI$_3$ [Fig.~\ref{fig8}(e)] and FAPbI$_3$ 
are primarily attributed to the distinct contributions of the MA and FA cations and the 
indirect coupling between the lattice dynamics of the organic and inorganic sublattices.
The phonon frequencies at the $\Gamma$ point for 300~K 
are $\w_{{\rm \, TO}_{1-4}} = 3.79-5.3$, $\w_{{\rm \, TO}_{5-6}} = 8.3-8.6$, $\w_{{\rm \, TO}_{7-8}} = 11.9-12.0$, 
$\w_{{\rm \, LO}_1} = 4.0$, $\w_{{\rm \, LO}_2} = 4.9$, $\w_{{\rm \, LO}_3} = 11.4$, and $\w_{{\rm \, LO}_4} = 15.5$~meV, as 
reported in Table~\ref{table.2}.
Analysis of phonon sidebands in PL spectra at 3.6~K of FAPbI$_3$ nanocrystals~\cite{Fu2018}, which have a cubic 
crystal structure and average sizes of 10-15~nm, yields LO phonons of 3.2, 7.8 and 15.4~meV. These values 
compare well with our calculations except the mode at 7.8~meV which differs from our calculated 
value of 11.4~meV. However, as reported in Ref.~[\onlinecite{Fu2018}], high-energy excitations 
result in temporal fluctuations of the intermediate phonon, spanning an energy range 
from 7.2 to 15~meV. Therefore, in Table~\ref{table.2}, we adopt 
the mean value of this range, 11.1~meV, as the representative intermediate phonon energy. Close to this value, Refs.~[\onlinecite{Fu2018}]
and [\onlinecite{Wright2016}] report an effective LO phonon of 10.7~meV and 11.5~meV, respectively, obtained by 
one-phonon Fr\"ohlich coupling model fits to temperature-dependent PL linewidths. 
Our calculations are also in good agreement with inelastic neutron scattering data~\cite{Ferreira2020} 
of 3.9, 9.8, and 15.1~meV, obtained for the low temperature phase of FAPbI$_3$.

Now we move to the effect of thermal effects on the electronic structure. 
As reported in Table~\ref{table.1}, our calculations with the ASDM accounting for anharmonic 
electron-phonon coupling and thermal expansion yield a band gap opening of $\Delta E_{\rm g, ref} (300 {\rm ~K})=0.39$~eV and 
$\Delta E_{\rm g, poly} (300 {\rm ~K})=0.13$~eV for the ZG reference and ZG polymorphous structures. 
The corresponding electron and hole effective mass enhancements, derived using the renormalization $\Delta m^*$
reported in Table~\ref{table.1}, are 
$\lambda_{\rm h, ref} (300 {\rm ~K}) = 2.3$, $\lambda_{\rm e, ref} (300 {\rm ~K}) = 1.7$,
$\lambda_{\rm h, poly} (300 {\rm ~K})= 0.6$, and $\lambda_{\rm e, poly} (300 {\rm ~K})= 0.4$.
These values show that local disorder leads to a considerable reduction to the thermal-induced 
renormalization of the electronic structure of FAPbI$_3$. 

In Fig.~\ref{fig9}(q) we compare the band gap of cubic FAPbI$_3$ as a function of 
temperature calculated for ZG polymorphous (black discs) and ZG reference (open squares) networks with measurements from 
Ref.~[\onlinecite{Chen2020}] (gray triangles). The DFT data are shifted by $\Delta_s =1.10$~eV 
close to our calculated PBE0 correction of 1.25~eV (see Table~\ref{table.1}). 
A linear fit to the data obtained for the ZG reference structures yields a slope of $dE_{\rm g} / dT = 7.5 \times 10^{-4}$~eV/K 
which overestimates the experimental value of $4.3 \times 10^{-4}$~eV/K by 75~\%.
Accounting for positional polymorphism improves significantly the agreement with experiment as it suppresses
both electron-phonon coupling and thermal expansion leading to a temperature coefficient 
$dE_{\rm g} / dT = 4.6 \times 10^{-4}$~eV/K. 
Figure~\ref{fig9}(q) also reports data calculated using only the ZG polymorphous network of the MSS structure (open circles). 
Remarkably, our values overlay with data obtained from Boltzmann-weighted averages over 10 ZG polymorphous structures, 
with the extracted coefficient being $4.5 \times 10^{-4}$~eV/K. 
As shown in Figs.~\ref{fig10}(a,b), electron-phonon coupling contributes more significantly than thermal expansion, 
accounting for 65\% and 70\% in our calculated band gap renormalization for the ZG reference and ZG polymorphous structures, 
respectively.

In Fig.~\ref{fig9}(u), we report hole and electron effective masses in the temperature range 220--350~K, 
revealing a thermal-driven increase for both the ZG reference (open squares) and ZG polymorphous structures (dark discs).
The slopes extracted from linear fits to the data are
$\lambda^T_{{\rm h, ref}}  =  2.8 \times 10^{-4}$, $\lambda^T_{{\rm e, ref}}  =  1.9 \times 10^{-4}$,
$\lambda^T_{{\rm h, poly}} = 3.3  \times 10^{-4}$, and $\lambda^T_{{\rm e, poly}} = 1.7  \times 10^{-4}$~$m_0 / {\rm K}$. 
Similar to the band gap renormalization, more than 75\% of $\lambda^T$ arises from electron-phonon coupling in all cases.
We also report that the reduced effective mass calculated for polymorphous cubic FAPbI$_3$ 
varies linearly being  $\mu_{\rm poly}(220~{\rm K}) = 0.063$ to $\mu_{\rm poly}(350~{\rm K}) = 0.078$.
Applying the $GW$ corrections~\cite{Wang2020} of 72\% and 93\% to the DFT hole and electron effective masses gives 
 $\mu_{{\rm poly} + GW}(220~{\rm K}) = 0.115$ to $\mu_{{\rm poly} + GW}(350~{\rm K}) = 0.143$. 
Finally, we note that the temperature dependence of the effective masses is well reproduced by 
a single ZG polymorphous configuration derived from the MSS structure, as indicated by the open circles in Fig.~\ref{fig9}(u).

\subsection{FAPbBr$_3$}

In this section, we describe our calculations for cubic FAPbBr$_3$. 
Table~\ref{table.1} reports a DFT-PBEsol average band gap of 0.46~eV for the reference structure.
Similar to FAPbI$_3$, positional polymorphism leads to a relatively small band gap opening of 0.24~eV, 
consistent with the relative small degree of structural disorder calculated for FA-based compounds, as shown 
in Figs.~\ref{fig2}(c)-(h). 
Combining the PBE0 functional and the reference structures, we obtain a band gap of
1.89~eV, which underestimates the experimental value of 2.29 at 300~K~\cite{Mannino2020}.
Using locally disordered polymorphous networks, the calculated band gap increases to 
2.13~eV, matching closer the experimental value.
Previous $GW$ calculations on ferroelectric monomorphous structures that exhibit an artificial Rashba splitting on 
the electronic structure yield band gaps of 2.20~eV and 2.26~eV~\cite{Bokdam2016,Muhammad2020}.  

Our calculated DFT-PBEsol effective masses for the reference structure are  $m^*_{\rm h, ref}=0.10$ and $m^*_{\rm e, ref}=0.10$.
Local disorder enhances the effective masses, resulting to $m^*_{\rm h, poly}=0.16$ 
and $m^*_{\rm e, poly}=0.15$, and thus to $\lambda_{\rm h} = 0.6$ and $\lambda_{\rm e} = 0.5$. 
These enhancements increase the reduced effective mass of cubic FAPbBr$_3$
from $\mu_{\rm ref}=0.050$ to $\mu_{\rm poly}= 0.077$ improving
the comparison with the experimental values for orthorhombic and tetragonal phases of 0.117 and 0.13~\cite{Galkowski2016}. 
To further improve consistency with experimental values, we apply $GW$ enhancements~\cite{Wang2020} of 72\% and 93\% 
to the DFT hole and electron effective masses, respectively, following the approach used for FAPbI$3$. 
This results in a reduced effective mass of $\mu_{{\rm poly}+GW} = 0.141$.

Now we study the anharmonic phonon dispersion of cubic FAPbBr$_3$.
We focus on temperatures around 270~K, where the cubic phase of FAPbBr$_3$ becomes stable~\cite{Schueller2017,Mannino2020}. 
In Fig.~\ref{fig8}(j), we present the converged ASDM phonon dispersion at 300 K (fourth iteration)
in the energy range 0--20~meV. As shown in Fig.~\ref{fig14}(b), this range is mostly associated with vibrations of the inorganic atoms. 
The phonons of the ferroelectric monomorphous structure exhibit strong instabilities [green in Fig.~\ref{fig8}(j)], 
which are alleviated by using a reference structure and including self-energy corrections with the ASDM. 
The stabilized phonon frequencies at the $\Gamma$ point are 
$\w_{{\rm \, TO}_{1-4}} = 5.4-7.1$, $\w_{{\rm \, TO}_{5-6}} = 9.7-10.0$, $\w_{{\rm \, TO}_{7-8}} = 13.3-13.7$, 
$\w_{{\rm \, LO}_1} = 5.5$, $\w_{{\rm \, LO}_2} = 6.53$, $\w_{{\rm \, LO}_3} = 12.9$, and $\w_{{\rm \, LO}_4} = 18.8$~meV, as 
reported in Table~\ref{table.2}.
These frequencies are comparable to those of MAPbBr$_3$ and CsPbBr$_3$ with main differences attributed to contributions 
from the A-site cations and their coupling with the inorganic sublattice. 
Our calculated LO phonon frequencies for cubic FAPbBr$_3$ compare well with the experimental values: 
(i) 5.5, 13.1, and 22.2~meV, obtained from inelastic neutron scattering measurements on the orthorhombic 
phase~\cite{Ferreira2020}, and (ii) 4.4 and 18.8~meV, derived from the analysis of photoluminescence (PL) sidebands 
of 4--17~nm nanocrystals at 5.5~K~\cite{Cho2021}.
Additionally, Ref.~[\onlinecite{Wright2016}] reports an effective phonon frequency of 15.3~meV, obtained by fitting 
an one-phonon Fröhlich coupling model to PL linewidths measured over the 10--370~K range. This value 
lies approximately midway between our calculated  $\w_{{\rm \, LO}_3} = 12.9$~meV and $\w_{{\rm \, LO}_4} = 18.8$~meV.  

In Table~\ref{table.1}, we report for the ZG reference and ZG polymorphous structures a thermal-induced band gap 
renormalization of $\Delta E_{\rm g, ref} (300 {\rm ~K})=0.13$~eV and 
$\Delta E_{\rm g, poly} (300 {\rm ~K})=0.10$~eV. Using the reported thermal-induced mass corrections, $\Delta m^*$, 
we obtain the following enhancements factors: $\lambda_{\rm h, ref} (300 {\rm ~K}) = 0.4$, 
$\lambda_{\rm e, ref} (300 {\rm ~K})=0.3$, $\lambda_{\rm h, poly} (300 {\rm ~K})= 0.3$, 
and $\lambda_{\rm e, poly} (300 {\rm ~K})= 0.3$. 
The small changes in both $\Delta E_{\rm g}$ and $\lambda$ 
are consistent with the relatively small degree of local disorder in cubic FAPbBr$_3$. 

In Fig.~\ref{fig9}(r), we present temperature-dependent band gaps of the ZG reference (open squares)
and ZG polymorphous (black discs) structures of cubic FAPbBr$_3$ evaluated as Boltzmann-weighted 
averages over 10 configurations.
Our DFT calculations are corrected by $\Delta_s = 1.50$~eV to align with the experimental 
data from Ref.~[\onlinecite{Mannino2020}] (gray triangles); $\Delta_s$ 
is close our PBE0 correction of 1.43~eV obtained for the polymorphous structure (see Table~\ref{table.1}). 
Using ZG reference structures leads to a temperature-induced band gap widening, but the calculated values 
consistently underestimate the experimental band gap by more than 0.20~eV across all temperatures. 
This discrepancy is resolved by employing ZG polymorphous structures. The temperature coefficient of 
the band gap, calculated as $dE_{\rm g} / dT = 6.6 \times 10^{-4}$~eV/K for the reference structures, overestimates 
the extracted experimental value of $4.9 \times 10^{-4}$~eV/K by approximately 35\%.
As shown in Figs.~\ref{fig10}(a) and (b), both $dE_{\rm g} / dT|_{\rm EPC}$ and $dE_{\rm g} / dT|_{\rm TE}$ contribute to the band gap 
opening with temperature, with electron-phonon coupling accounting for 54\% of the total effect in the reference case. 
The use of ZG polymorphous networks reduces both contributions, particularly that from electron-phonon interactions, 
leading now to a dominant thermal expansion contribution of 66\%.
The total slope obtained with ZG polymorphous structures is improved 
to $dE_{\rm g} / dT = 3.6 \times 10^{-4}$~eV/K, underestimating the experimental value by 27\%. 
Notably, the single ZG polymorphous configuration of the MSS structure [open circles in Fig.~\ref{fig9}(r)] 
yields excellent agreement with experiment, with an extracted slope of $4.9 \times 10^{-4}$~eV/K.

Figure~\ref{fig9}(v) presents the temperature dependence of hole and electron effective masses in the range 275--400~K, 
calculated for both the ZG reference (open squares) and ZG polymorphous (black discs) networks of FAPbBr$_3$. 
The extracted slopes are $\lambda^T_{\rm h, ref} = 1.8 \times 10^{-4}$, $\lambda^T_{\rm e, ref} = 1.5 \times 10^{-4}$, 
$\lambda^T_{\rm h, poly} = 1.9 \times 10^{-4}$, and $\lambda^T_{\rm e, poly} = 1.5 \times 10^{-4}$/K. 
These values indicate that local disorder has small influence on the temperature dependence of the effective masses.
Unlike the case of the band gap, the temperature-induced enhancement of the effective masses is primarily driven by 
electron-phonon coupling, which accounts for over 75\% of the total effect in both structural models. 
For polymorphous cubic FAPbBr$_3$, the reduced effective mass increases from 
$\mu_{\rm poly}(275~{\rm K}) = 0.096$ to $\mu_{\rm poly}(400~{\rm K}) = 0.107$.
Applying $GW$ correction factors of 72\% and 93\% to the DFT hole and electron effective 
masses~\cite{Wang2020} yields corrected values of 
 $\mu_{{\rm poly} + GW}(275~{\rm K}) = 0.175$ and $\mu_{{\rm poly} + GW}(400~{\rm K}) = 0.195$. 
Finally, we note that a single ZG polymorphous configuration derived from the MSS structure [open circles in Fig.~\ref{fig9}(v)] 
provides a reasonable approximation for estimating the effective masses of FAPbBr$_3$.

\subsection{FASnI$_3$} \label{sec.FASnI3}

We next discuss our results for cubic FASnI$_3$. 
Similar to CsSnI$_3$ and MASnI$_3$, our DFT-PBEsol calculations yield a semimetallic behavior 
for the reference structures of FASnI$_3$, with a band gap of 0.11~eV as reported in Table~\ref{table.1}. However, 
in this case we did not observe an exchange in orbital character between the band extrema.
Including local disorder leads to a band gap opening of 0.31~eV of similar magnitude with other FA-based compounds. 
This value is consistent with the 0.22~eV band gap opening previously reported by some of us for a single 
locally disordered configuration of cubic FASnI$_3$, calculated without including spin-orbit coupling~\cite{Dirin2023}.
Experimental measurements on FASnI$_3$ nanocrystals with cubic symmetry (space group Pm$\bar{3}$m) 
have reported a band gap increase of 190~meV compared to the corresponding bulk cubic 
phase~\cite{Dirin2023}. In addition to quantum confinement, this increase has been attributed to local structural 
disorder, specifically to an average Sn$-$I$-$Sn bond angle reduction of 13$^\circ$ relative to the idealized cubic configuration. 
This angular deviation is 3$^\circ$ larger than that found in our locally disordered bulk structures, corresponding to an 
additional band gap renormalization of approximately 72~meV, based on the 24~meV/$^\circ$ slope shown 
in Fig.~\ref{fig2}(d). We therefore attribute roughly 72~meV of the total 190~meV band gap increase to enhanced 
local disorder in cubic FASnI$_3$ nanocrystals, with the remaining contribution arising from quantum confinement 
or other effects. Our PBE0 calculations for the bulk reference structures give a band gap
of 1.13 eV which compares well with the experimental value of 1.38~eV at 300~K~\cite{Kahmann2020}, 
both reported in Table~\ref{table.1}. Employing locally disordered networks results to a band gap of
1.48~eV, which agrees with the experimental value. We note that $GW$ calculations using
an unphysical distorted monomorphous structure~\cite{Bokdam2016} report a band gap of 1.24~eV.  

The DFT-PBEsol hole and electron effective masses of the reference structures are extremely 
small: $m^*_{\rm h, ref} = 0.03$ and $m^*_{\rm e, ref} = 0.03$. 
Local disorder increase the hole and electron effective masses 
to $m^*_{\rm h, poly} = 0.08$ and $m^*_{\rm e, poly} = 0.11$, which remain small close to 
those reported for MASnI$_3$. The adjustments due to local disorder correspond 
to effective mass enhancement factors of $\lambda_{\rm h} = 1.7$ and $\lambda_{\rm e} = 2.7$  
and an increase in the reduced effective mass from $\mu_{\rm ref}=0.015$ to $\mu_{\rm poly}= 0.046$.

Now we examine the anharmonic phonon dispersion of cubic FASnI$_3$. 
The cubic phase of FASnI$_3$ is observed above 250~K~\cite{Schueller2017,Kahmann2020} 
and therefore we perform calculations around this value. 
In Fig.~\ref{fig8}(k), we show the anharmonic phonon dispersion at 300~K in the energy range of 0--15~meV, 
which is found to converge by the third ASDM iteration, alleviating all phonon instabilities present in the 
monomorphous structure. 
The stabilized phonon frequencies at the $\Gamma$ point are 
$\w_{{\rm \, TO}_{1-4}} = 4.93-5.8$, $\w_{{\rm \, TO}_{5-6}} = 6.7-6.8$, $\w_{{\rm \, TO}_{7-8}} = 12.1-12.6$, 
$\w_{{\rm \, LO}_1} = 4.98$, $\w_{{\rm \, LO}_2} = 5.24$, $\w_{{\rm \, LO}_3} = 11.3$, and $\w_{{\rm \, LO}_4} = 15.0$~meV, as 
reported in Table~\ref{table.2}.
The calculated frequencies are comparable to those obtained for CsSnI$_3$ and MASnI$_3$ [Figs.~\ref{fig8}(c) and (g)], 
with the primary differences arising from the influence of the A-site cations and their interactions 
with the inorganic sublattice.
Furthermore, the phonon dispersions of FASnI$_3$ [Fig.~\ref{fig8}(k)] and FAPbI$_3$ [Fig.~\ref{fig8}(i)] are 
qualitatively and even quantitatively similar across large regions of the Brillouin zone. 
We attribute this to the fact that many vibrational modes are dominated by the motion of I atoms,
despite the difference in B-site cations.
This also supports the good agreement between our calculated phonon frequencies for cubic FASnI$_3$ and experimental 
measurements reported for FAPbI$_3$~\cite{Fu2018,Ferreira2020,Wright2016}, reported in Table~\ref{table.2}.


In Fig.~\ref{fig9}(s), we show the band gap of cubic FASnI$_3$ as a function of temperature. 
The black discs and open squares correspond to Boltzmann-averaged band gaps calculated from 10 
distinct ZG polymorphous configurations and their respective ZG reference structures. Temperature-dependent band gaps 
obtained from a single ZG polymorphous configuration, derived from the MSS structure, 
are shown as open circles.
Our DFT values are shifted by $\Delta_s = 0.85$~eV to match the experimental data~\cite{Kahmann2020} (gray triangles). 
This shift is close to our computed PBE0 correction of 1.06~eV obtained for the polymorphous structures and  
the $GW$ correction of 0.79~eV obtained for tetragonal MASnI$_3$ in Ref.~[\onlinecite{Umari2014}].
The band gaps of the ZG reference structures increase with temperature, yielding a temperature coefficient of 
$dE_{\rm g} / dT = 7.9 \times 10^{-4}$~eV/K. This value compares reasonably well with the experimental coefficient of 
$10.6 \times 10^{-4}$~eV/K but underestimates it by approximately 25\%. A similar level of underestimation is observed 
when ZG polymorphous structures are used, with the extracted coefficient being $dE_{\rm g} / dT = 8.0 \times 10^{-4}$~eV/K. 
Notably, a single ZG polymorphous configuration derived from the MSS structure yields results nearly identical 
to those obtained by averaging over 10 ZG configurations.

As shown in Figs.~\ref{fig10}(a) and (b), both $dE_{\rm g} / dT|_{\rm EPC}$ and $dE_{\rm g} / dT|_{\rm TE}$ 
contribute to the temperature-induced band gap widening, with electron-phonon coupling being the dominant mechanism 
in the reference structures, accounting for 62\%.
Interestingly, local disorder reverses this trend, suppressing the electron-phonon 
coupling contribution while enhancing that of thermal expansion, which now dominates with 73\%.
An enhancement of thermal expansion contribution is also observed for MASnI$_3$.
Although thermal expansion leads to an overall increase in average bond lengths, which is associated with 
the band gap opening in halide perovskites [Fig.~\ref{fig2}(c)], its effect is modulated by changes in 
the metal-halide-metal bond angles due to local disorder. For example, in cubic FAPbI$_3$, 
our analysis suggests that local disorder leads to the increase of the Pb-I-Pb angles towards 180$^\circ$ with thermal expansion, 
partially offsetting the band gap opening expected from bond length increase. In contrast, 
in cubic FASnI$_3$, local disorder leads to a larger deviation of Sn-I-Sn angles away from 180$^\circ$, 
enhancing the band gap opening and, thus the thermal expansion contribution. 
This contrasting behavior can be attributed to the more pronounced stereochemical activity of the Sn 5s$^2$ 
lone pairs~\cite{Fabini2020,Balvanz2024}, which favors asymmetric bonding under lattice expansion.

Figure~\ref{fig9}(w) shows the temperature dependence of the hole and electron effective masses of 
cubic FASnI$_3$ in the range 140 -- 400~K. 
Linear fits to the data for the ZG reference (open squares) and ZG polymorphous (red discs)
structures are $\lambda^T_{{\rm h, ref}}  =  1.8 \times 10^{-4}$,
and $\lambda^T_{{\rm e, ref}}  =  2.4 \times 10^{-4}$,
$\lambda^T_{{\rm h, poly}} = 1.9  \times 10^{-4}$,
  and $\lambda^T_{{\rm e, poly}} = 2.4 \times 10^{-4}$~$m_0 / {\rm K}$. 
A single ZG polymorphous configuration obtained from the MSS structure [open circles in Fig.~\ref{fig9}(w)] yields 
similar slopes but underestimates the values obtained from Boltzmann-weighted averages by about 0.01--0.02~$m_0$.
Electron-phonon coupling in reference structures is the leading thermal effect, 
contributing to about 72\% to the effective masses enhancements. This contribution is
reduced to 50\% when local disorder is taken into account with thermal expansion contributing equally.
Similar to the band gap behavior, the increased thermal expansion contribution to the effective mass enhancement in 
locally disordered structures is attributed to the enhanced lone pair activity in Sn-based compounds, promoting 
greater Sn-I-Sn bond angle bending under lattice expansion.
Finally we report that for ZG polymorphous structures, the reduced effective mass exhibits a linear temperature dependence, increasing 
from $\mu_{\rm poly}(200~{\rm K}) = 0.053$ to $\mu_{\rm poly}(400~{\rm K}) = 0.075$.


\begin{figure*}[htb!]
 \begin{center}
\includegraphics[width=0.85\textwidth]{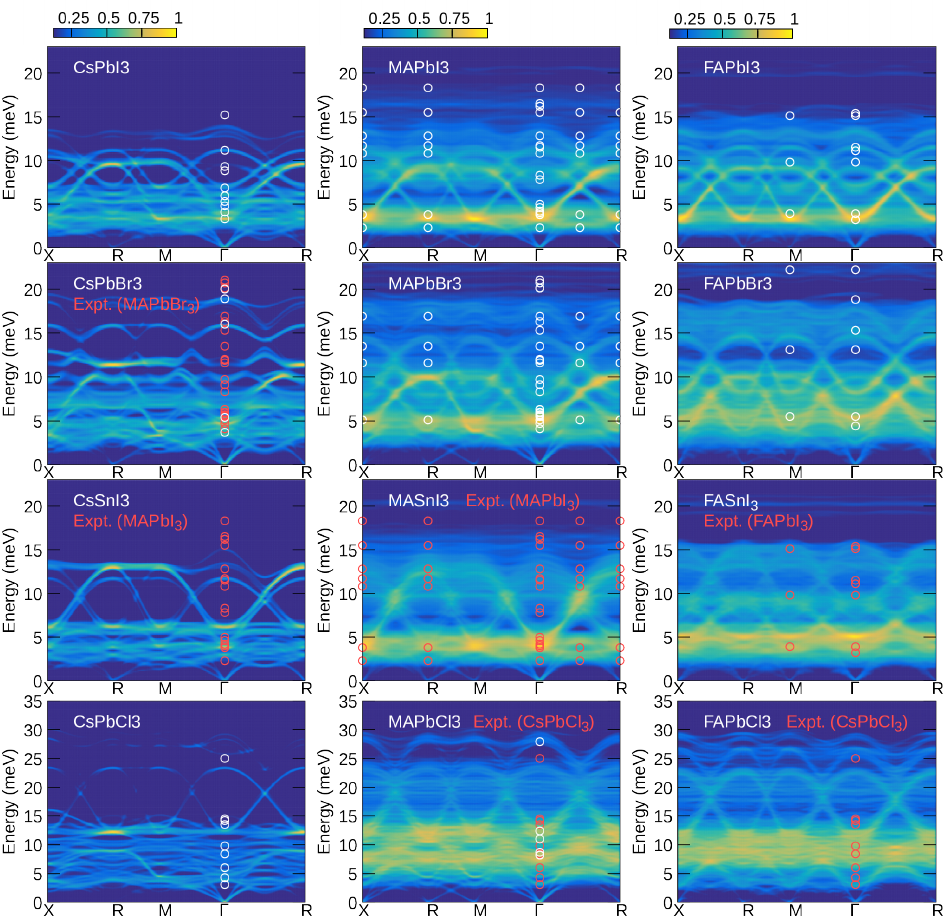}
 \end{center} 
\caption{ (a) Phonon spectral functions (color map) of inorganic and hybrid halide perovskites computed by phonon band 
unfolding~\cite{Zacharias2023npj}. Calculations represent averages over spectral functions computed for 10 
polymorphous configurations of each compound. 
Open white circles denote experimental data from various techniques, as listed in Table~\ref{table.2}. 
Open red circles indicate experimental data from other compounds, as specified in each plot.
For MAPbI$_3$, MAPbBr$_3$, FAPbI$_3$ and FAPbBr$_3$ we include phonon energies at the X, R, M, and 
mid $\Gamma$-R high-symmetry points obtained from neutron scattering measurements~\cite{Ferreira2020}. 
\label{fig11} 
}
\end{figure*}

\subsection{FAPbCl$_3$}

Here, we present our findings for cubic FAPbCl$_3$. 
As reported in Table~\ref{table.1}, the average DFT-PBEsol band gap for the reference structures is 
0.89~eV. Incorporating local disorder results in a moderate band gap widening of 0.28~eV. Hybrid functional 
calculations using PBE0 yield a band gap of 2.51~eV for the reference structures, which underestimates 
the experimental value of 2.91~eV~\cite{Lpez2025}. 
The use of locally disordered networks improves the agreement, yielding a PBE0 band gap of 2.79~eV.
Assuming also that the band gap of FAPbCl$_3$ is similar to that of 
MAPbCl$_3$~\cite{Bokdam2016}, our calculated value of 2.79~eV
compares well with the experimental band gaps of 2.88~eV and $3.07$~eV at 300~K reported for MAPbCl$_3$ in Refs.~[\onlinecite{Maculan2015,Hsu2019}].

The effective masses calculated for the reference structures
are $m^*_{\rm h, ref}=0.17$ and $m^*_{\rm e, ref}=0.18$. Using locally disordered structures,
the effective masses increase to $m^*_{\rm h, poly}=0.23$  and $m^*_{\rm e, poly}=0.25$. 
The associated hole and electron 
effective mass enhancement factors due to local disorder are $\lambda_{\rm h} = 0.35$ and $\lambda_{\rm e} = 0.39$, 
 which are slightly lower than those of MAPbCl$_3$ due to the reduced degree of local disorder in FAPbCl$_3$.
Corrections arising from polymorphism increase the reduced effective mass from $\mu_{\rm ref} = 0.087$ to $\mu_{\rm poly} = 0.120$. 
The latter value compares well with the average $GW$ value of 0.159 reported for tetragonal 
MAPbCl$_3$, falling within the expected accuracy of the calculations~\cite{Mosconi2016}.

We next investigate phonon anharmonicity in FAPbCl$_3$. 
The cubic phase of FAPbCl$_3$ is stable above 200~K~\cite{Govinda2018}; accordingly, 
all calculations are performed at temperatures above this value.
Figure~\ref{fig8}(l) presents the ASDM phonon dispersion at 300~K, converged by the third iteration, within 
the 0--30~meV energy range characteristic of vibrations in the inorganic lattice framework.
The anharmonic phonon dispersion of FAPbCl$_3$ has some similar features to 
those reported for MAPbCl$_3$ [Fig.~\ref{fig8}(h)] and CsPbCl$_3$ [Fig.~\ref{fig8}(d)]; 
for example, all span comparable energy ranges and exhibit an LO-TO splitting above 
10~meV (see $\w_{{\rm \, LO}_4}$ values in Table~\ref{table.2}). 
Differences primarily stem from A-site cations contributions and how they modulate the lattice dynamics through 
interactions with the metal-halide network.
Table~\ref{table.2} reports the frequencies at the $\Gamma$ point of FAPbCl$_3$: 
$\w_{{\rm \, TO}_{1-2}}  = 7.56-7.62$, $\w_{{\rm \, TO}_{3-4}} = 10.1-11.4$, $\w_{{\rm \, TO}_{5-6}}  = 12.4-12.9$, 
 $\w_{{\rm \, TO}_{7-8}} = 15.8-16.6$, 
$\w_{{\rm \, LO}_1} = 7.64$, $\w_{{\rm \, LO}_2} = 10.2$, $\w_{{\rm \, LO}_3} = 15.5$, and $\w_{{\rm \, LO}_4} = 25.8$~meV.
These values are in good agreement with room-temperature infrared and THz-TDS spectroscopy measurements for MAPbCl$_3$~\cite{Sendner2016,Nagai2018},
which report TO modes at 8.2 and 11.0~meV and LO modes at 8.7 and 27.9~meV, as well as LO modes 
at 8.2 and 12.4~meV. 


Now we examine the effect of electron-phonon coupling and thermal expansion on the electronic structure. 
Table~\ref{table.1} reports the band gap renormalization calculated using ZG reference and 
ZG polymorphous networks which are $\Delta E_{\rm g, ref}(300 {\rm ~K}) = 0.09$~eV
and $\Delta E_{\rm g, poly}(300 {\rm ~K}) = 0.03$~eV. The reported thermal-induced mass corrections, $\Delta m^*$,
lead to the following enhancements factors: 
$\lambda_{\rm h, ref} (300 {\rm ~K}) = 0.24$, $\lambda_{\rm e, ref} (300 {\rm ~K})=0.22$, 
$\lambda_{\rm h, poly} (300 {\rm ~K})= 0.22$ and $\lambda_{\rm e, poly} (300 {\rm ~K})= 0.20$, slightly smaller
than those reported for MAPbCl$_3$. 

In Fig.~\ref{fig9}(t), we present temperature-dependent band gaps of cubic FAPbCl$_3$ in the range 250--400~K
calculated using 10 ZG reference (open squares) and their corresponding ZG polymorphous (black discs) structures.
We also include band gaps computed with a single ZG polymorphous configuration obtained from the MSS structure (open cirles).
To match the experimental value of 2.91~eV~\cite{Lpez2025} at room temperature, we shift all DFT data  
by about $\Delta_s = 1.71$~eV, similar to the PBE0 correction of 1.82~eV obtained for the polymorphous structures (see Table~\ref{table.1}).
Our data for the ZG reference structures show a band gap widening with temperature, with a temperature coefficient 
of $dE_{\rm g} / dT = 5.3 \times 10^{-4}$~eV/K. 
Figures~\ref{fig10}(a) and (b) show that both electron-phonon coupling and thermal expansion lead to an opening of 
the band gap of the ZG reference structures, with the thermal expansion being the leading contribution with 60\%. 
When ZG polymorphous networks are employed, electron-phonon coupling is suppressed, consistent with trends observed in other 
FA-based compounds, while the thermal expansion contribution remains nearly unchanged. As a result, thermal expansion 
contribution to the total band gap renormalization is increased to 81\%. The calculated temperature coefficient of the band gap for 
the ZG polymorphous structures is $dE_{\rm g} / dT = 4.1 \times 10^{-4}$~eV/K, which aligns well with the 
experimental value of $3.2 \times 10^{-4}$~eV/K reported for CsPbCl$_3$~\cite{He2021}. In contrast, our calculations for 
FAPbCl$_3$ do not reproduce the nearly temperature independent band gap behavior observed in MAPbCl$_3$. 
Direct experimental measurements for FAPbCl$_3$ are needed to validate our predictions.
We remark that using a single ZG polymorphous structure offers a reasonable approximation for evaluating the 
temperature dependence of the band gap in FAPbCl$_3$, giving a temperature coefficient of $5.3 \times 10^{-4}$~eV/K. 

Figure~\ref{fig9}(x) shows the temperature dependence of the hole and electron effective masses of
the ZG reference (open squares) and ZG polymorphous structures (black discs). In both cases, the effective masses
exhibit a linear increase with temperature with gradients: 
$\lambda^T_{{\rm h, ref}}  =  1.9 \times 10^{-4}$, $\lambda^T_{{\rm e, ref}}  =  1.8 \times 10^{-4}$, 
$\lambda^T_{{\rm h, poly}} = 1.9  \times 10^{-4}$, and $\lambda^T_{{\rm e, poly}} = 2.4  \times 10^{-4}$~$m_0 / {\rm K}$, 
indicating that local disorder has little impact on the temperature variation of the effective masses. 
We remark that now electron-phonon coupling contribution dominates 
thermal induced renormalization in all cases, being more than 63\%. 
We also report that the reduced effective mass calculated for polymorphous cubic FAPbCl$_3$ 
varies linearly with temperature, 
being  $\mu_{\rm poly}(250~{\rm K}) = 0.136$ and $\mu_{\rm poly}(400~{\rm K}) = 0.153$.
Finally, we note that unlike the band gap, the ZG polymorphous configuration of the MSS structure 
is not a good approximation to explain the temperature variation of the effective masses [open circles in Fig.~\ref{fig9}(x)].

\section{General comparison and discussion} \label{sec.general_comp}

\subsection{Electronic structure} \label{sec.general_comp_A}
In general, we find that the electronic structure modifications induced by polymorphism are critical for improving 
agreement between calculated and experimental values of band gaps and effective masses across all compounds. 
This conlusion aligns with previous work by Zhao {\it et al}~\cite{Zhao2020,Zhao2021,Zhao2024}, who emphasized that, in 
perovskite systems, including oxides, variations in band gap often arise from structural polymorphism rather than 
strong electronic correlations alone. 
Furthermore, our findings suggest that accounting for local disorder in cubic phases is 
essential to prevent spurious semimetallic 
behavior in DFT calculations, especially for Pb- and Sn-based iodide perovskites, where such 
artifacts severely hinder electronic structure properties.
Depending on the size of the A-site cation, which governs the available steric environment, 
local disorder induces an average band gap opening of 0.27, 0.46, and 0.61~eV; hole effective mass increase
of 0.06, 0.12 and 0.18~$m_0$; electron effective mass increase of 0.07, 0.11, and 0.19~$m_0$  in FA-, MA-, and Cs-based 
compounds, respectively. We, thus, identify local disorder as one of the main factors determining the band gap and effective mass 
ordering in APbI$_3$ and APbBr$_3$ halide perovskites, where Cs-based compounds exhibit the largest band gaps and effective masses, 
followed by MA- and FA-based systems. In contrast, for Sn-based and Cl-based compounds, additional factors, such as thermal expansion 
as well as lone pair expression and excitonic effects, respectively, play an important role. 

Improved treatment of electronic correlations using PBE0 calculations combined with polymorphous structures yields 
overall good agreement with experimental band gap values, as shown in Table~\ref{table.1}. The calculated band gaps exhibit an average 
absolute deviation of 0.11~eV from experimental values. Including thermal induced corrections [$\Delta E_{\rm g, poly}(T)$] further 
improves agreement in some cases, but increases the discrepancy in others, resulting in a slightly higher average absolute deviation 
of 0.12~eV. Moreover, our PBE0 values compare well with $GW$ calculations for the tetragonal 
and orthorhombic phases, which incorporate the effect of octahedral tilting. 
In future work, it would be interesting to explore how the combination of state-of-the-art $GW$ calculations using cubic 
polymorphous structures and thermal induced corrections compares with experiments.
We emphasize that polymorphism is also essential for improving agreement with experimental values of the reduced effective masses, 
as shown in Table~\ref{table.1}. 
In this case, $GW$ calculations are also expected to play an important role in achieving better quantitative accuracy.
We observe that for the reduced effective masses of Pb-based compounds, the agreement with experiment decreases 
from Cs-based to MA-based, and further to FA-based compounds. For example, our values underestimate 
experimental data for CsPbI$_3$, MAPbI$_3$, and FAPbI$_3$ by 18\%, 31\%, and 48\%, respectively. 
This trend aligns with the fact that compounds with larger cations exhibit smaller degree of polymorphism, 
and thus stronger electronic correlations. 

\subsection{Band gap across different polymorphs}

As discussed in Secs.~\ref{sec.case_by_case} and~\ref{sec.general_comp_A}, the band gaps obtained for the polymorphous cubic structures 
agree well with prior calculations on lower-symmetry polymorphs that explicitly incorporate octahedral tilts, i.e. the 
tetragonal and orthorhombic phases. Considering the monomorphous phases of CsPbBr$_3$, our
calculations yield a DFT-PBEsol band gap of $E_{\mathrm g} =0.83$, $0.69$, and $0.24$~eV for the orthorhombic (Pnma), 
tetragonal (P4/mbm), and cubic ($\mathrm{Pm}\bar{3}\mathrm m$) phases, respectively. Assuming a monomorphous picture this trend 
is reasonable: larger octahedral tilts in the 
orthorhombic phase widen the gap relative to the tetragonal phase (a$^0$a$^0$c$^{\pm}$; no apical tilting) and, ultimately, to the 
untilted cubic phase. However, this ordering contradicts the measured band gap sequence in CsPbBr$_3$~\cite{Mannino2020}, 
where $E_{\mathrm g}(\mathrm{Pm}\bar{3}\mathrm m) > E_{\mathrm g}(\mathrm{P4/mbm}) > E_{\mathrm g}(\mathrm{Pnma})$.
The corresponding calculated band gaps using polymorphous structures are: $E_{\mathrm g}=0.83$, $0.86$, and $0.85$~eV 
for the orthorhombic, tetragonal, and cubic phases, respectively.
This nearly recovers the expected trend (orthorhombic $<$ tetragonal $\simeq$ cubic), compressing the spread 
to 0.03 eV. As shown in Ref.~[\onlinecite{Zacharias2023npj}], adopting polymorphous rather 
than monomorphous structures and including thermal effects is crucial to capture the continuous increase 
of the band gap with temperature across different phases, avoiding spurious discontinuities at the phase boundaries. 
This suggests that reproducing the correct band gap ordering requires explicitly accounting for local structural disorder 
and finite-temperature effects, i.e. thermal lattice expansion and electron-phonon coupling.

A similar trend holds for MAPbBr$_3$~\cite{Mannino2020}. Using monomorphous orthorhombic, reference tetragonal, and reference cubic structures 
of MAPbBr$_3$, our DFT-PBEsol calculations yield $E_{\mathrm g }=0.83$, $0.46$, and $0.39$~eV, respectively. In contrast, polymorphous
tetragonal and cubic structures that capture local disorder give
an almost phase-independent gap of $E_{\mathrm g} \simeq 0.83$~eV.
Interestingly, representing the finite-temperature band gap with a ZG orthorhombic MAPbBr$_3$ snapshot already 
provides a better description of the band gap at high temperatures than using ZG reference cubic cells~\cite{short}.

For MAPbI$_3$, the picture differs slightly. Monomorphous orthorhombic, reference tetragonal, and reference cubic structures give
$E_{\mathrm g}=0.61$, $0.29$, and $0.03$~eV, respectively, having a spread 
of 0.58 eV. Employing polymorphous tetragonal and cubic structures the corresponding band gaps are $E_{\mathrm g}=0.61$, $0.51$, and $0.52$~eV, 
reducing the spread to 0.1~eV; this residual offset reflects the gap drop at the 
orthorhombic to tetragonal transition, not observed in MAPbBr$_3$~\cite{Mannino2020}. 
This result is consistent with experimental observations for MAPbI$_3$~\cite{Milot2015,Dar2016} which report a band gap drop at the orthorhombic 
to tetragonal phase transition of about 0.1~eV and a nearly smooth evolution from the tetragonal to cubic phase transition. 

\subsection{LO-TO splitting}
In the inorganic monomorphous Pm$\bar{3}$m structures, the 12 optical phonons at the Brillouin zone 
center are grouped into four triply degenerate modes~\cite{Even2015} prior to the inclusion of long-range dipole-dipole interactions. 
These comprise three infrared-active modes ($\Gamma_4^-$) and one silent mode ($\Gamma_5^-$). 
When LO-TO splitting is taken into account, our calculations show that three TO modes are doubly degenerate, 
while $\omega_{\,{\rm TO}_2}$ retains its triply degenerate character, consistent with its assignment as the silent mode. 

In Table~\ref{table.2}, we report in parenthesis the frequency of LO modes before LO-TO splitting for all compounds, 
each matching a value of the TO modes. 
Since the LO-TO splitting scales inversely with the atomic mass, materials incorporating lighter halides, such as Cl instead of Br or I, 
exhibit a more pronounced splitting related to the high energy LO mode. This effect arises from the stronger long-range restoring 
forces acting on the lighter ions, which elevate the LO frequency relative to the TO. As a result, compounds like CsPbCl$_3$ show a 
larger LO-TO separation, e.g 10.8~meV for $\w_{{\rm \, LO}_3}$, compared to their heavier halide counterparts, such as CsPbI$_3$, CsPbBr$_3$, and CsSnI$_3$,
which show a splitting of 4.9, 6.5, and 6.2~meV, respectively. A similar picture holds for MA and FA-based compounds with different X-site anions. 

The LO-TO splitting associated with the high energy LO mode is the smallest in FA-based perovskites. 
For example, the LO-TO splitting in FAPbI$_3$ is 3.6~meV, in MAPbI$_3$ is 5.8~meV, and in CsPbI$_3$ is 4.9~meV.  
To rationalize this result we first use the simplified expression for the non-analytic LO-TO splitting, 
$\Delta_{\rm LO-TO} \propto |Z^{*}|^2/ L^3 \epsilon^{\infty}$, where $Z^{*}$ indicates the Born effective charges in units of elementary charge, 
$\epsilon^{\infty}$ is the high frequency dielectric constant, and $L^3$ is the volume of the cubic system. In view of the expression 
for $\Delta_{\rm LO-TO}$, one can expect FAPbI$_3$ to exhibit a similar if not larger splitting among the A-site variants. This is based on the fact that 
FAPbI$_3$ has relatively larger Born effective charges (e.g., $Z_{\rm Pb}^*=4.72$, $Z_{\rm I}^*=4.68$), a similar
 high-frequency dielectric constant ($\epsilon^{\infty}=6.13$), and a similar lattice constant ($L=6.36$~\AA) compared to MAPbI$_3$ 
(with $Z_{\rm Pb}^*=4.59$, $Z_{\rm I}^*=4.44$, $\epsilon^{\infty}=5.87$, and $L=6.31$~\AA) and CsPbI$_3$ (with $Z_{\rm Pb}^*=4.49$, $Z_{\rm I}^*=4.17$, 
$\epsilon^{\infty}=5.75$, and $L=6.25$~\AA). 
Nonetheless, our density functional perturbation theory calculations show that FAPbI$_3$ exhibits the smallest LO-TO splitting among the three compounds. 
This contradiction indicates that the simplified expression does not capture the full microscopic picture~\cite{Gonze1997}, as it neglects the phonon mode character, dipole 
directionality, and mode mixing. We ascribe the reduced LO-TO splitting in FAPbI$_3$ to enhanced mode mixing between polar optical phonons 
and internal molecular vibrations [Fig.~\ref{fig14}(c)], as well as directional coupling effects that diminish the effective dipole strength of the LO mode.

\subsection{Overdamped phonon dynamics} \label{Sec.Overdamped_ph}
In general, the calculated phonon energies at the zone center, reported in Table~\ref{table.2}, show good agreement 
with measurements obtained from various experimental techniques. However, our anharmonic phonon dispersions (Fig.~\ref{fig8}) 
are based on reference structures and the quasiparticle approximation, and therefore neglect vibrational correlations arising 
from local structural disorder. 
As demonstrated in Ref.~[\onlinecite{Zacharias2023npj}] for inorganic halide perovskites, local disorder gives rise to strongly overdamped 
and interacting vibrational modes, with only acoustic phonons near the $\Gamma$ point remaining clearly identifiable.
This picture, for example, aligns with several experimental observations: (i) inelastic neutron scattering measurements in hybrid halide perovskites 
revealing the absence of well-defined optical phonon modes across the entire energy range, attributed to strong overdamping 
at all momenta~\cite{Ferreira2020}; (ii) Raman spectra of both inorganic and hybrid halide perovskites displaying a pronounced central 
peak over a broad frequency range, indicative of correlated polar vibrations~\cite{Yaffe2017}; (iii) infrared transmission measurements 
showing highly overdamped vibrational modes lacking distinct quasiparticle features~\cite{Anikeeva2023}; and 
(iv) THz time-domain spectroscopy pointing to damping of LO phonons due to mode coupling~\cite{Nagai2018}.

In Fig.~\ref{fig11}, we present phonon spectral functions of all compounds calculated as an average over 10 
polymorphous structures in $2\times2\times2$ supercells in conjunction with the phonon unfolding technique; for more computational details 
see Sec.~\ref{sec_ph_spectral_calc}. We focus on energy windows associated mostly with vibrations of the inorganic sublattice. 
Similar to findings of Ref.~[\onlinecite{Zacharias2023npj}], the phonon spectral functions of all cubic phases 
reveal that acoustic phonons emerge from a sea of strongly correlated optical vibrations across the entire Brillouin zone. 
This picture, emerging from our quasistatic polymorphous approach, is inconsistent with the translational diffusion characteristics 
of a liquid~\cite{Lahnsteiner2022}, as it preserves well-defined transverse acoustic phonons despite the presence of strong anharmonicity.
Qualitatively, Fig.~\ref{fig11} shows that the degree of phonon overdamping varies with the A-site cation, being strongest 
in FA-based compounds, followed by MA-based, and weakest in Cs-based halide perovskites.
This conclusion is also supported by low frequency Raman spectra reported for CsPbBr$_3$ and MAPbBr$_3$~\cite{Yaffe2017}.
Furthermore, hybrid halide perovskites exhibit strongly correlated vibrational dynamics extending across a broad energy range, 
including high-energy optical modes. In contrast, overdamped phonons in Cs-based compounds are primarily confined to lower energies, 
typically below 10~meV. We attribute the greater extent of phonon overdamping in hybrid halide perovskites to the presence of molecular 
vibrational modes and their coupling to the vibrations of the inorganic sublattice accross a larger energy range. 

In Fig.~\ref{fig11}, we also compare our calculated phonon spectral functions with measurements of phonon energies at the $\Gamma$ point
from various techniques and samples, as reported in Table~\ref{table.2}. We additionally include neutron scattering data 
for MAPbI$_3$, MAPbBr$_3$, FAPbI$_3$ and FAPbBr$_3$ at the X, R, M, and mid $\Gamma$-R high-symmetry points from Ref.~[\onlinecite{Ferreira2020}]. 
These data reveal nearly dispersionless optical vibrations with energies that remain constant across the Brillouin zone, an observation 
that is supported by our calculations. 
Furthermore, the highest spectral weights in the calculated phonon spectral functions are distributed around 5~meV for I-based and Br-based compounds, 
and around 10~meV for Cl-based compounds, in good agreement with various experimental phonon energies in this range.
In general, we observe that incorporating structural disorder allows us to more accurately capture the variations observed in experimental data 
across different sample types, such as thin films, single crystals, polycrystalline pellets, nanocrystals, and nanocrystal films, where deviations 
from idealized reference structures, 
significantly influence phonon behavior leading to extensive broadening and phonon frequencies shifts.
We remark that in some cases, especially for high energy modes, the spectra function peaks either overestimate or underestimate 
experimental data which we attribute to (i) the DFT functional used for computing IFCs, (ii) different lattice constants,
(iii) the supercell size used to simulate local disorder, (iv) the temperature and perovskite phases considered in the experiment, and (iv) 
the sample used for the measurements, e.g. thin films, single crystals, pellets, and nanocrystals.
We note that, for example, in the case of nanocrystals other factors affect the phonon energy measurements, such as quantum confinement,  dielectric
screening, shape and surface effects, as well as the enhanced local disordered discussed in Sec.~\ref{sec.FASnI3}.

\subsection{Thermal induced renormalization}
We have seen that thermal induced corrections are strongly modified for most cubic compounds in the presence of local disorder. 
As shown in Fig.~\ref{fig10}(a), electron-phonon contribution is strongly suppressed for all compounds, with the largest 
reductions found for Cs-based compounds which exhibit the largest degree of local disorder. 
The only case where electron-phonon coupling contributes negatively to the temperature coefficient of the 
band gap leading to a band gap closing is MAPbCl$_3$. 
We attribute this to the fact that MAPbCl$_3$ exhibits one of the highest degrees of local disorder among the 
hybrid halide perovskites, reaching levels comparable to those of Cs-based compounds [Fig.~\ref{fig2}(b) and Fig.~2(c) of Ref.~[\onlinecite{short}]]. 
The opposing effects of electron-phonon coupling and thermal expansion in polymorphous cubic MAPbCl$_3$ lead to 
the nearly constant variation of the band gap with temperature, consistent with experiments. 

Thermal expansion contribution to the band gap renormalization is also reduced due to polymorphism for most compounds, 
especially for Cs-based compounds [Fig.~\ref{fig10}(b)]. Thermal expansion increases the bond lengths, which tends to 
open the band gap in halide perovskites as shown in Fig.~\ref{fig2}(c). However, this effect is also influenced by how 
the bond angles change upon allowing the structure to explore local disorder. For example, in cubic FAPbI$_3$ and most 
of the compounds studied here, local disorder causes the Pb-I-Pb angles to open toward 180$^\circ$, which reduces the 
band gap opening due to thermal expansion. In contrast, in FASnI$_3$, the Sn-I-Sn angles bend more, enhancing the gap 
opening, as shown in Fig.~\ref{fig10}(b). This difference comes from the stronger lone pair stereochemical expression of Sn as compared to Pb.

In general, we observe that the electron-phonon coupling percentage contribution depends on the choice of the 
A-site cation and decreases from Cs-based to MA-based, and further to FA-based compounds.
In particular, electron-phonon coupling is the dominant contribution to the band gap renormalization in 
Cs-based and MA-based polymorphous compounds, except from MAPbCl$_3$, reaching as high as 97\%. 
For FA-based compounds, except for FAPbI$_3$, the dominant contribution to the band gap renormalization is thermal 
expansion reaching as high as 81\%. We attribute this to the relatively larger thermal expansion coefficients 
for the FA-based cubic phases~\cite{Dang2016,Schueller2017,Handa2020} varying from $0.58 - 0.68 \times 10^{-4}$  K$^{-1}$, while 
for MA-based compounds~\cite{Mashiyama2003,Rakita2015,Whitfield2016,Dang2016,Schueller2017,Handa2020} 
from $0.33-0.54 \times 10^{-4}$  K$^{-1}$ and for Cs-based compounds~\cite{Rakita2015,Kontos2018,Marronnier2018,He2021} 
from $0.29 - 0.51 \times 10^{-4}$  K$^{-1}$.

Our results for the total temperature coefficients of the band gap using ZG polymorphous structures compare well with the 
experimental values, as shown quantitatively in Fig.~\ref{fig10}(c) and Figs.~\ref{fig9}(a)-(d), (i)-(l), and (q)-(t). 
Importantly, we have seen that even using a single ZG polymorphous configuration obtained from the MSS structure 
is a good approximation for capturing the temperature variation of the band gap. 
We also find that thermal-induced corrections lead to a linear variation of the effective masses with temperature 
[Figs.~\ref{fig9}(e)-(h), (m)-(p), and (u)-(x)]. In APbI$_3$ and APbBr$_3$ compounds, the hole effective masses remain 
larger than the electron effective masses, whereas in ASnI$_3$ and APbCl$_3$ compounds, electron effective masses remain 
larger than hole effective masses. We note that our finite-temperature calculations: 
 (i) rely on a constant temperature phonon approximation, (ii) neglect non-adiabatic effects in electron-phonon 
renormalization of the band structure~\cite{Ponc2015,Miglio2020,Engel2022}, (iii) neglect many-body quasiparticle corrections 
to the electronic structure~\cite{Antonius2014,Marini2015,Li2024}, and (iv) omit phonon renormalization 
due to thermal expansion. While including these effects may improve quantitative agreement with experiment, we do not expect 
them to alter the qualitative trends, as our current results already show good overall agreement. 



\begin{figure}[htb!]
 \begin{center}
\includegraphics[width=0.30\textwidth]{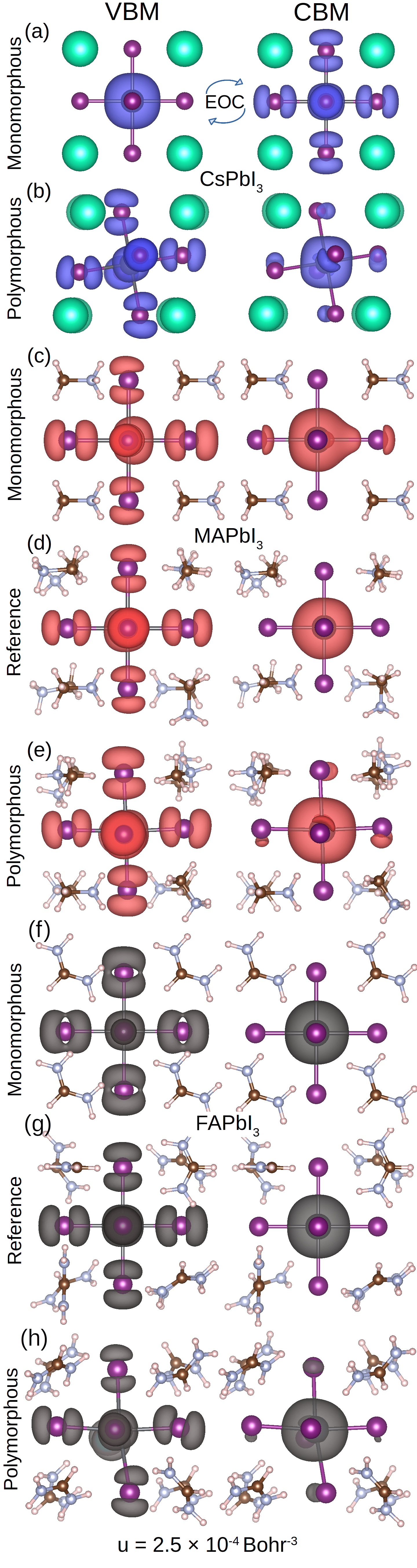}
 \end{center} 
\caption{(a-h) Charge densities at the valence band maximum (VBM) and conduction band minimum (CBM) of a single motif of the 
monomorphous, reference, and polymorphous 
CsPbI$_3$ (a,b), MAPbI$_3$ (c-e), and FAPbI$_3$ (f-h). The isosurface value $u$ is indicated at the bottom. 
\label{fig12} 
}
\end{figure}

\begin{figure*}[htb!]
 \begin{center}
\includegraphics[width=0.68\textwidth]{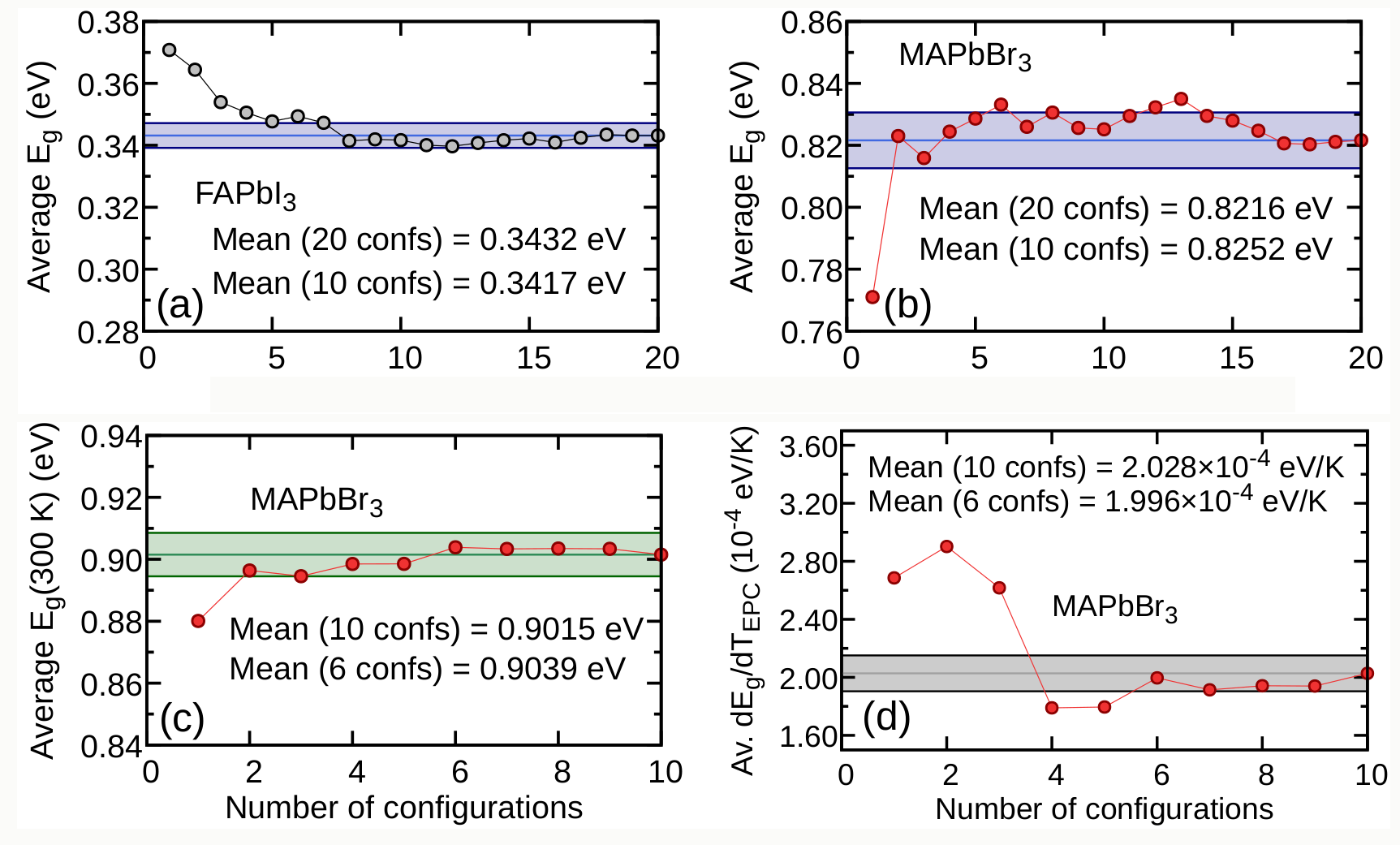}
 \end{center} 
\caption{(a,b) Convergence test of the average DFT-PBEsol band gap of cubic FAPbI$_3$ (a) and MAPbBr$_3$ (b) as a function 
of the number of polymorphous configurations. (c) Convergence test of the average DFT-PBEsol band gap of cubic 
MAPbBr$_3$ at 300~K [Eq.~\eqref{eq11}] as a function of the number of ZG polymorphous configurations. 
Average values for 10 configurations are 
converged within 5~meV. (d) Convergence test of the average band gap temperature coefficient due to electron-phonon coupling 
as a function of the number of ZG polymorphous configurations used to compute the band gap of  MAPbBr$_3$ at each temperature. 
The temperature coefficient is calculated by fitting data points for the 5 temperatures shown in Fig.~\ref{fig7}(c).
The shaded regions represent the standard error of the mean on either side of the lines.
\label{fig13} 
}
\end{figure*}

\section{Computational details} \label{computdet}

\subsection{Electronic structure calculations}

DFT calculations were performed using {\tt Quantum Espresso} (QE)~\cite{QE,QE_2}. 
We employed optimized norm-conserving Vanderbilt pseudopotentials from
the PseudoDojo library~\cite{Haman_2013,vanSetten2018} and the Perdew-Burke-Ernzerhof exchange-correlation functional 
revised for solids (PBEsol)~\cite{Perdew_2008}. 
The lattice constants ($L$) of all cubic phases, reported in Table~\ref{table.1}, were fixed close to their experimental 
values~\cite{Mashiyama2003,Rakita2015,Whitfield2016,Dang2016,Schueller2017,Kontos2018,Marronnier2018,Handa2020,He2021}. 
Calculations for the monomorphous structures (5 atoms for inorganic and 12 atoms for 
hybrid halide perovskites) were performed using a plane wave cutoff energy of 120~Ry and $6\times6\times6$ uniform ${\bf k}$-grids. 
For exploring reference and polymorphous cubic structures with the ASDM as described 
in Sec.~\ref{sec.Poly_explore}, we performed geometry optimization
in $2\times2\times2$ supercells (40 atoms for inorganic and 96 atoms for
hybrid halide perovskites), until residual forces per atom were below 3$\times$10$^{-4}$~eV/\AA \,
using a plane wave cutoff energy of 120~Ry and $3\times3\times3$ uniform ${\bf k}$-grids. 
During geometry optimization the lattice constants were kept fixed. We note that for the 
reference structures of hybrid halide perovskites only the molecules were relaxed, while for the unique reference
(monomorphous) structures of inorganic compounds no relaxation was performed as the residual forces were zero. 
For each inorganic halide perovskite, we generated 1 high-symmetry monomorphous structure and 10 polymorphous structures.
For each hybrid halide perovskite, we generated 1 MSS reference structure and its corresponding polymorphous structure, as well as 
10 reference structures obtained by relaxing molecules from random initial orientations and their corresponding polymorphous 
structures.
To construct MSS structures for hybrid halide perovskites, we employed a $2\times2\times2$ supercell and initially fixed the orientation 
of one molecular cation within a single unit cell. Under this supercell constraint, the remaining seven molecules were placed such that
 a subset of the original space group symmetries, corresponding to the P422 space group, was preserved.
The band gap of each structure was obtained by including SOC effects via fully relativstic pseudopotentials 
and evaluating the energy difference between the conduction band minimum and valence band maximum at the R point.
Convergence tests of the band gap of FAPbI$_3$ and MAPbBr$_3$ with respect to the number of 
polymorphous configurations are shown in Figs.~\ref{fig13}(a) and (b).

To check whether DFT-SOC calculations for each structure yield an exchange of orbital character (EOC) we computed 
the charge densities of the band edges at the R point. 
In Figs.~\ref{fig12}(a)-(h), we present the computed charge densities for a single octahedron in the monomorphous, 
reference, and polymorphous cubic structures of CsPbI$_3$, MAPbI$_3$, and FAPbI$_3$.
We find that using a monomorphous network of CsPbI$_3$ yields an EOC which is corrected upon 
using a polymorphous structure [Figs.~\ref{fig12}(a,b)]. A similar behavior is observed for monomorphous 
CsSnI$_3$ as well as for monomorphous and reference MASnI$_3$.
In Fig.~\ref{fig12}(c), we observe for MAPbI$_3$ that using a monomorphous network with a single 
orientation for the MA cation results in an anisotropic charge density, 
especially around the Pb atom, in both the valence band maximum and conduction band minimum.
This spurious directional distortion in the electronic charge density arises from the 
net dipole moment of the MA molecule. Using a reference structure within a $2\times2\times2$ supercell and allowing 
the molecules to relax starting from random orientations alleviates the anisotropy in the charge density 
considerably, as shown in Fig.~\ref{fig12}(d). The same picture is maintained when a polymorphous structure 
is used [Fig.~\ref{fig12}(e)]. 
In Fig.~\ref{fig12}(f), our calculations for monomorphous FAPbI$_3$ show a reduced anisotropy in the charge density
compared to monomorphous MAPbI$_3$ due to the smaller dipole moment of the FA molecule. However, at the 
valence band maximum, the charge is more concentrated around the equatorial I atoms 
(those lying in the same plane as the FA molecules) compared to the apical ones. 
Using the reference and polymorphous structures [Figs.~\ref{fig12}(g) and (h)], we observe a more uniform distribution 
of charge density around the I atoms.

PBE0 hybrid functional calculations~\cite{Perdew1996,Adamo1999} for the reference and polymorphous cubic
structures were performed using the VASP code~\cite{Kresse1996} within the projector augmented-wave 
formalism~\cite{Kresse1999}. A plane-wave energy cut-off of 400 eV was used, 
and the Brillouin zone was sampled with a 2$\times$2$\times$2 uniform {\bf k}-grid.

 Calculations for the monomorphous orthorhombic (Pnma) and tetragonal (P4/mbm) structures of CsPbBr$_3$ were performed 
using the unit cells containing 20 and 10 atoms, respectively, with uniform {\bf k}-grids of
 3$\times$3$\times$2 and 4$\times$4$\times$6. We used lattice constants of (a = 7.97, b = 8.40, and c = 11.64~\AA) and 
(a = b = 5.73, c = 5.96~\AA)~\cite{Zacharias2023npj}. To obtain a polymorphous structure of tetragonal 
CsPbBr$_3$, we relaxed the atoms in a $2\times2\times2$ supercell (80 atoms), which lowered the energy 
by 16~meV / f.u. relative to the monomorphous model. As shown in Ref.~[\onlinecite{Zacharias2023npj}], 
exploring local disorder in the orthorhombic phase yields a negligible stabilization, consistent with the fact that the 
structure is already being at a minimum in the PES. Calculations for the orthorhombic (Pnma) MAPbI$_3$ and MAPbBr$_3$ were performed using 
the unit cell (48 atoms) and a 4$\times$4$\times$3 uniform {\bf k}-grid. We generated 10 reference and 10 polymorphous tetragonal 
structures of MAPbI$_3$ and MAPbBr$_3$ employing a $1\times2\times1$ supercell (96 atoms) and a 3$\times$2$\times$3 uniform {\bf k}-grid.
Lattice constants were fixed to their experimental values~\cite{Lehmann2019}.

\begin{figure*}[htb!]
 \begin{center}
\includegraphics[width=0.98\textwidth]{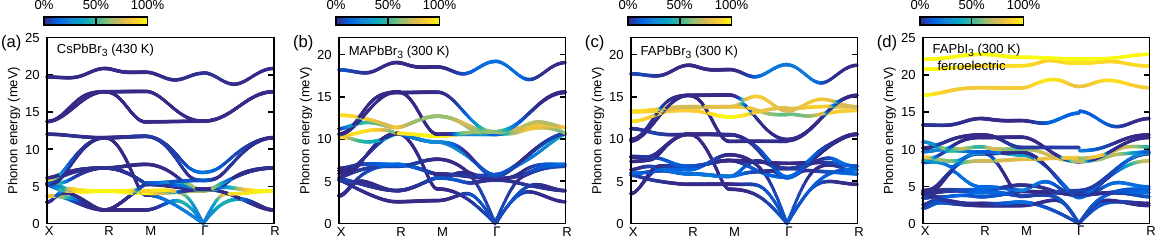}
 \end{center} 
\caption{ (a) Percentage contribution of Cs atom to the ASDM phonon dispersion of CsPbBr$_3$ 
calculated using the high-symmetry monomorphous structure.
(b,c) Percentage contribution of the organic sublattice to the ASDM phonon dispersion of MAPbBr$_3$ (b) and FAPbBr$_3$ (c)
in the range 0--20 meV calculated using MSS reference structures.
(d)  Percentage contribution of the organic sublattice to the ASDM phonon dispersion of FAPbI$_3$
in the range 0--25 meV calculated using the ferroelectric monomorphous network. 
All ASDM calculations refer to $2\times2\times2$ supercells and the temperatures specified in each plot. 
All phonon dispersions include long-range dipole-dipole interaction corrections, leading to LO-TO splitting.
\label{fig14} 
}
\end{figure*}

\subsection{Phonon calculations} \label{sec.phonons_calc}
Harmonic phonons, high-frequency dielectric constants, and Born effective charges of the monomorphous structures in the unit cell 
were computed via density functional perturbation theory (DFPT)~\cite{Baroni2001,QE} using a $2\times2\times2$ uniform ${\bf q}$-grid. 
To compute anharmonic effective IFCs within the SCP theory, we employed $2\times2\times2$ MSS reference structures 
and the ASDM in conjunction with finite differences of 0.01~\AA\, as 
implemented in {\tt ZG.x} of {\tt EPW} code~\cite{Zacharias2023,Lee2023}. 
We employed a plane wave energy cutoff of 100~Ry and a $3\times3\times3$ uniform ${\bf k}$-grid.
Full anharmonic phonon dispersions were obtained by diagonalizing the dynamical matrices at 400 
equally spaced $\bq$-points along the X-R-M-$\Gamma$-R path of the Brillouin zone, using Fourier-interpolated IFCs.
To account for corrections on the anharmonic phonon dispersions due to long-range dipole-dipole interactions,
we used high-frequency dielectric constants and Born effective charges computed for the polymorphous structures 
and the mixed-space approach described in Ref.~[\onlinecite{Wang2010}]. 
The number of iterations at which fully converged ASDM phonon dispersions were obtained for the MSS reference structures 
are indicated in Fig.~\ref{fig8}. 
Anharmonic phonons up to ASDM-2 were computed for the reference configurations of hybrid halide perovskites 
with random molecular orientations (80 configurations in total).
These phonons were used to generate anharmonic special displacements in each 
reference and polymorphous structure for incorporating the effect of electron-phonon coupling (see Sec.~\ref{sec.el-ph_calc}).

To determine the percentage contribution of the molecule (PCM) to the ASDM phonon dispersions shown in Fig.~\ref{fig14}, we used the 
computed phonon polarization vectors at each $\bq$-point and phonon branch $\nu$ with the following expression:
\begin{eqnarray}\label{eq12}
   {\rm PCM}(\nu,\bq) = \sum_{\k \in {\rm mol}, \a} |e_{\k \a,\nu}(\bq)|^2 / \sum_{\k, \a} |e_{\k \a,\nu}(\bq)|^2 \times 100. \nonumber \\
\end{eqnarray}

\begin{figure}[t!]
 \begin{center}
\includegraphics[width=0.4\textwidth]{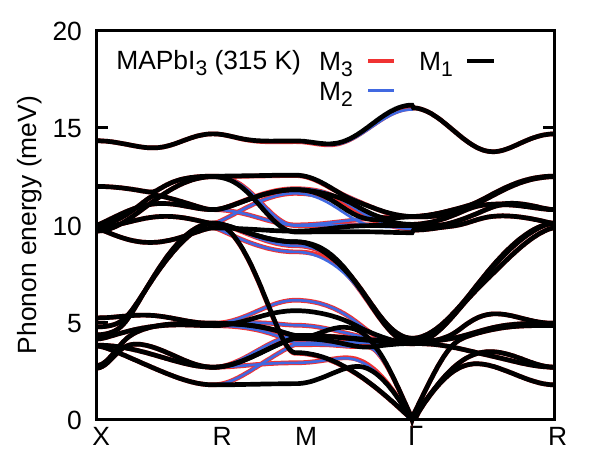}
 \end{center} 
\caption{ ASDM phonon dispersion of the MSS reference structure of MAPbI$_3$ calculated for the three inequivalent 
M points: M$_1$ = [ 0.5 0.5 0.0], M$_2$ = [ 0.5 0.0 0.5], M$_3$ = [ 0.0 0.5 0.5].  
\label{fig15} 
}
\end{figure}

We note that the orientation of the first molecule in the MSS structure affects the phonon dispersion, as certain high-symmetry 
points in the Brillouin zone become inequivalent. To assess this effect, Fig.~\ref{fig15} shows the phonon dispersion of MAPbI$_3$ 
along paths passing through the three inequivalent M points. While the overall dispersions are similar, subtle variations are observed, 
with the most pronounced differences occurring for the ultrasoft flat band along the R-M direction. We note that this low energy band 
does not play an important role in the electron-phonon renormalization of the band gap~\cite{short}.

\subsection{Diffuse scattering} \label{sec.disca_calc}
One-phonon diffuse scattering maps of MAPbBr$_3$, shown in Fig.~\ref{fig6}, were computed using {\tt disca.x} of 
EPW~\cite{Zacharias2021PRB,Zacharias2021PRL,Lee2023}. The Debye-Waller factors at each scattering wavevector $\bQ$ 
were evaluated for a 30$\times$30$\times$30 $\bq$-grid using 
phonon polarization vectors and frequencies obtained by 
Fourier interpolation of the effective IFCs computed at 300~K with ASDM for 2$\times$2$\times$2 supercells.  
A 30$\times$30$\times$1 uniform $\bQ$-grid per
Brillouin zone together with the computed Debye-Waller factors were used to calculate the one-phonon diffuse scattering 
intensity in reciprocal lattice planes perpendicular to the Cartesian $z$-axis. 
Parameters from Ref.~[\onlinecite{Peng_book}] were used to model the atomic scattering amplitudes as a sum of Gaussians.

\subsection{Effective masses} \label{sec.el-ph_calc}
Effective masses were computed at the DFT level including SOC and by evaluating the diagonal components of the 
effective mass tensor via finite differences~\cite{FG_Book}, followed by an isotropic average. 
That is we used the second-order finite difference formula based on the energies at the R point and nearby $\bk$-points
($|\Delta \bk|$ = 0.01 $2\pi$/$\a$) along each Cartesian direction. 
The effective mass enhancement factors due to polymorphism 
were determined as $\lambda = m^*_{\rm poly}/ m^*_{\rm ref} - 1$ and due to thermal effects 
as $\lambda (T) = [m^* + \Delta m^* (T)]/ m^* - 1$. The reduced effective masses were evaluated 
as $\mu = m^*_{\rm h} m^*_{\rm e} / (m^*_{\rm h} + m^*_{\rm e})$.
The slopes describing the temperature variation of the effective masses  are denoted by $\lambda^T$.

\subsection{Electron-phonon calculations} \label{sec.el-ph_calc}

To account for electron-phonon coupling effects at various temperatures, we employed special displacements 
(or ZG displacements)~\cite{Zacharias2020} generated via Eq.~\eqref{eq6} for 4$\times$4$\times$4 supercells. 
We considered a constant temperature approximation for the phonons, i.e. anharmonic phonons computed at a specific 
temperature for each material (specified in Fig.~\ref{fig8}), and the resulting vibrational modes were thermally 
populated at various temperatures to generate special displacements.
We applied the special displacements on the nuclei of both the reference and polymorphous structures leading to the ZG reference and 
ZG polymorphous structures. We note that the initial 4$\times$4$\times$4 reference and polymorphous structures 
(containing 320 atoms for inorganic and 768 atoms for hybrid halide perovskites) were constructed by replicating 
the corresponding structures obtained in 2$\times$2$\times$2 supercells. SOC was included in all electron-phonon coupling calculations.  
To reduce computational cost, the plane wave energy cutoff was set to 60~Ry. Convergence tests performed on a subset 
of materials (MAPbBr$_3$, FAPbBr$_3$, FAPbI$_3$, and MASnI$_3$) using 2$\times$2$\times$2 supercells showed that the band gap values 
differ by less than 1~meV compared to calculations with a 120~Ry cutoff. Furthermore, the Brillouin zone of 
the 4$\times$4$\times$4 supercell was sampled using a 1$\times$1$\times$1 $\bk$-grid. In this case, we found for the 
same subset of materials that the band gap varied by less than 10~meV and the thermal induced renormalization by less than 1~meV.
Figure~\ref{fig13}(c) further shows that the band gap of MAPbBr$_3$ at 300~K is already well converged with
6 ZG polymorphous configurations.

To determine the contribution of electron-phonon coupling to the band gap renormalization and 
effective masses ($dE_{\rm g} / dT|_{\rm EPC}$ and $dm^* / dT|_{\rm EPC}$), we applied linear fits to the 
data points of each material computed at various temperatures without considering 
the effect of thermal expansion. 
Each data point was obtained as a Boltzmann-weighted average [Eq.~\eqref{eq11}] over the band gap of 10 ZG 
reference or 10 ZG polymorphous structures.
Figure~\ref{fig13}(d) shows the convergence of $dE_{\rm g} / dT|_{\rm EPC}$ computed for MAPbBr$_3$ 
as a function of the number of ZG polymorphous configurations used to compute the band gap for each temperature.

Electron-phonon renormalized band gaps of MAPbBr$_3$ in the orthorhombic phase, reported in the companion paper~\cite{short},
were computed using the special displacement method in conjunction with harmonic phonons. Calculations employed a 
4$\times$2$\times$2 supercell of the orthorhombic unit cell (768 atoms), a cutoff of 80 Ry, 
and a 1$\times$1$\times$1 $\bk$-grid, with SOC effects included.

\subsection{Thermal expansion contribution} \label{Sec_TE_comp_details} 

To evaluate the contribution of thermal expansion to the temperature dependence of the band gap, $dE_{\rm g} / dT|_{\rm TE}$,
and effective masses, $dm^* / dT|_{\rm TE}$, we employed the following strategy for each material:
(i) The lattice constants of the initial 10 reference and 10 polymorphous 2$\times$2$\times$2 supercells were 
varied by $\Delta L$ = $\pm$0.04~\AA.
(ii) For each modified lattice, geometry optimizations were performed while keeping the lattice constants fixed, 
yielding new reference and polymorphous configurations.
(iii) Band gaps and effective masses were computed for each individual structure, and averaged over the set 
of configurations corresponding to the same lattice constant.
(iv) The derivatives $dE_{\rm g} / dL$ and $dm^* / dL$ were estimated by fitting a linear function to the 
computed averaged values at $L-\Delta L$, $L$, and $L+\Delta L$, 
equivalent to a central finite difference approximation.
(v) Experimental values of the linear thermal expansion coefficient, $\a =1/L \, d L / d T$, of each material were taken from 
the literature~\cite{Mashiyama2003,Rakita2015,Whitfield2016,Dang2016,Schueller2017,Kontos2018,Marronnier2018,Handa2020,He2021}. 
Due to the absence of experimental data for FAPbCl$_3$, we used $\a = 0.6 \times 10^{-4}$  K$^{-1}$ close to the value 
reported for MAPbCl$_3$~\cite{Handa2020}. 
(vi) Finally, the thermal expansion contributions for each material were evaluated using the relations: 
\begin{eqnarray}\label{eq13}
    \frac{dE_{\rm g}}{dT}\Big|_{\rm TE} &=&  \frac{dE_{\rm g}} {dL} \, \a \, L  \, {\rm \,\,\,\,\, and}\\ 
    \frac{dm^*}{dT}\Big|_{\rm TE} &=&  \frac{dm^*} {dL} \, \a \, L.
\end{eqnarray}

\begin{figure}[b!]
 \begin{center}
\includegraphics[width=0.49\textwidth]{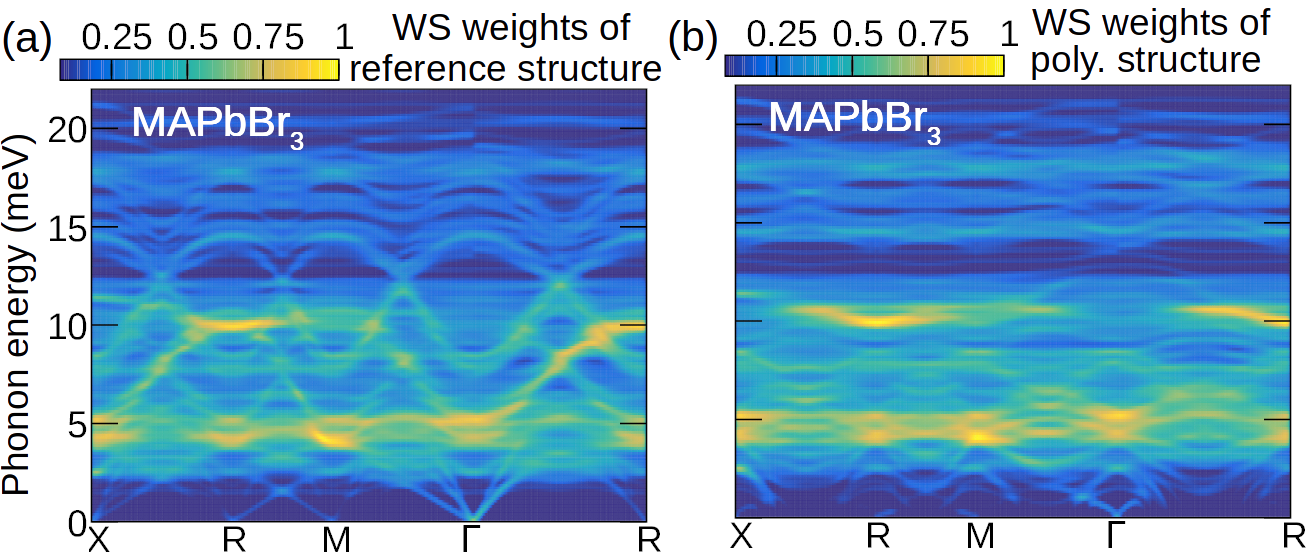}
 \end{center} 
\caption{ (a,b) Phonon spectral functions of one polymorphous configuration of MAPbBr$_3$ calculated using phonon unfolding in 
conjunction with Wigner-Seitz (WS) weights computed for the reference (a) and polymorphous (b) structures.  
\label{fig16} 
}
\end{figure}

\subsection{Phonon spectral functions} \label{sec_ph_spectral_calc}
Phonon spectral functions of the polymorphous structures shown in Fig.~\ref{fig11} were obtained by phonon unfolding 
using the approach described in Ref.~[\onlinecite{Zacharias2023npj}] and the {\tt ZG.x} code of EPW~\cite{Lee2023}. 
We used the IFCs calculated for the polymorphous structures without enforcing any translational or rotational 
symmetries of the unit cell or the reference structures. 
Based on the computed eigenvectors, which reflect spatial correlations introduced by local disorder, we determined 
the spectral weights of each configuration using 394 equally-spaced $\bq$-points along the 
the X-R-M-$\Gamma$-R path of the Brillouin zone. Convergence of the spectral weights was achieved by
using a 12$\times$12$\times$12 $\bg$-grid of reciprocal lattice vectors.  

To obtain the phonon spectral functions of the polymorphous structures using the computed IFCs,
we employed Wigner-Seitz (WS) weights calculated for the reference structures.
These weights are used to determine whether a given interatomic vector lies within the WS cell, ensuring that each 
real-space interaction is counted once and contributes correctly to the phonon properties.
Using the WS weights of the reference structure ensures that each atomic 
interaction is correctly weighted according to the macroscopic symmetry of the lattice. 
In Figs.~\ref{fig16}(a) and (b), we compare the phonon spectral function of polymorphous cubic MAPbBr$_3$ calculated using WS 
weights of the reference and polymorphous structure, respectively. The results show that using the WS weights of the reference structure (i) 
maintains the overdamped and correlated character of the optical phonons and (ii) plays 
an important role in stabilizing the acoustic phonons, particularly near the $\Gamma$ point, where long-wavelength behavior is 
sensitive to translational invariance. In other words, using the WS weights of the reference structure ensures that 
the acoustic sum rule is satisfied and accurately capture bulk-like lattice dynamics without 
compromising the overall behavior of the optical modes. This approach is particularly suitable for systems where 
disorder represents localized perturbations rather than fundamental changes to the macroscopic structure.


\section{Conclusions and outlook}\label{sec.Conclusions}

Based on the ASDM, we developed a unified methodology for electronic structure, anharmonicity, and electron-phonon coupling calculations 
in locally disordered inorganic and hybrid halide perovskites. We carried out a comparative high-throughput study of 12 cubic compounds and analyzed 
the modifications due to local disorder to their electronic structures with and without thermal effects. In addition to generating polymorphous structures, we also 
investigated reference configurations obtained by relaxing only the molecular sublattice, in order to minimize artifacts arising from net dipole moments, 
particularly relevant for MA-based compounds, and to enable systematic comparisons with the fully polymorphous structures. Our calculations reveal strong correlations 
between band gap widening and effective mass enhancement with structural distortions, such as changes in bond angles and lengths, induced by positional polymorphism. 
We further demonstrate that the extent of positional polymorphism is strongly influenced by the size of the A-site cation, a trend that correlates with the 
tolerance factor~\cite{short}, where lower values typically indicate increased structural flexibility and a greater propensity for local distortions.
 We also show the critical role of local disorder in explaining the experimentally observed band gap variations 
across different phases.

We calculated phonon quasiparticle dispersions of MSS reference structures at finite temperatures 
within the SCP theory, and validated our results through comparison with experimental measurements of $\Gamma$ point phonon 
frequencies and phonon-induced diffuse scattering. We examined the influence of A-site cation size on LO-TO splitting, showing that larger 
molecular cations, like FA, result in a less pronounced LO-TO splitting. This arises due to the enhanced 
coupling between the organic and inorganic sublattices. 
Furthermore, we demonstrated the breakdown of the 
phonon quasiparticle picture due to local disorder, leading to strongly correlated and dispersionless optical vibrations, 
consistent with state-of-the-art inelastic neutron scattering, Raman, and THz spectroscopy measurements~\cite{Yaffe2017,Nagai2018,Andrianov2019,Ferreira2020,Anikeeva2023}. 
Hybrid halide perovskites exhibit more strongly overdamped phonons over a broader energy range than their fully inorganic counterparts, which we attribute to the enhanced 
coupling between the molecular and inorganic network.

We evaluated the temperature dependence of band gaps and effective masses using both ZG reference structures 
and ZG polymorphous structures for all cubic compounds, taking into account configurational entropy effects. 
For the band gaps, we achieved excellent agreement with experimental values, highlighting 
the importance of incorporating corrections due to positional polymorphism, exchange-correlation treatments beyond standard DFT, anharmonic electron-phonon coupling, 
and thermal expansion in order to accurately reproduce experimental trends. We further quantified the contributions of electron-phonon coupling and 
thermal expansion to the temperature-induced band gap renormalization, showing that electron-phonon coupling dominates in Cs-based and MA-based compounds, 
whereas thermal expansion plays the leading role in FA-based systems. This behavior is attributed to the larger volumetric thermal expansion coefficients of FA-based compounds. 
In the case of locally disordered FASnI$_3$, we found that the stereochemically active Sn lone pair enhances the thermal expansion contribution, 
which primarily accounts for the largest temperature coefficient of the band gap among the materials studied. Finally, we showed that even 
a single ZG polymorphous configuration, generated from the MSS structure and ASDM phonons, yields results comparable to Boltzmann-weighted averages over 10 ZG polymorphous configurations.

The strong correlations observed between local disorder and electronic properties indicate that band gap tuning and charge transport 
in halide perovskites are intrinsically linked to local structural fluctuations. This finding suggests that engineering local disorder, 
through strategies such as A-site cation selection or strain, could enable systematic optimization of the optoelectronic performance of these materials. 
We emphasize that our approach is readily applicable to mixed-halide perovskites and will be valuable in future studies exploring how A-site, B-site, or X-site 
mixing influences positional polymorphism and, in turn, electronic structure. 
We anticipate that our methodology will complement recent 
experimental efforts~\cite{Dirin2023,Weadock2023,Balvanz2024,He2024,Sabisch2025,Dubajic2025} aimed at elucidating local structural features not only 
in three-dimensional halide perovskites, but also in layered hybrid perovskites and nanocrystals.
Furthermore, our theoretical framework will aid in gaining a deeper understanding of
the fundamental limitations of the carrier transport (lifetimes and diffusion lengths) at elevated temperatures.
It will also help understanding the decoherence limits for the spins of single charge carriers, particularly for bulk materials, or
for excitons, which are more relevant for quantum dots. 
Both phenomena may be crucial: on one hand, for spin transport in spintronic applications and on the other, for single-photon emission in 
quantum optics applications~\cite{Xu2024,Boehme2025}.

To our knowledge, this is the first report of anharmonic quasiparticle phonon dispersions in hybrid halide perovskites, validated against 
experimental measurements. The anharmonic, temperature-dependent interatomic force constants (IFCs), along with the corresponding reference 
and polymorphous structures, which are available in the NOMAD repository~\cite{Zacharias_NOMAD_2025}, provide a foundation for state-of-the-art perturbative 
electron-phonon calculations~\cite{Baroni2001,Giustino2017,Sangalli_2019,Gonze2020,Engel2020,Zhou2021,Yang2022,Lee2023,Li2024} in 
cubic hybrid halide perovskites. These calculations can help elucidate key phenomena such as charge carrier mobilities~\cite{Ponc2019,Ranalli2024}, 
polaron formation~\cite{LafuenteBartolome2024}, as well as non-equilibrium dynamics~\cite{Caruso2021}. Furthermore, our finding that a single ZG 
polymorphous configuration within a $4 \times 4 \times 4$ supercell provides an excellent approximation for nonperturbative electron-phonon 
calculations holds promise for extending this approach to large-scale, high-throughput studies across other hybrid halide perovskite systems.

We emphasize that machine learning force fields (MLFFs)~\cite{Jinnouchi2019,Unke2021,Lahnsteiner2022,Ranalli2024} 
offer significant potential for advancing anharmonic phonon calculations and the exploration of polymorphous structures. 
However, the development of more systematic and efficient training protocols may be necessary to capture the complexity and 
long-range correlated positional polymorphism in hybrid halide perovskites using larger supercells. Furthermore, MLFFs can help 
effectively assess approximations in our electron-phonon calculations, such as the constant temperature 
phonon anharmonicity and the neglect of phonon renormalization due to thermal expansion.

In addition, our calculations revealing the breakdown of the phonon quasiparticle picture in both inorganic and hybrid halide perovskites 
are expected to be of significant interest to recent experimental spectroscopy and scattering 
studies~\cite{Yaffe2017,Nagai2018,Marronnier2018,Fu2018,Andrianov2019,Ferreira2020,Anikeeva2023,Manoli2025}. Our results 
demonstrate that structural disorder, whether originating from surface effects, grain boundaries, or finite-size confinement, 
is intrinsic to single crystals, thin films, nanocrystals, polycrystalline pellets, and nanocrystal-based films, and must be taken 
into account to accurately describe the experimentally observed phonon behavior in 
these systems. Spectroscopic and scattering techniques, when complemented by first-principles calculations that explicitly 
include local disorder, offer a powerful framework not only for understanding the extremely overdamped phonons in ultrasoft halide perovskites, 
but also for elucidating the microscopic origins of their ultralow thermal conductivity~\cite{Pisoni2014,Lee2017,Haeger2020}.

Our comparative study underscores the critical role of data in materials science, particularly when include
thermal effects and positional polymorphism. Capturing such effects is essential for generating datasets that 
reflect a more realistic material behavior which is vital for  
training machine learning models to discover reliable descriptors and trends for 
finite-temperature properties. 
These insights also open a path toward more predictive and transferable models 
for data-driven materials discovery.


\vspace*{0.3cm}

A database of monomorphous, reference, and polymorphous structures of halide perovskites, together with input 
and output files that led to the results of this study, is available in the NOMAD repository~\cite{Zacharias_NOMAD_2025}.

\acknowledgments
This research was funded by the European Union (project ULTRA-2DPK / HORIZON-MSCA-2022-PF-01 /
Grant Agreement No. 101106654). Views and opinions expressed are however those of the authors only and do not necessarily 
reflect those of the European Union or the European Commission. Neither the European Union nor the 
granting authority can be held responsible for them.
We thank D. R. Ceratti for graciously providing data on 
temperature-dependent band gaps of Br-based hybrid halide perovskite single crystals.
J.E. acknowledges financial support from the Institut Universitaire de France.
F.G. was supported by the the Robert A. Welch Foundation under Award No. F-2139-20230405 and by the National Science Foundation under DMREF Grant No. 2119555.
G.V. acknowledges funding from the ANR through the CPJ program and the SURFIN project (ANR-23-CE09-0001), the ALSATIAN project (ANR-23-CE50-0030), and 
the ANR under the France 2030 programme, MINOTAURE project (ANR-22-PETA-0015).
We acknowledge that the results of this research have been achieved using computational resources from the EuroHPC Joint Undertaking 
and supercomputer LUMI [https://lumi-supercomputer.eu/], hosted by CSC (Finland) and the LUMI consortium through a EuroHPC Extreme Scale Access call.
Preliminary calculations were performed at FOTON Institute using HPC resources of TGCC/CINES under the allocation A0140911434, A0160911434, and 
A0180911434 made by GENCI. 

\bibliography{references}{}

\end{document}